\documentclass[11pt]{article}

\usepackage[final]{acl}

\usepackage{times}
\usepackage{amsmath}
\usepackage{latexsym}
\usepackage{adjustbox}
\usepackage{booktabs}
\usepackage{multirow} 
\usepackage{booktabs}
\usepackage{makecell}
\usepackage{tabularx}
\usepackage{siunitx,xcolor}
\usepackage{adjustbox}
\usepackage{amssymb}
\usepackage{pifont}
\usepackage[table,dvipsnames]{xcolor} 
\usepackage{tcolorbox}
\usepackage{subcaption}

\usepackage[T1]{fontenc}

\usepackage[utf8]{inputenc}

\usepackage{microtype}

\usepackage{inconsolata}

\usepackage{graphicx}

\title{TEMPO: A Realistic Multi-Domain Benchmark for Temporal Reasoning-Intensive Retrieval}

\author{
  \textbf{Abdelrahman Abdallah$^{1}$, Mohammed Ali$^{1}$, Muhammad Abdul-Mageed$^{2}$, Adam Jatowt$^{1}$} \\
  $^{1}$University of Innsbruck \qquad $^{2}$University of British Columbia \\
  \texttt{\{abdelrahman.abdallah,mohammed.ali,adam.jatowt\}@uibk.ac.at} \\
  \texttt{muhammad.mageed@ubc.ca}
}

\begin{document}
    \maketitle
\begin{abstract}

Existing temporal QA benchmarks focus on simple fact-seeking queries from news corpora, while reasoning-intensive retrieval benchmarks lack temporal grounding. However, real-world information needs often require reasoning about temporal evolution and synthesizing evidence across time periods. We introduce \textbf{TEMPO}, the first benchmark combining temporal reasoning with reasoning-intensive retrieval across 13 domains. TEMPO features: (1) 1,730 complex queries requiring deep temporal reasoning such as tracking changes, identifying trends, or comparing cross-period evidence; (2) step-wise retrieval planning with 3,976 decomposed steps and gold documents mapped to each step for multi-hop evaluation; and (3) novel temporal metrics including Temporal Coverage@k and Temporal Precision@k measuring whether results span required time periods. Evaluation of 12 retrieval systems reveals substantial challenges: the best model (DiVeR) achieves only 32.0 NDCG@10 and 71.4\% Temporal Coverage@10, demonstrating difficulty in retrieving temporally complete evidence. We believe TEMPO provides a challenging benchmark for improving temporal reasoning in retrieval and RAG systems\footnote{Our code and data are available at \url{https://github.com/tempo-bench/Tempo}. See also our official website: \url{https://tempo-bench.github.io/}.}. 
\end{abstract}

\section{Introduction}

Information retrieval is a fundamental technology that assists users in locating relevant information from extensive corpora~\citep{ali2025sustainableqa,thakur2021beir,abdallah2025dear,abdallah-etal-2025-good,nguyen2016ms}. In real-world applications, many information needs inherently involve temporal dimensions, including understanding how phenomena evolve over time, comparing historical baselines with current states, or tracking changes across multiple time periods~\citep{campos2014survey,joho2014ntcir}. For instance, a cryptocurrency developer might ask "How have block reorganizations changed since 2017?", a legal researcher might need to understand "How has privacy law evolved after GDPR?", or an economist might investigate "What were the trends in quantitative easing before and after the 2008 financial crisis?". In these scenarios, the temporal aspects are not merely supplementary; they carry essential information that fundamentally changes what constitutes a relevant document.

\begin{figure}[t]
    \centering
    \includegraphics[width=0.50\textwidth]{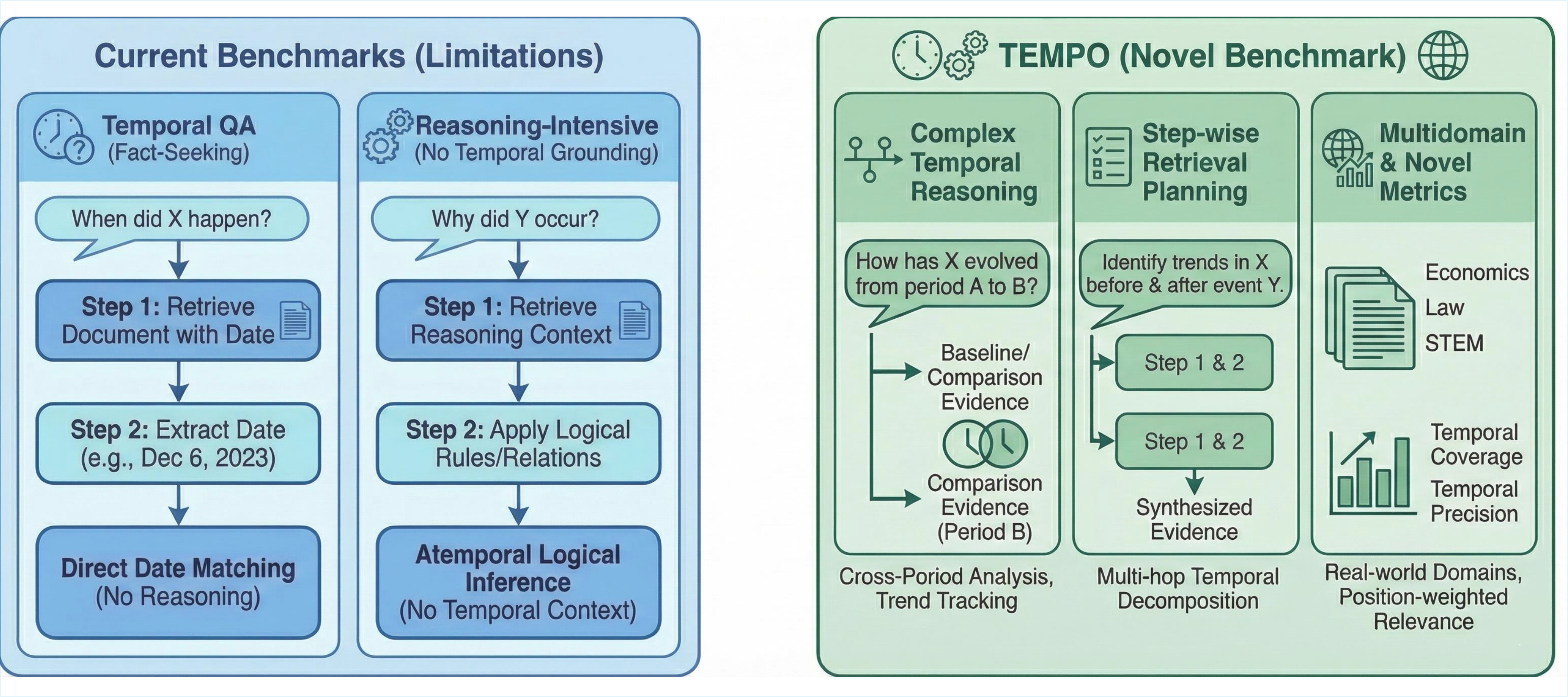}
    \caption{
   \textbf{TEMPO combines temporal reasoning with reasoning-intensive retrieval.}  TEMPO introduce  complex temporal reasoning with cross-period analysis, step-wise retrieval planning, and specialized temporal metrics. 
    }
    \label{fig:intro}
\end{figure}

Despite the prevalence of such complex temporal information needs, existing benchmarks remain inadequate. Temporal QA benchmarks~\citep{chen2025question,wei2025time,mandal2025historybankqa} largely emphasize fact-seeking questions where identifying a date or timestamp is sufficient, whereas reasoning-intensive retrieval benchmarks such as BRIGHT~\citep{su2024bright} and RAR-b~\citep{xiao2024rar} do not explicitly model or evaluate \emph{temporal grounding}---i.e., whether retrieved evidence is aligned with the required time periods and supports cross-period comparison (\autoref{fig:intro}). This leaves a critical gap: how do retrieval systems perform on queries that are \emph{simultaneously} reasoning-intensive and temporally grounded, requiring retrieval of topically relevant evidence \emph{and} temporally appropriate coverage across periods?

\newcommand{\cmark}{\ding{51}}
\newcommand{\xmark}{\ding{55}}

\begin{table}[t]
\centering
\scriptsize
\setlength{\tabcolsep}{2.2pt}
\renewcommand{\arraystretch}{1.08}

\begin{adjustbox}{width=\linewidth,center}
\begin{tabular}{lcccccccc}
\toprule
\textbf{Benchmark} &
\textbf{\#Q} &
\textbf{\#D} &
\textbf{Src.} &
\textbf{Temp.} &
\textbf{Reason.} &
\textbf{Expert} &
\textbf{Step} &
\textbf{Cross} \\
\midrule
\rowcolor{gray!12}\multicolumn{9}{c}{\textit{Reasoning-Intensive Retrieval Benchmarks}} \\
\midrule
BRIGHT              & 1,384  & 12   & Mixed        & \xmark & \cmark & \cmark & \xmark & \xmark \\
RAR-b               & 45,745 & 17   & Mixed        & \xmark & \cmark & \cmark & \xmark & \xmark \\
\midrule
\rowcolor{gray!12}\multicolumn{9}{c}{\textit{Temporal IR Benchmarks}} \\
\midrule
NTCIR Temporalia    & 100    & Open & News/Blogs   & \cmark & \xmark & \xmark & \xmark & \xmark \\
\midrule
\rowcolor{gray!12}\multicolumn{9}{c}{\textit{Temporal QA Benchmarks}} \\
\midrule
TempQuestions       & 1,271  & Open & Freebase     & \cmark & \xmark & \xmark & \cmark & \xmark \\
ChronoQA            & 5,176  & Open & News (CN)    & \cmark & \xmark & \xmark & \xmark & \xmark \\
TIME                & 38,522 & 3    & Wiki/News/D  & \cmark & \xmark & \xmark & \xmark & \xmark \\
HistoryBankQA       & 535K   & 10   & Wikipedia    & \cmark & \xmark & \xmark & \xmark & \xmark \\
ComplexTempQA       & 100M+  & Open & Wikipedia    & \cmark & \xmark & \xmark & \cmark & \xmark \\
\midrule
\textbf{TEMPO (Ours)} & \textbf{1,730} & \textbf{13} & \textbf{Stack Exch.} &
\cmark & \cmark & \cmark & \cmark & \cmark \\
\bottomrule
\end{tabular}
\end{adjustbox}
\caption{
Comparison of TEMPO with existing temporal reasoning and retrieval benchmarks. TEMPO combines temporal reasoning, complex retrieval, and step-wise evaluation in different domains.
\textbf{Column legend:} Src.=Source Data; Temp.=Temporal Reasoning; Reason.=Reasoning-Intensive; Expert=Technical/Expert; Step=Multi-Hop/Step-Wise; IR=Retrieval Task; Cross=Cross-Period Analysis.
}
\label{tab:tempo-benchmark-comparison}
\end{table}

In this work, we address this gap by introducing \textbf{TEMPO}, a benchmark for \emph{reasoning-intensive retrieval with explicit temporal requirements} across 13 domains. Prior temporal QA benchmarks primarily evaluate answer generation and often reduce to locating a date, while existing reasoning-intensive retrieval benchmarks (e.g., BRIGHT, RAR-b) do not require temporal alignment or cross-period evidence. TEMPO targets queries that are \emph{simultaneously} reasoning-intensive and temporally grounded, emphasizing retrieval of topically relevant evidence that is also appropriate across the required time periods.
It consists of 1,730 naturally occurring Stack Exchange queries spanning blockchain (Bitcoin, Cardano, IOTA, Monero), social sciences (Economics, Law, Politics, History), applied domains (Quantitative Finance, Travel, Workplace, Genealogy), and STEM (History of Science and Mathematics).

To evaluate systems, we define two retrieval tasks:
\textbf{(1) Query $\to$ Documents}: Traditional temporal retrieval with 1,730 queries, evaluating whether systems can retrieve temporally relevant documents that address complex temporal information needs;
\textbf{(2) Query $\to$ Step $\to$ Documents}: Multi-step temporal reasoning with 1,605 queries decomposed into 3,976 retrieval steps, testing whether systems can follow step-wise retrieval plans where each step targets specific time periods or aspects of the query.

\begin{figure*}[t!]
  \centering
  \includegraphics[width=0.85\textwidth]{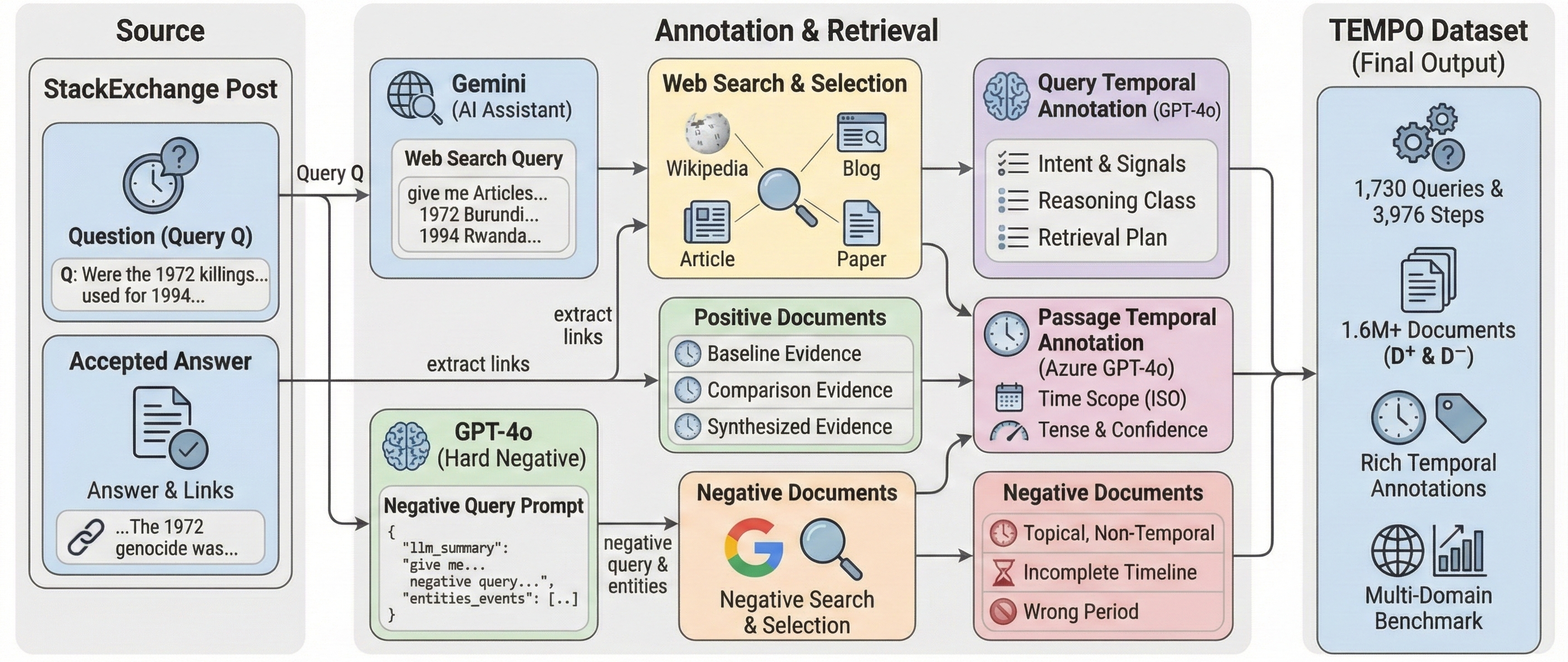}
  \caption{
  Overview of TEMPO construction. We collect temporally grounded Stack Exchange queries, curate and verify positive evidence (from answer links and Gemini-assisted web search), and mine hard negatives via GPT-4o queries targeting topically similar but temporally mismatched documents.
  }
  \label{fig:annotation}
\end{figure*}


Our key contributions include three novel components that distinguish TEMPO from prior work. First, we provide \textbf{comprehensive temporal annotations} at three levels: (i) \textit{query-level with 10 fine-grained temporal reasoning classes} (e.g., trend changes and cross-period, event analysis and localization, causation analysis), temporal intent classification, temporal signals, and key time anchors; (ii) \textit{step-wise retrieval planning} that decomposes queries into sequential steps mapped to gold documents, enabling multi-hop temporal evaluation; and (iii) \textit{passage-level annotations} with temporal signals, events, and ISO-formatted time scopes. Second, we introduce new temporal evaluation metrics designed to capture temporal reasoning aspects missed by traditional IR metrics: Temporal Precision@k uses LLM-as-judge evaluation to measure temporal relevance quality; Temporal Coverage@k assesses whether top-k results span required baseline and comparison time periods for cross-period queries. Third, our \textbf{real-world complex queries} average approximately 300 words, 
featuring relevant documents and challenging hard negatives mined through multi-LLM query reformulation and web search.

\section{Related Work}

Early work on temporal query classification~\citep{joho2014ntcir,joho2016building,campos2014survey} produced benchmarks like NTCIR Temporalia. They rely on news/blog corpora where timestamps and basic temporal expressions suffice.
As illustrated in Table~\ref{tab:tempo-benchmark-comparison}, existing temporal QA benchmarks have made progress but remain limited for evaluating temporally grounded retrieval. 
Recent efforts include TempQuestions~\citep{jia2018tempquestions}, ComplexTempQA~\citep{gruber-etal-2025-complextempqa}, TIME~\citep{wei2025time}, HistoryBankQA~\citep{mandal2025historybankqa}, and ChronoQA~\citep{chen2025question}. 
While these datasets go beyond simple timestamp lookup (e.g., temporal constraints, ordering, and cross-time comparisons), they are primarily \emph{answer-generation} benchmarks~\cite{piryani2025s,wallat2025study,abdallah2025tempretriever,qian2024timer4,xu2024crp,brown2025systematic} and do not explicitly evaluate whether retrieval results provide \emph{temporally aligned} evidence that covers all required time periods for cross-period analysis.
We provide more description of related work in Appendix \ref{app:benchmark_comparison} and \ref{app:benchmark_comparison_1}.



\section{TEMPO Dataset}
\label{sec:construct}

In this section, we first formulate the task (\S\ref{sec:formulation}), then detail the data collection and annotation process from Stack Exchange (\S\ref{sec:collection}). Data statistics are presented in Table~\ref{tab:dataset_stats}.
\subsection{Task Formulation}
\label{sec:formulation}

Given a temporal query $Q = Q_{\text{text}}$ and a retrieval corpus $\mathcal{D} = \{D_1, \ldots, D_n\}$, retrievers are tasked to find temporally relevant documents $\mathcal{D}^+_Q = \{D_{Q,1}^+, \ldots, D_{Q,m}^+\} \subset \mathcal{D}$ where $m \ll n$. Negative documents are defined as $\mathcal{D}_Q^- = \mathcal{D} \setminus \mathcal{D}_Q^+$. In temporal reasoning-intensive retrieval, the relevant document set $\mathcal{D}^+_Q$ is connected to query $Q$ through temporal reasoning traces involving temporal evolution understanding, cross-period analysis, and temporal dependency resolution, rather than simple keyword matching or date filtering.

\textbf{Query-Level Annotations.} Each query $Q$ is annotated with a tuple $\mathcal{A}_Q = (\tau, \mathcal{S}_Q, \mathcal{E}_Q, \rho, \mathcal{P}, \mathcal{T})$ where: $\tau \in \mathcal{I}$ denotes temporal intent from the set $\mathcal{I} = \{when, duration, order, before\_after, \allowbreak ongoing\_status, period\_definition, timeline\}$; $\mathcal{S}_Q = \{s_1, \ldots, s_k\}$ is the set of temporal signals (e.g., ``since 2017'', ``before the war''); $\mathcal{E}_Q = \{e_1, \ldots, e_l\}$ is the set of temporal events; $\rho \in \mathcal{R}$ is the primary temporal reasoning class from 10 categories (see Appendix~\ref{sec:reasoning_class_appendix}); $\mathcal{P} = \{(p_1, \mathcal{D}_{p_1}^+), \ldots, (p_j, \mathcal{D}_{p_j}^+)\}$ is the step-wise retrieval plan mapping each step $p_i$ to its gold documents $\mathcal{D}_{p_i}^+$; and $\mathcal{T} = \{t_1, \ldots, t_h\}$ is the set of key time anchors.

\textbf{Passage-Level Annotations.} Each document $D \in \mathcal{D}$ is annotated with $\mathcal{A}_D = (\mathcal{S}_D, \mathcal{E}_D, [t_s, t_e], \phi)$ where: $\mathcal{S}_D$ and $\mathcal{E}_D$ denote temporal signals and events in the passage; $[t_s, t_e]$ represents the temporal scope as ISO-formatted start and end dates; $\phi \in \{\texttt{past}, \texttt{present}, \texttt{future}, \texttt{mixed}\}$ indicates the dominant tense.



\subsection{Data Collection from Stack Exchange}
\label{sec:collection}

StackExchange\footnote{\url{https://stackexchange.com/}} is a community-driven platform where domain experts ask and answer complex technical questions. We select 13 diverse domains spanning blockchain (Bitcoin, Cardano, IOTA, Monero), social sciences (Economics, Law, Politics, History), applied fields (Quantitative Finance, Travel, Workplace, Genealogy), and STEM (History of Science and Mathematics). StackExchange posts we collect contain detailed temporal descriptions requiring reasoning about how phenomena evolved, changed over time, or differ across periods. We construct query-document pairs based on user posts and documents referenced in answers (Figure~\ref{fig:annotation}).

\textbf{Selecting posts.} Human annotators\footnote{Five PhD and two master students.} browse posts from newest to oldest and select posts that: (1) have at least one answer that is either accepted by the user or receives $>10$ votes, and (2) require \emph{temporal reasoning} as defined by our query-level taxonomy $\rho$ (Appendix~\ref{sec:reasoning_class_appendix}), e.g., event localization, time-period contextualization, and cross-period comparison/trend analysis. The distribution of selected queries across these temporal reasoning categories is shown in Figure~\ref{fig:reasoning_distribution}, and we later analyze retrieval difficulty by category (Figure~\ref{fig:reasoning_class_bars}), where cross-period comparison queries are consistently more challenging than single-period temporal localization. Detailed examples in Appendix~\ref{app:dataset_examples}

\textbf{Constructing query and positive documents.} For each selected post, annotators use the title and body text to form the query $Q$. Annotators visit web pages linked in the answers and use Gemini (Google's AI assistant) to \textit{return relevant web documents (which are not AI-generated)} by prompting: \textit{"Give me articles from the internet to answer this query: [post content]"}. For each discovered web page, annotators extract passages that provide critical temporal information for answering the query. Sources include Wikipedia, technical blogs, research articles, official documentation, and news sites. If no temporally relevant documents are found, the post is discarded.

\textbf{Constructing hard negative documents.} To prevent models from relying on simple semantic matching, we ensure negative documents are topically related but temporally incomplete or irrelevant. We use GPT-4o to analyze each post and generate a search query designed to find hard negatives, along with entities and events mentioned in the post (prompt details in Appendix~\ref{app:negative_mining_prompt}). Annotators use the generated query to search Google and collect hard negative passages per query. They extract passages that are topically related but do not provide the temporal reasoning steps or time-period coverage needed to answer the query.
\begin{table}[t!]

\centering
\resizebox{0.42\textwidth}{0.12\textheight}{
\begin{tabular}{l|rrr|rr|c}
\toprule
& \multicolumn{3}{c|}{\textbf{Total Number}} & \multicolumn{2}{c|}{\textbf{Avg. Length}} & \textbf{Avg.}\\
\cmidrule{2-6}
\textbf{Dataset} &
\multicolumn{1}{c}{$\mathbf{Q}$} &
\multicolumn{1}{c}{$\boldsymbol{\mathcal{D}}$} &
\multicolumn{1}{c}{$\boldsymbol{\mathcal{D}^+}$} &
\multicolumn{1}{c}{$\mathbf{Q}$} &
\multicolumn{1}{c|}{$\boldsymbol{\mathcal{D}}$} &
\textbf{Steps} \\
\midrule
\rowcolor{gray!12}\multicolumn{7}{c}{\textit{\textbf{Blockchain}}} \\
\midrule
Bitcoin  & 100 & 153{,}291 & 3.3  & 222.0 & 596.9 & 2.93 \\
Cardano  &  51 &  87{,}201 & 2.5  & 161.1 & 647.2 & 2.84 \\
Iota     &  10 &  10{,}372 & 3.8  & 148.6 & 1{,}036.5 & 3.20 \\
Monero   &  65 &  85{,}093 & 2.6  & 171.8 & 703.3 & 2.72 \\
\midrule
\rowcolor{gray!12}\multicolumn{7}{c}{\textit{\textbf{Social Sciences}}} \\
\midrule
Economics &  83 &  93{,}756 & 3.6  & 290.2 & 495.9 & 3.08 \\
Law       &  35 &  43{,}288 & 3.0  & 258.5 & 500.7 & 3.23 \\
Politics  & 150 & 183{,}394 & 2.7  & 343.2 & 476.3 & 3.35 \\
History   & 801 & 356{,}493 & 4.5  & 374.2 & 682.7 & 3.42 \\
\midrule
\rowcolor{gray!12}\multicolumn{7}{c}{\textit{\textbf{Applied}}} \\
\midrule
Quant      &  34 &  28{,}785 & 2.4  & 422.5 & 477.1 & 2.68 \\
Travel     & 100 & 177{,}677 & 2.6  & 264.5 & 374.8 & 3.11 \\
Workplace  &  36 &  64{,}659 & 2.8  & 291.8 & 368.7 & 2.42 \\
Genealogy  & 115 & 156{,}228 & 2.8  & 359.6 & 629.2 & 3.78 \\
\midrule
\rowcolor{gray!12}\multicolumn{7}{c}{\textit{\textbf{STEM}}} \\
\midrule
HSM & 150 & 213{,}818 & 2.5  & 303.5 & 563.6 & 3.25 \\
\midrule
\textbf{Total} & \textbf{1{,}730} & \textbf{1{,}654{,}055}  & -- & -- & -- & -- \\
\bottomrule
\end{tabular}
}
\caption{TEMPO dataset statistics across 13 domains. $\mathbf{Q}$: number of queries; $\boldsymbol{\mathcal{D}}$: corpus size (total documents); $\boldsymbol{\mathcal{D}^+}$: average number of positive (relevant) documents per query; Avg.\ Length: average token count for queries and documents; Steps: average number of steps. Domains are grouped into four categories: Blockchain, Social Sciences, Applied, and STEM.}

\label{tab:dataset_stats}
\end{table}


\subsection{Temporal Annotation and Quality Control}
\label{sec:annotation}
\begin{figure}[t!]
    \centering
    \includegraphics[width=0.80\columnwidth]{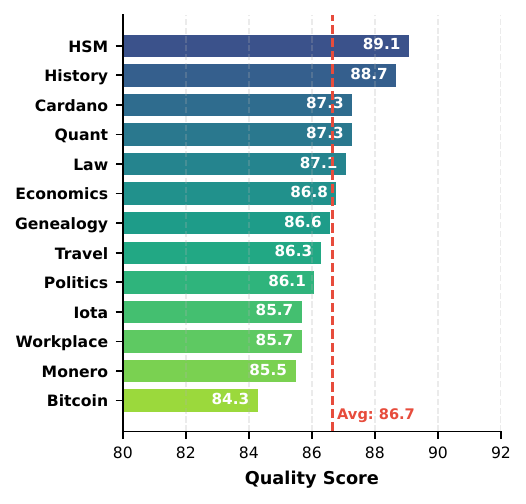}
    \caption{Dataset quality validation using Qwen-72B as LLM judge. 
    }
    \label{fig:quality_validation}
\end{figure}

\textbf{Temporal annotations.} For comprehensive temporal evaluation, we annotate queries and passages at multiple levels using GPT-4o, with human annotators reviewing a sample to ensure annotation quality. Specifically, two expert annotators independently verified a random sample of 200 queries (11.6\%) and their associated annotations, measuring inter-annotator agreement using Cohen's Kappa. We achieved $\kappa = 0.82$ for temporal reasoning class assignment, $\kappa = 0.78$ for temporal intent classification, and $\kappa = 0.85$ for gold document relevance judgments. Additionally, we employ Qwen-72B as an independent LLM judge to evaluate alignment between queries, retrieval steps, and gold documents on a 0--100 scale. As shown in Figure~\ref{fig:quality_validation}, TEMPO achieves an average quality score of  86.7 across all domains (range: 84.3--89.1), further validating annotation quality.

\begin{figure}[h!]
  \centering
  \includegraphics[width=.8\columnwidth]{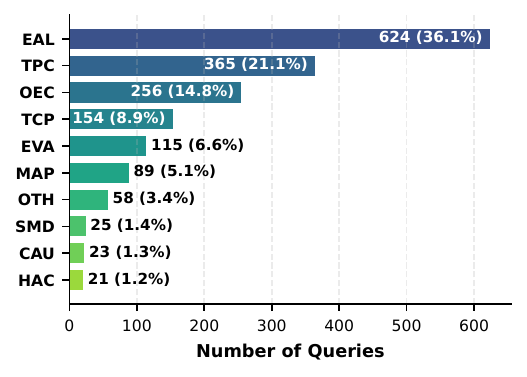}
  \caption{Distribution of temporal reasoning classes. 
  See Appendix~\ref{sec:reasoning_class_appendix} for class definitions.}
  \label{fig:reasoning_distribution}
\end{figure}

At the \textit{query level}, we extract temporal intent (when/duration/order/before\_after/ongoing\_status/\allowbreak period\_definition/timeline/none), temporal signals (phrases like "since 2017", "nowadays"), temporal events, temporal reasoning class (10 fine-grained categories), retrieval plan with sequential steps, key time anchors, expected granularity, and quality checks (prompt in Appendix~\ref{app:query_annotation_prompt}). At the \textit{passage level}, we extract temporal signals, temporal events, time mentions, time scope (start/end ISO dates with granularity), tense, and confidence score (prompt in Appendix~\ref{app:passage_annotation_prompt}). This multi-level annotation enables fine-grained temporal evaluation beyond traditional IR metrics.
\textbf{Step-wise retrieval planning.} For \textbf{Task 2}, we use the step-wise retrieval plans described above, mapping each step to step-specific gold documents. Each step describes a specific retrieval action (e.g., "Retrieve historical baseline statistics from 2013-2017", "Retrieve current statistics from 2020-present") and is mapped to gold documents that satisfy that step. This enables evaluation of multi-hop temporal reasoning where systems must retrieve evidence from multiple time periods in sequence.


\textbf{Temporal reasoning classification.} We categorize queries into 10 fine-grained temporal reasoning classes based on the type of temporal inference required (Figure~\ref{fig:reasoning_distribution}; see Appendix~\ref{sec:reasoning_class_appendix} for full definitions). The dataset is dominated by \textit{Event Analysis \& Localization} (EAL; 624 queries, 36.1\%), which requires pinpointing when events occurred and understanding their temporal context, followed by \textit{Time Period Contextualization} (TPC; 365 queries, 21.1\%), which situates phenomena within specific historical periods. \textit{Origins \& Evolution Comparative} (OEC; 256 queries, 14.8\%) and \textit{Trends \& Cross-Period Comparison} (TCP; 154 queries, 8.9\%) queries require tracking how concepts evolved over time or comparing states across periods. The remaining categories, including Event Verification (EVA), Materials \& Artifacts Provenance (MAP), Sources \& Methods Documentation (SMD), Causation Analysis (CAU), and Historical Attribution \& Context (HAC) represent more specialized temporal reasoning patterns. We analyze retrieval performance across these reasoning classes in \S\ref{sec:reasoning_class_analysis}.

\begin{table*}[t!]
\centering
\small

\resizebox{0.8\textwidth}{!}{%
\begin{tabular}{lcccccccccccc}
\toprule
\textbf{Domain} & \textbf{BM25} & \textbf{BGE} & \textbf{Contriever} & \textbf{DiVeR} & \textbf{E5} & \textbf{GritLM} & \textbf{Inst-L} & \textbf{Qwen} & \textbf{Rader} & \textbf{ReasonIR} &  \textbf{SBERT} & \textbf{SFR} \\
\midrule
\rowcolor{gray!12}\multicolumn{13}{c}{\textit{Blockchain}} \\
\midrule
\textbf{Cardano} & 13.4 & 13.1 & 12.1 & \underline{29.3} & \textbf{35.7} & 21.7 & 14.6 & 20.6 & 18.6 & 22.9 & 21.4 & 28.1 \\
\textbf{Iota} & 9.7 & 36.1 & \underline{38.3} & 38.2 & \textbf{41.7} & 36.6 & 34.3 & 28.6 & 19.2 & \textbf{41.7} & 33.2 & 37.1 \\
\textbf{Monero} & 2.8 & 14.5 & 9.9 & 20.3 & 20.0 & 14.7 & 16.9 & 11.0 & \underline{21.0} & 19.6 & 15.1 & \textbf{23.7} \\
\textbf{Bitcoin} & 6.2 & 14.4 & 13.3 & 17.4 & 16.3 & \textbf{19.1} & 15.7 & 11.4 & 14.9 & 16.3 & 14.3 & \underline{17.6} \\
\midrule
\rowcolor{gray!12}\multicolumn{13}{c}{\textit{Social Sci.}} \\
\midrule
\textbf{Economics} & 5.8 & 12.6 & 16.3 & \textbf{27.8} & \underline{25.0} & 17.2 & 17.5 & 17.1 & 22.7 & 20.0 & 15.3 & 21.9 \\
\textbf{Law} & 12.7 & 31.9 & 28.1 & \underline{40.4} & 34.0 & 38.3 & 37.3 & 32.0 & 33.5 & 37.9 & 33.8 & \textbf{40.8} \\
\textbf{Politics} & 32.7 & 28.2 & 31.6 & \underline{45.5} & \textbf{47.9} & 41.4 & 32.6 & 38.1 & 32.4 & 35.4 & 34.6 & 44.9 \\
\textbf{History} & 9.2 & 27.4 & 26.5 & \textbf{34.5} & 28.7 & 27.3 & 28.5 & 25.6 & 25.8 & 34.3 & 28.7 & \underline{32.4} \\
\midrule
\rowcolor{gray!12}\multicolumn{13}{c}{\textit{Applied}} \\
\midrule
\textbf{Quant} & 2.5 & 11.7 & 11.1 & \underline{27.2} & 13.8 & 21.6 & 14.6 & 12.7 & \textbf{27.8} & 19.5 & 15.7 & 16.8 \\
\textbf{Travel} & 4.6 & 23.8 & 23.7 & 26.8 & \underline{28.3} & 25.0 & 25.0 & 22.0 & 26.1 & 21.4 & 27.3 & \textbf{29.7} \\
\textbf{Workplace} & 6.2 & 27.2 & 23.9 & \textbf{42.6} & 32.9 & 30.8 & 36.2 & 30.3 & \underline{36.6} & 30.0 & 34.6 & 31.6 \\
\textbf{Genealogy} & 13.3 & 22.0 & 24.9 & \textbf{35.6} & \underline{33.5} & 26.9 & 24.6 & 25.3 & 18.7 & 30.3 & 23.5 & 31.7 \\
\midrule
\rowcolor{gray!12}\multicolumn{13}{c}{\textit{STEM}} \\
\midrule
\textbf{HSM} & 21.2 & 23.2 & 18.9 & 31.0 & \textbf{37.7} & 33.4 & 24.4 & 21.3 & 16.9 & 24.7 & 26.1 & \underline{33.5} \\
\midrule
\midrule
\textbf{Avg.} & 10.8 & 22.0 & 21.4 & \textbf{32.0} & \underline{30.4} & 27.2 & 24.8 & 22.8 & 24.2 & 27.2 & 24.9 & 30.0 \\
\bottomrule
\end{tabular}}
\caption{NDCG@10 performance of retrieval models on TEMPO across 13 domains: Blockchain (Bitcoin, Cardano, Iota, Monero), Social Sciences (Economics, Law, Politics, History), Applied (Quant, Travel, Workplace, Genealogy), and STEM (HSM: History of Science and Mathematics). Avg. denotes the average score across all domains. The best score on each domain is shown in \textbf{bold} and the second best is \underline{underlined}. 
}\label{tab:results_all_domains}
\end{table*}

\begin{figure}[t]
    \centering
    \includegraphics[width=0.8\linewidth]{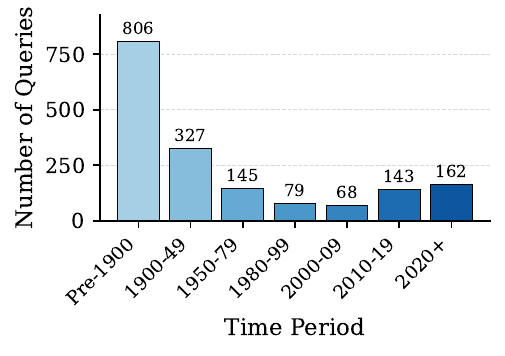}
    \caption{Overall distribution of query temporal anchors across time periods in TEMPO.}
    \label{fig:total_dist}
\end{figure}

\textbf{Temporal Distribution.} TEMPO spans Pre-1900 to 2020+ (Figure \ref{fig:total_dist}), emphasizing cross-period reasoning over simple news retrieval. The distribution features historical queries: 806 queries from Pre-1900 and 327 from 1900–49. However, it maintains strong modern representation, including 143 queries for 2010–19 and 162 for 2020+. This breadth facilitates evaluating both long-term evolutionary patterns and contemporary dynamics. See Appendix \ref{app:domain_temporal} for domain-specific breakdowns.

\begin{table}[t]
\centering
\small

\resizebox{0.90\columnwidth}{0.13\textheight}{%
\begin{tabular}{lcccc}
\toprule
\textbf{Model} & \textbf{TP@10} & \textbf{TR@10} & \textbf{TC@10} & \textbf{NDCG|FC@10} \\
\midrule
\rowcolor{gray!12}\multicolumn{5}{c}{\textit{Sparse model}} \\
\midrule

BM25 & 24.0 & 11.1 & 32.8 & 22.5 \\

\midrule
\rowcolor{gray!12}\multicolumn{5}{c}{\textit{Open-sourced models (<1B)}} \\
\midrule
BGE & 53.7 & 34.0 & 66.1 & 25.3 \\
Contriever & 49.6 & 30.3 & 64.1 & 26.2 \\
Inst-L & 53.1 & 33.8 & 66.9 & 25.8 \\
SBERT & 54.1 & 35.0 & 67.2 & 26.9 \\

\midrule
\rowcolor{gray!12}\multicolumn{5}{c}{\textit{Open-sourced models (>1B)
}} \\
\midrule
E5 & 53.5 & 31.1 & 63.2 & \textbf{33.4} \\
GritLM & 53.8 & 35.0 & 69.1 & 27.2 \\
Qwen & 46.2 & 28.8 & 61.0 & 29.9 \\
\midrule
\rowcolor{gray!12}\multicolumn{5}{c}{\textit{Reasoning models (>1B)
}} \\
\midrule
DiVeR & \textbf{62.0} & \textbf{41.3} & \underline{71.4} & \underline{32.5} \\
Rader & 50.6 & 32.3 & 66.1 & 26.1 \\
ReasonIR & \underline{57.4} & \underline{38.2} & \textbf{72.4} & 28.8 \\
\bottomrule
\end{tabular}}

\caption{Temporal evaluation metrics at rank 10 averaged across all domains. TP@10: Temporal Precision (position-weighted precision of temporally relevant documents); TR@10: Temporal Relevance; TC@10: Temporal Coverage; NDCG|FC@10: NDCG conditioned on full temporal coverage. 
}
\label{tab:temporal_avg_metrics}
\end{table}

\begin{table}[t]
\centering

\resizebox{0.85\columnwidth}{0.13\textheight}{%
\begin{tabular}{lcccc}
\toprule
\textbf{Model} & \textbf{Baseline} & \textbf{Strip} & \textbf{Temporal-Only} & \textbf{Normalized} \\
\midrule
\rowcolor{gray!12}\multicolumn{5}{c}{\textit{Sparse}} \\
\midrule
BM25 & 10.8 & 10.4 & 11.0 & 10.9 \\
\midrule
\rowcolor{gray!12}\multicolumn{5}{c}{\textit{Dense (<1B)}} \\
\midrule
BGE & 22.0 & 19.5 & 9.6 & 22.1 \\
Contriever & 21.4 & 18.9 & 10.2 & 22.4 \\
Inst-L & 24.8 & 23.2 & 13.5 & 24.7 \\
SBERT & 24.9 & 23.3 & 8.7 & 23.5 \\
\midrule
\rowcolor{gray!12}\multicolumn{5}{c}{\textit{Dense (>1B)}} \\
\midrule
E5 & \underline{30.4} & 27.0 & 14.2 & 29.1 \\
GritLM & 27.2 & 24.9 & 15.2 & 27.1 \\
Qwen & 22.8 & 21.2 & \underline{17.2} & 23.3 \\
SFR & 30.0 & \underline{27.8} & 16.3 & 29.5 \\
\midrule
\rowcolor{gray!12}\multicolumn{5}{c}{\textit{Reasoning}} \\
\midrule
DiVeR & \textbf{32.0} & \textbf{29.9} & \textbf{17.7} & \underline{30.2} \\
Rader & 24.2 & 21.8 & 9.4 & 21.6 \\
ReasonIR & 27.3 & 25.1 & 13.8 & \textbf{35.3} \\
\bottomrule
\end{tabular}}
\caption{Study on temporal query variants (NDCG@10 averaged across all domains). Baseline: original queries; Strip: temporal signals removed; Temporal-Only: only temporal information retained; Normalized: explicit temporal intent tags added. 
}\label{tab:ablation_variants}
\end{table}

\section{Temporal Evaluation Metrics}
\label{sec:metrics_main}

Traditional IR metrics (e.g., NDCG@k) do not measure whether retrieved evidence is temporally appropriate or spans the required time periods. We therefore report four temporal metrics computed using an LLM-as-judge that labels (i) whether a retrieved document is temporally relevant to the query and (ii) which required period(s) it supports (Appendix~\ref{app:temporal_metrics}). \textbf{TP@k (Temporal Precision@k).} A position-weighted metric that rewards ranking temporally relevant documents earlier in the top-$k$.  
\textbf{TR@k (Temporal Relevance@k).} The fraction of the top-$k$ documents judged temporally relevant.  
\textbf{TC@k (Temporal Coverage@k).} The fraction of required time periods covered by at least one document in the top-$k$ (we use two periods in our main experiments).  
\textbf{NDCG|FC@k.} NDCG@k computed only on queries where full temporal coverage is achieved (i.e., TC@k $=1$).






\section{Experimental}
\label{sec:results}

\subsection{Experimental Setup}

We evaluate 12 representative retrieval models spanning sparse, dense, and reasoning-enhanced architectures. All experiments use NDCG@10 as the primary metric following prior IR benchmarks~\citep{abdallah2025rankify,thakur2021beir,nguyen2016ms,su2024bright}. While we focus on NDCG@10 for ranking evaluation, a comprehensive set of additional metrics—including Precision, Recall, Mean Average Precision (MAP), and Mean Reciprocal Rank (MRR)—is provided in Appendix~\ref{app:additional_metrices}.

\textbf{Sparse retrieval.} We use BM25~\citep{robertson2009probabilistic} as our lexical baseline, which remains competitive on traditional retrieval benchmarks despite its simplicity.
\textbf{Dense retrieval.} We evaluate BGE~\citep{bge-m3}, Contriever~\citep{izacard2021unsupervised}, E5-Mistral~\citep{wang2022text}, GritLM~\citep{muennighoff2024generative}, Inst-L~\citep{su2024bright}, Qwen~\citep{li2023towards}, SBERT~\citep{reimers2019sentencebertsentenceembeddingsusing}, and SFR~\citep{meng2024sfrembedding}.
\textbf{Reasoning-enhanced retrievers.} Given TEMPO's emphasis on temporal reasoning, we evaluate three specialized models designed to incorporate logical inference during retrieval: DiVeR~\citep{diver}, Rader~\citep{das2025rader}, and ReasonIR~\citep{shao2025reasonir}.

\textbf{Temporal metrics.} We report TP@10, TR@10, TC@10, and NDCG|FC@10; formal definitions and the LLM judging prompts are provided in Appendix~\ref{app:temporal_metrics}. In our evaluation we instantiate TC@k with $M{=}2$ periods (baseline vs.\ comparison), but the definition generalizes to $M{>}2$.

\paragraph{Step-wise Evaluation}
Each Task~2 query $q$ has $S_q$ steps $\{p_{q,i}\}_{i=1}^{S_q}$ with step-specific gold documents $\mathcal{D}^{+}_{q,i}$. We evaluate at the step level by retrieving for each step and computing NDCG@10 against $\mathcal{D}^{+}_{q,i}$, then macro-averaging over steps and queries. Step-Only uses only $p_{q,i}$, Query+Step uses $q\oplus p_{q,i}$, and Query+All retrieves once with $q\oplus p_{q,1}\oplus\cdots\oplus p_{q,S_q}$ and evaluates the same ranking against every step.



\begin{figure}[h]
    \centering
    \includegraphics[width=.5\textwidth]{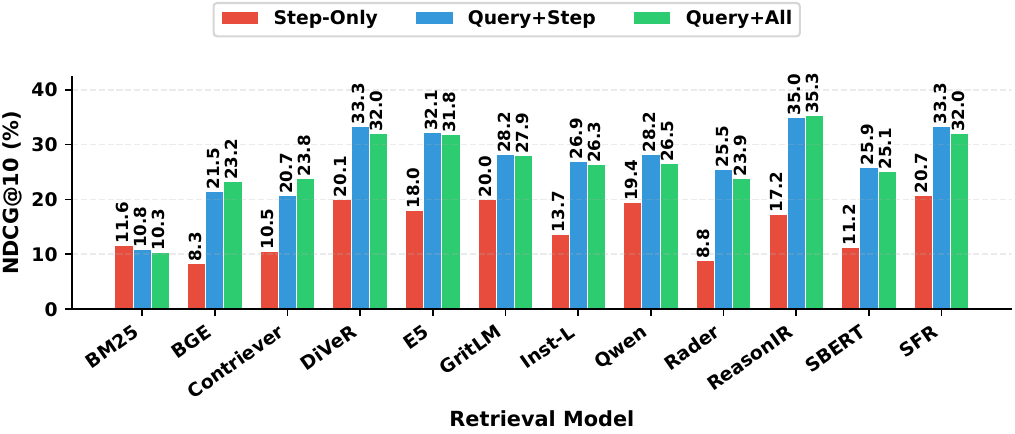}
    \caption{Comparison of step-wise retrieval strategies across 12 retrieval models. 
    }
    \label{fig:stepwise_comparison}
\end{figure}

\subsection{Main Results}

\paragraph{Existing retrieval systems perform poorly on TEMPO:} Table \ref{tab:results_all_domains} presents NDCG@10 performance across all 13 domains. Reasoning-enhanced models achieve the best results, with DiVeR leading at 32.0 NDCG@10, followed by E5 (30.4) and SFR (30.0). The sparse baseline BM25 achieves only 10.8, demonstrating the inadequacy of keyword matching for retrieval that requires temporal reasoning. Dense retrievers show substantial improvements (21.4--30.4 range), while reasoning-enhanced models reach 27.3--32.0, highlighting the value of explicit reasoning mechanisms. Domain difficulty varies substantially. History 
achieves only 34.5 NDCG@10 with DiVeR. Politics and Law reach higher scores (47.9 and, 40.8 respectively), suggesting differences in temporal reasoning complexity. Blockchain domains show mixed results: Iota achieves 41.7 (ReasonIR), while Monero proves considerably harder at 23.7 (SFR).
\begin{figure}[h]
    \centering
    \includegraphics[width=.5\textwidth]{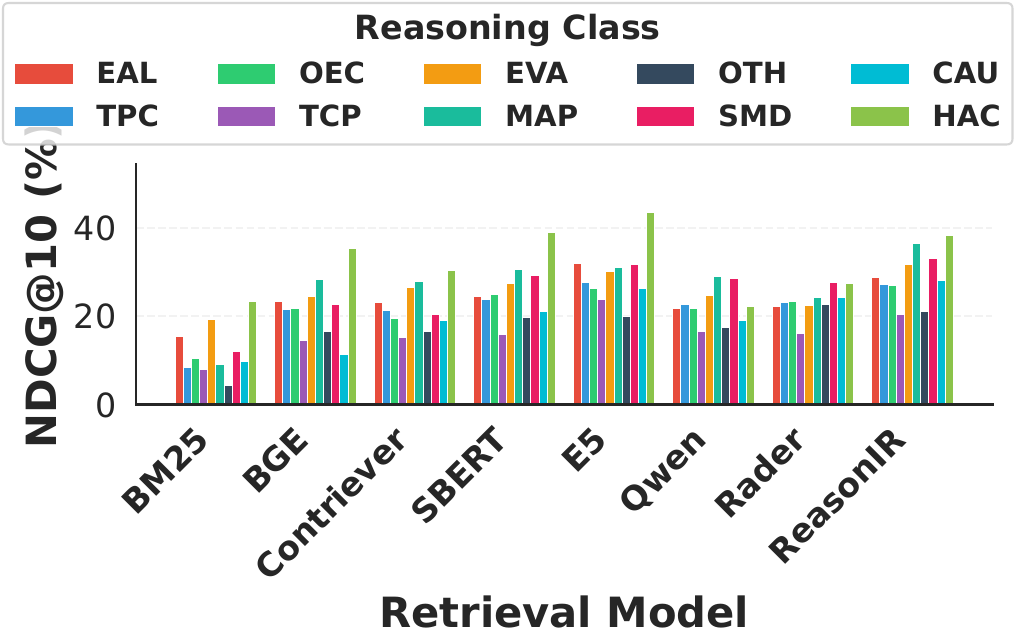}
    \caption{NDCG@10 performance across temporal reasoning classes. 
    }
    \label{fig:reasoning_class_bars}
\end{figure}

\paragraph{Temporal Precision and Relevance}

Table \ref{tab:temporal_avg_metrics} reveals temporal reasoning capabilities through specialized metrics. DiVeR achieves the highest temporal precision at 62.0\%, indicating superior ranking of temporally relevant documents. ReasonIR follows at 57.4\%, while standard dense retrievers cluster around 53--54\%. BM25 achieves only 24.0\%, confirming keyword matching cannot effectively identify temporal relevance. For TR@10, 
DiVeR leads at 41.3\%, followed by ReasonIR (38.2\%). BM25 manages only 11.1\%, highlighting the gap between sparse and neural approaches. Per-domain analysis (Appendix~\ref{sec:detailed_temporal_metrics}) shows the Quant domain is particularly challenging, with DiVeR achieving only 10.8\% temporal relevance, suggesting financial queries may require specialized temporal reasoning patterns.
\paragraph{Step-wise Retrieval Planning}
We evaluate next whether decomposing temporal queries into explicit reasoning steps improves retrieval (Task 2). Figure~\ref{fig:stepwise_comparison} compares three strategies: \textit{Step-Only} (retrieving with individual steps), \textit{Query+Step} (query concatenated with each step sequentially), and \textit{Query+All} (query with all steps combined). Step-Only yields substantially lower performance (avg. 14.6 NDCG@10), confirming that isolated reasoning steps lack sufficient context for effective retrieval. Both query-augmented strategies achieve comparable results, with Query+Step marginally outperforming Query+All (avg. 26.4 vs. 25.9). ReasonIR benefits most from step-wise planning, improving from 17.2 (Step-Only) to 35.0--35.3, a gain of over 18 points. Dense retrievers show similar patterns, BGE improves from 8.3 to 21.5--23.2 while BM25 remains largely unaffected (10.3--11.6 across strategies). Full per-domain results are provided in Appendix~\ref{sec:stepwise_retrieval}.

\section{Additional Analyses}

\paragraph{Temporal signals show modest impact on retrieval performance}
\label{sec:ablation}
We compare four query variants: \textbf{Baseline} (original), \textbf{Strip} (temporal signals removed), \textbf{Temporal-Only} (only temporal information retained), and \textbf{Normalized} (explicit temporal intent tags added). Table~\ref{tab:ablation_variants} shows that removing temporal signals causes modest degradation (avg. 2.2 points), with BM25 virtually unchanged and reasoning models being robust. The \textbf{Temporal-Only} variant reveals temporal information alone is insufficient---DiVeR drops from 32.0 to 17.7 NDCG@10, demonstrating that temporal reasoning requires topical grounding. Notably, the \textbf{Normalized} variant produces dramatic improvement for ReasonIR (35.3 NDCG@10, +8.0 points), establishing it as best-performing when provided structured temporal metadata.
\begin{figure}[h!]
    \centering
    \includegraphics[width=\columnwidth]{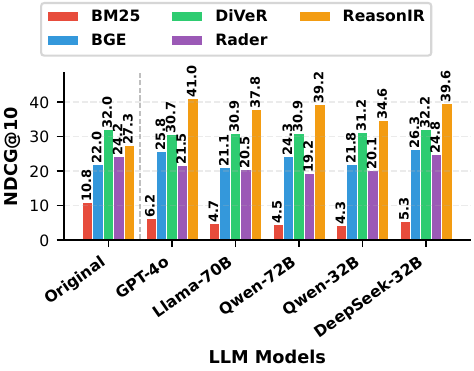}
    \caption{Impact of LLM-based query reformulation on retrieval performance.
    }
    \label{fig:reformulation_bars}
\end{figure}


\paragraph{Performance varies substantially across reasoning classes}
\label{sec:reasoning_class_analysis}

To understand which types of temporal reasoning pose the greatest challenges, we evaluate all retrieval models across 10 reasoning classes derived from our taxonomy. Figure~\ref{fig:reasoning_class_bars} reveals substantial variation in difficulty across reasoning types. \textbf{Trends \& Cross-Period Comparison (TCP)} emerges as the most challenging class, with even DiVeR achieving only 23.9 NDCG@10, as this task requires synthesizing information across multiple time periods. In contrast, \textbf{Historical Attribution \& Context (HAC)} proves comparatively easier (SFR: 51.8), likely due to more explicit answers in the corpus. Model performance profiles differ markedly across reasoning types. DiVeR demonstrates the most balanced performance, leading in 6 of 10 classes. SFR excels at fact-oriented classes (HAC: 51.8) but underperforms on temporal synthesis tasks, while ReasonIR shows strength in \textbf{Causation Analysis (CAU)} at 28.1. BM25's performance varies dramatically (4.3--23.3 range), confirming that lexical matching depends heavily on term overlap rather than temporal understanding. 
Full results are given in Appendix~\ref{sec:reasoning_class_appendix}.

\vspace{-3mm}
\paragraph{LLM-generated reasoning enhances reasoning-aware retrievers.}
\label{sec:reformulation_analysis}
Inspired by reasoning-augmented retrieval~\cite{su2024bright}, we prompt six LLMs to generate step-by-step reasoning, then concatenate with original queries. Figure~\ref{fig:reformulation_bars} reveals striking asymmetry: ReasonIR benefits dramatically (+13.7 NDCG@10 with GPT-4o), while BM25 consistently degrades (10.8$\rightarrow$4.3--6.2) as reasoning text dilutes lexical signals. DiVeR remains stable (30.7--32.2), indicating internalized reasoning that neither benefits from nor requires external augmentation. Full results in Appendix~\ref{sec:reformulation_appendix}.

\paragraph{Retrieval-Augmented Generation Evaluation}
\label{sec:rag_evaluation}

We finally evaluate whether improved retrieval translates to better downstream QA performance. We use Llama-3-70B-Instruct as the generator and GPT-4o to score answers (0--100) based on reference coverage (details in Appendix~\ref{app:rag_eval_prompt}). Table~\ref{tab:rag_results_main} reveals that current retrievers fail to improve over parametric knowledge: the no-retrieval baseline (77.3) outperforms all retrieval-augmented configurations, with BM25 causing the largest degradation ($-3.5$ points). Oracle retrieval achieves 80.5 (+3.2), demonstrating substantial headroom. We hypothesise that temporally incomplete retrieved documents actively mislead the generator. 
Similar to observations in BRIGHT \citep{su2024bright}, this suggests that QA results may not always perfectly capture retrieval performance. This gap often occurs because generator models may struggle to integrate retrieved evidence effectively, or evaluators may fail to recognize nuances in open-ended answers. We further hypothesize that temporally incomplete documents in TEMPO actively mislead the generator; for complex temporal queries, retrieving wrong temporal evidence is worse than retrieving nothing.

\begin{table}[t]
\centering
\small

\begin{tabular}{lc}
\toprule
\textbf{Retriever} & \textbf{Avg. Score} \\
\midrule
None (No Retrieval) & 77.3  \\
Oracle (Gold Docs) & \textbf{80.5}  \\
\midrule
DiVeR & 75.6  \\
BGE & 74.8  \\
BM25 & 73.8  \\
\bottomrule
\end{tabular}
\caption{RAG performance (answer correctness, 0--100) averaged across 13 domains. 
}
\label{tab:rag_results_main}
\end{table}

\section{Conclusion}

We introduced TEMPO, the first benchmark combining temporal reasoning with reasoning-intensive retrieval across 13 domains. TEMPO features 1,730 complex queries with step-wise retrieval planning and novel temporal metrics measuring cross-period coverage. Evaluation of 12 retrieval systems reveals substantial challenges: the best model achieves only 32.0 NDCG@10 and 71.4\% temporal coverage, demonstrating that current systems struggle to retrieve temporally complete evidence. We believe TEMPO provides a testbed for advancing temporal reasoning in retrieval and RAG.

\section*{Limitations}

\paragraph{Domain and Language Coverage.} TEMPO focuses on 13 English-language domains from Stack Exchange. While these domains span blockchain, social sciences, applied fields, and STEM, they may not fully represent all temporal reasoning scenarios. Future work could extend to other languages and domains such as medicine, legal case law, or scientific literature.

\paragraph{Temporal Scope.} Our queries reflect the temporal distribution naturally occurring in Stack Exchange posts, which may over-represent recent time periods. Historical queries from decades or centuries ago are less frequent, potentially limiting evaluation of long-range temporal reasoning.

\paragraph{Annotation Methodology.} While we employ LLM-assisted annotation with human verification, the temporal annotations (reasoning classes, retrieval steps) may contain errors. The LLM-as-judge temporal metrics, though validated on samples, inherit limitations of current language models in temporal understanding.

\bibliography{custom}
\clearpage
\appendix

\newcommand{\apxline}[6]{%
  \par\noindent
  \begin{minipage}[b]{\dimexpr\textwidth-#2\relax}
    \hspace{#2}%
    \makebox[#3][l]{#4}%
    \textbf{#5}\dotfill\makebox[2em][r]{#6}%
  \end{minipage}\vspace{0.2em}
}

\section*{Appendix Contents}
\vspace{-0.6em}

\begingroup
\setlength{\parindent}{0pt}
\setlength{\parskip}{0pt}
\noindent
\begin{minipage}[t]{0.95\textwidth}

\apxline{0}{0em}{6.0em}{\textbf{Appendix A}}{Benchmark Comparison}{\pageref{app:benchmark_comparison}}

\vspace{0.35em}
\apxline{0}{0em}{6.0em}{\textbf{Appendix B}}{Dataset Construction Prompts}{\pageref{app:negative_mining_prompt}}
\apxline{1}{1.6em}{3.2em}{B.1}{Hard Negative Mining Prompt}{\pageref{app:negative_mining_prompt}}
\apxline{1}{1.6em}{3.2em}{B.2}{Query-Level Annotation Prompt}{\pageref{app:query_annotation_prompt}}
\apxline{1}{1.6em}{3.2em}{B.3}{Passage-Level Annotation Prompt}{\pageref{app:passage_annotation_prompt}}

\vspace{0.35em}
\apxline{0}{0em}{6.0em}{\textbf{Appendix C}}{Temporal Metrics \& Evaluation Prompts}{\pageref{app:temporal_metrics}}
\apxline{1}{1.6em}{3.2em}{C.1}{Metric Definitions}{\pageref{app:temporal_metrics}}
\apxline{1}{1.6em}{3.2em}{C.2}{Temporal Intent Detection Prompt}{\pageref{app:temporal_intent_prompt}}
\apxline{1}{1.6em}{3.2em}{C.3}{Temporal Relevance Judgment Prompt}{\pageref{app:temporal_relevance_prompt}}
\apxline{1}{1.6em}{3.2em}{C.4}{Temporal Evidence Judgment Prompt}{\pageref{app:temporal_evidence_prompt}}
\apxline{1}{1.6em}{3.2em}{C.5}{Detailed Temporal Metrics by Domain}{\pageref{sec:detailed_temporal_metrics}}
\apxline{1}{1.6em}{3.2em}{C.6}{Metric Validation}{\pageref{app:metric_validation}}

\vspace{0.35em}
\apxline{0}{0em}{6.0em}{\textbf{Appendix D}}{Detailed Analysis \& Results}{\pageref{app:DetailedAnalysis}}
\apxline{1}{1.6em}{3.2em}{D.1}{Step-wise Retrieval Results}{\pageref{sec:stepwise_retrieval}}
\apxline{1}{1.6em}{3.2em}{D.2}{Reasoning Class Definitions \& Analysis}{\pageref{sec:reasoning_class_appendix}}
\apxline{1}{1.6em}{3.2em}{D.3}{Query Reformulation Analysis}{\pageref{sec:reformulation_appendix}}
\apxline{1}{1.6em}{3.2em}{D.4}{Additional Retrieval Metrics}{\pageref{app:additional_metrices}}

\vspace{0.35em}
\apxline{0}{0em}{6.0em}{\textbf{Appendix E}}{Quality Assessment \& RAG}{\pageref{app:Quality}}
\apxline{1}{1.6em}{3.2em}{E.1}{Dataset Quality Validation}{\pageref{app:quality_validation}}
\apxline{1}{1.6em}{3.2em}{E.2}{Domain Temporal Distribution}{\pageref{app:domain_temporal}}
\apxline{1}{1.6em}{3.2em}{E.3}{RAG Evaluation Details \& Results}{\pageref{app:rag_eval_prompt}}

\vspace{0.35em}
\apxline{0}{0em}{6.0em}{\textbf{Appendix H}}{Extended Related Work}{\pageref{app:benchmark_comparison_1}}

\vspace{0.35em}
\apxline{0}{0em}{6.0em}{\textbf{Appendix I}}{Dataset Examples}{\pageref{app:dataset_examples}}
\vspace{0.35em}
\apxline{0}{0em}{6.0em}{\textbf{Appendix J}}{Annotation Guidelines}{\pageref{app:annotation_guidelines}}


\end{minipage}
\endgroup

\clearpage

\section{Benchmark Comparison}
\label{app:benchmark_comparison}

Table~\ref{tab:tempo-benchmark-comparison} provides a comprehensive comparison of TEMPO with existing temporal reasoning and retrieval benchmarks. We describe each benchmark category and highlight TEMPO's distinguishing characteristics.

\paragraph{Reasoning-Intensive Retrieval Benchmarks.}
BRIGHT~\cite{su2024bright} introduced the first benchmark requiring intensive reasoning for document retrieval, featuring 1,384 queries across 12 domains including economics, psychology, and coding. While state-of-the-art models achieve 59.0 nDCG@10 on standard benchmarks, they score only 18.3 on BRIGHT, revealing significant gaps in reasoning capabilities. RAR-b~\cite{xiao2024rar} extends this with 45,745 queries across 17 domains, framing reasoning tasks as retrieval problems to evaluate whether retrievers can solve reasoning problems directly. However, both benchmarks lack temporal grounding---queries do not require reasoning about time periods, temporal evolution, or cross-period analysis.

\paragraph{Temporal IR Benchmarks.}
NTCIR Temporalia~\cite{joho2014ntcir} established foundational benchmarks for temporal information retrieval through two subtasks: Temporal Query Intent Classification (classifying queries as past/recency/future/atemporal) and Temporal Information Retrieval (ranking documents by temporal relevance). Built on the LivingKnowledge News/Blog Corpus containing 3.8 million timestamped documents, Temporalia pioneered systematic temporal IR evaluation. However, it relies on news/blog corpora with simple temporal queries where document timestamps and basic temporal expressions suffice, lacking the reasoning complexity required for technical domain queries.

\paragraph{Temporal QA Benchmarks.}
TempQuestions~\cite{jia2018tempquestions} provides 1,271 temporal questions over Freebase, introducing a formal definition of temporal questions covering explicit (``in 2008'') and implicit (``during the Cold War'') temporal expressions. Questions require decomposition into sub-questions, but focus on knowledge base QA rather than document retrieval.

ChronoQA~\cite{chen2025question} is a Chinese benchmark with 5,176 questions constructed from 300,000+ news articles (2019--2024), designed for evaluating temporal reasoning in RAG systems. It covers absolute, aggregate, and relative temporal types but is limited to Chinese and news domain.

TIME~\cite{wei2025time} offers 38,522 QA pairs across three sub-datasets: TIME-Wiki (Wikipedia), TIME-News (news articles), and TIME-Dial (multi-session dialogues), covering 11 fine-grained temporal reasoning subtasks. While comprehensive in task coverage, it focuses on LLM evaluation rather than retrieval system assessment.

HistoryBankQA~\cite{mandal2025historybankqa} presents a multilingual database of 10M+ historical events from Wikipedia, generating 535K questions across 10 languages. It evaluates temporal reasoning over historical events but focuses on factual recall from encyclopedic content rather than reasoning-intensive retrieval.

ComplexTempQA~\cite{gruber-etal-2025-complextempqa} is the largest temporal QA dataset with 100M+ synthetic question-answer pairs from Wikipedia and Wikidata, spanning 1987--2023. It introduces a taxonomy of attribute, comparison, and counting questions requiring multi-hop reasoning. Despite its scale, questions are synthetically generated and focus on answer generation rather than document retrieval.

\paragraph{TEMPO's Distinguishing Characteristics.}
Unlike existing benchmarks, TEMPO uniquely combines: (1) \textbf{Temporal + Reasoning-Intensive}: Queries require both temporal reasoning (tracking changes, cross-period comparison) and deep domain understanding, addressing the gap between temporal QA and reasoning-intensive retrieval; (2) \textbf{ Domains}: 13 specialized domains from Stack Exchange where domain experts pose naturally complex questions, rather than news or encyclopedia sources; (3) \textbf{Step-Wise Retrieval}: 3,976 decomposed retrieval steps with gold documents mapped to each step, enabling multi-hop temporal evaluation; (4) \textbf{Cross-Period Analysis}: Novel metrics (Temporal Coverage@k) measuring whether systems retrieve evidence spanning both baseline and comparison time periods; (5) \textbf{Retrieval Focus}: Emphasizes document retrieval over answer generation, aligning with real-world RAG system requirements.

\section{Dataset Construction Prompts}

\subsection{Hard Negative Mining Prompt}
\label{app:negative_mining_prompt}

We use GPT-4o to analyze Stack Exchange posts and generate search queries designed to find challenging negative documents that are topically related but temporally incomplete or irrelevant. The complete prompt is shown in Figure~\ref{prompt:hard_negative_mining}.

\begin{figure*}[!t]
\centering
\begin{tcolorbox}[
  colback=gray!5,
  colframe=gray!75,
  title=Prompt for Hard Negative Mining,
  width=\textwidth,
]
\small\ttfamily
You are a document annotator specializing in hard negative mining for information retrieval.

Your task: Analyze the following Stack Exchange question and extract key information for finding challenging negative passages from web search.

\textbf{Question Title:} \{title\}

\textbf{Question Body:} \{clean\_body\}...

\textbf{Tags:} \{', '.join(tags)\}

Generate a structured analysis with:
\begin{enumerate}
  \item give me a query for hard negative to use it to search on google
  \item All entities (people, places, organizations, concepts, technologies, etc.)
  \item All events, actions, or processes mentioned
\end{enumerate}

Output ONLY a valid JSON object with this exact structure:

\{\\
~~"llm\_summary": "give me a query for hard negative to use it to search on google...",\\
~~"entities\_events": ["entity1", "entity2", "event1", "event2", ...]\\
\}

\textbf{Rules:}
\begin{itemize}
  \item Summary must be EXACTLY 32 words (count carefully)
  \item List ALL relevant entities and events
  \item Include technical terms, concepts, and domain-specific vocabulary
  \item Extract named entities (people, places, companies, technologies)
  \item Include temporal events and processes
  \item Output ONLY valid JSON, no explanations
\end{itemize}
\end{tcolorbox}
\caption{Prompt used for mining hard negative documents. GPT-4o generates a search query and extracts entities to find topically similar but temporally distinct content.}
\label{prompt:hard_negative_mining}
\end{figure*}

The LLM generates a search query designed to retrieve topically similar but temporally incomplete content, along with entities and events that help construct effective negative search queries. Annotators use the generated query to collect hard negative passages from Google search.

\subsection{Query-Level Temporal Annotation Prompt}
\label{app:query_annotation_prompt}

We use GPT-4o to extract temporal characteristics from queries. The complete prompt is shown in ~\ref{prompt:promptQueryLevel}.

\begin{figure*}[!t]
\centering
\begin{tcolorbox}[
  colback=gray!5,
  colframe=gray!75,
  title= Prompt for Query-Level Temporal Annotation,
  width=1\textwidth,
]
\small\ttfamily
You are an expert in \textbf{temporal} information retrieval. Analyze ONLY the QUERY below and produce retrieval guidance and categories.

\textbf{Goals:}
\begin{enumerate}
  \item Determine whether the query is temporal and classify its intent. (1) temporal\_intent: one of ["when","duration","order","before\_after","ongoing\_status",\\
    ~~~~"period\_definition","timeline","none"] (2) query\_temporal\_signals: phrases indicating time (e.g., "in 1914", "during", "after", \\
    ~~~~"first", "since", "today", "in the 18th century") (3) query\_temporal\_events: ONLY time-bound events (e.g., "Battle of Hastings", \\
    ~~~~"signing of Treaty X", "election of Y"). Exclude generic actions unless anchored in time.
  \item Provide a compact, specific plan to retrieve \textbf{temporal} evidence.
  \item Identify time anchors, expected granularity, and sanity checks.
\end{enumerate}

\textbf{Allowed temporal reasoning classes} (choose one primary, optional secondaries):
\begin{itemize}
  \item "event\_analysis\_and\_localization"
  \item "time\_period\_contextualization"
  \item "event\_verification\_and\_authenticity"
  \item "sources\_methods\_and\_documentation"
  \item "materials\_artifacts\_and\_provenance"
  \item "trends\_changes\_and\_cross\_period"
  \item "origins\_evolution\_comparative\_analysis"
  \item "historical\_misinterpretation\_or\_reenactment"
  \item "causation\_analysis"
  \item "artifact\_verification"
  \item "historical\_attribution\_and\_context"
\end{itemize}

\textbf{CRUCIAL RULES:}
(1) Use only the QUERY content (do NOT assume any passage). (2) All arrays must be present even if empty. Use "" for missing strings. (3) Return ONLY one JSON object with EXACT keys and value types below.

\textbf{JSON schema to output:}

\{\\
~~"is\_temporal\_query": true,\\
~~"temporal\_intent": "when",\\
~~"query\_temporal\_signals": ["..."],\\
~~"query\_temporal\_events": ["..."],\\
~~"query\_summary": "summary of the query <=50 words",\\
~~"temporal\_reasoning\_class\_primary": "time\_period\_contextualization",\\
~~"temporal\_reasoning\_class\_secondary": ["materials\_artifacts\_and\_provenance"],\\
~~"retrieval\_reasoning": "explanation of how to retrieve temporal evidence",\\
~~"retrieval\_plan": [\\
~~~~\{"step": 1, "action": ".."\},\\
~~~~\{"step": 2, "action": ".."\}\\
~~],\\
~~"key\_time\_anchors": ["..."],\\
~~"expected\_granularity": "date",\\
~~"quality\_checks": ["cross-check dates from multiple sources", "prefer primary/authoritative sources"]\\
\}

\textbf{QUERY:}\\
\{query\}
\end{tcolorbox}
\caption{Query-level temporal annotation prompt. The model identifies temporal intent, signals, and reasoning classes, and generates a step-wise retrieval plan.}
\label{prompt:promptQueryLevel}
\end{figure*}

\noindent GPT-4o processes each query and outputs a JSON object containing temporal intent classification, temporal signals, reasoning class, and a step-wise retrieval plan. This annotation enables fine-grained evaluation of temporal reasoning capabilities.

\subsection{Passage-Level Temporal Annotation Prompt}
\label{app:passage_annotation_prompt}

We use GPT-4o to extract temporal information from passages. The complete prompt is shown in Figure~\ref{prompt:passage_annotation}.

\begin{figure*}[!t]
\centering
\begin{tcolorbox}[
  colback=gray!5,
  colframe=gray!75,
  title=Prompt for Passage-Level Temporal Annotation,
  width=\textwidth,
]
\small\ttfamily
You are an expert annotator for \textbf{temporal} information retrieval.

Given a QUERY and a PASSAGE, do BOTH of the following:

\textbf{1) Extract TEMPORAL info from the PASSAGE only:}
\begin{itemize}
  \item passage\_temporal\_signals: time cues (e.g., "in 1914", "during the 18th century", \\
  ~~~~"after the treaty")
  \item passage\_temporal\_events: ONLY time-bound events (battle/treaty/reign/election/founding). \\
  ~~~~Exclude non-temporal events.
  \item time\_mentions: explicit or relative expressions (years, dates, centuries, eras, "after X", \\
  ~~~~"during Y")
  \item time\_scope\_guess:
  \begin{itemize}
    \item start\_iso: ISO-like if visible (YYYY or YYYY-MM or YYYY-MM-DD), else ""
    \item end\_iso: same format; "" if none
    \item granularity: one of ["date","month","year","decade","century","multicentury","unknown"]
  \end{itemize}
  \item tense\_guess: one of ["past","present","future","mixed","unknown"]
  \item confidence: 0.0--1.0
\end{itemize}

\textbf{CRUCIAL RULES:}
\begin{itemize}
  \item Do NOT output any query-level fields here (no is\_temporal\_query, temporal\_intent, etc.).
  \item Return empty lists (not null) when nothing is found.
  \item Return ONLY one JSON object with EXACT keys and value types below.
\end{itemize}

\textbf{JSON schema to output:}

\{\\
~~"passage\_temporal\_signals": ["..."],\\
~~"passage\_temporal\_events": ["..."],\\
~~"time\_mentions": ["..."],\\
~~"time\_scope\_guess": \{"start\_iso": "", "end\_iso": "", "granularity": "unknown"\},\\
~~"tense\_guess": "past",\\
~~"confidence": 0.0\\
\}

\textbf{QUERY:}\\
\{query\}

\textbf{PASSAGE:}\\
\{passage\}
\end{tcolorbox}
\caption{Passage-level temporal annotation prompt. The model extracts temporal signals, events, and estimates the ISO time scope of the retrieved passage.}
\label{prompt:passage_annotation}
\end{figure*}

GPT-4o processes each query-passage pair and outputs a JSON object with temporal signals, events, time scope with ISO-formatted dates, and temporal granularity. This passage-level annotation enables evaluation of whether retrieved documents contain appropriate temporal evidence.

\section{Temporal Evaluation Metrics: Definitions and Prompts}\label{app:temporal_metrics}

\subsection{Temporal Evaluation Metrics}
\label{sec:metrics}

Traditional IR metrics like NDCG and Recall measure whether relevant documents are retrieved, but they do not capture temporal reasoning quality, specifically, whether retrieved documents contain appropriate temporal information and cover required time periods. We introduce novel temporal metrics that address these limitations, implemented using LLM-as-judge evaluation.

\textbf{Temporal Intent Detection.} Before evaluating temporal relevance, we use LLM-as-judge to determine whether a query requires temporal reasoning. LLM classifies queries into temporal intents (when/duration/order/before\_after/ongoing\_status/\allowbreak period\_definition/timeline/none) and identifies temporal keywords. Only queries with detected temporal intent proceed to temporal evaluation (prompt in Appendix~\ref{app:temporal_intent_prompt}).

\textbf{Temporal Precision@k (TP@k).} Position-weighted precision measuring the quality of temporally relevant documents using LLM judgment. For each document $d_i$ at rank $i \in \{1, \ldots, k\}$, we obtain a binary verdict $v_i \in \{0,1\}$ from the LLM indicating temporal relevance (prompt in Appendix~\ref{app:temporal_relevance_prompt}). Let $R = \{r_1, r_2, \ldots, r_m\} \subseteq \{1, \ldots, k\}$ denote the set of rank positions where documents are temporally relevant (i.e., $v_{r_i} = 1$). Then:
\begin{equation}
\text{TP@k} = \frac{1}{|R|} \sum_{r \in R} \frac{|\{j \in R : j \leq r\}|}{r}
\end{equation}
where $|\{j \in R : j \leq r\}|$ counts the number of relevant documents at or before rank $r$. 

This rewards documents that are both temporally relevant and highly ranked, with higher weight given to relevant documents appearing earlier in the ranking.

\textbf{Temporal Relevance@k (TR@k).} Simple proportion of temporally relevant documents in top-$k$:

\begin{equation}
\text{TR@k} = \frac{1}{k} \sum_{i=1}^{k} v_i
\end{equation}

where $v_i \in \{0,1\}$ is the LLM's verdict for document at rank $i$.

\textbf{Temporal Coverage@k (TC@k).}
For queries that require evidence spanning multiple time periods, we define a set of required periods
$\mathcal{P}_Q = \{P_1, \ldots, P_M\}$, where each $P_m$ is a time interval (e.g., a baseline period, a comparison period, or additional historical intervals).
For each retrieved document $d_i$ at rank $i$, an LLM judge predicts a binary coverage vector
$\mathbf{c}_i \in \{0,1\}^M$, where $c_{i,m}=1$ if $d_i$ contains evidence relevant to period $P_m$.
We compute cumulative period coverage up to rank $k$:
\begin{equation}
C_m(k) = \max_{i \le k} c_{i,m}.
\end{equation}
Temporal Coverage@k is the fraction of required periods covered in the top-$k$ results:
\begin{equation}
\text{TC@k} = \frac{1}{M}\sum_{m=1}^{M} C_m(k).
\end{equation}
This metric ranges from 0 (no period covered) to 1 (all required periods covered).
When $M=2$, TC@k reduces to the baseline/comparison formulation used in prior work and in our main experiments.

\textbf{NDCG | Full Coverage.}
We compute NDCG@k conditioned on full temporal coverage across all required periods:
\begin{equation}
\text{NDCG|FC@k} = 
\begin{cases}
\text{NDCG@k} & \text{if } \text{TC@k} = 1.0 \\
\text{NaN} & \text{otherwise.}
\end{cases}
\end{equation}

To ensure the reliability of our proposed temporal metrics, we conducted a meta-evaluation using a set of control queries and documents. Table \ref{tab:intent_val} demonstrates that our intent classifier correctly identifies temporal grounding without false positives on atemporal queries. Furthermore, Table \ref{tab:tc_matrix} and \ref{tab:math_weight} confirm that our metrics correctly enforce cross-period synthesis requirements and maintain rank-sensitivity, respectively.

\subsection{Temporal Evaluation Prompts}
\label{sec:detailed_temporal_metrics}
\subsubsection{Temporal Intent Detection Prompt}
\label{app:temporal_intent_prompt}

We use GPT-4o to classify whether queries require temporal reasoning and identify temporal characteristics. The complete prompt is shown in Figure~\ref{prompt:intent_detection}:

\begin{figure*}[!t]
\centering
\begin{tcolorbox}[
  colback=gray!5,
  colframe=gray!75,
  title=Prompt for Temporal Intent Detection,
  width=\textwidth,
]
\small\ttfamily
Analyze if this query requires temporal reasoning to answer correctly.

\textbf{Query:} "\{query\}"

Temporal queries ask about:
\begin{itemize}
  \item WHEN something happened (specific time/date)
  \item HOW LONG something takes/lasts (duration)
  \item RECENT events or changes (recency)
  \item Changes OVER TIME (temporal evolution)
  \item BEFORE/AFTER relationships (temporal ordering)
\end{itemize}

Respond ONLY with valid JSON in this exact format:

\{\\
~~~~"has\_temporal\_intent": true/false,\\
~~~~"temporal\_keywords": ["keyword1", "keyword2"],\\
~~~~"temporal\_focus": "duration" or "specific\_time" or "recency" or "change\_over\_time" or "none"\\
\}

\textbf{Examples:}

Query: "When did Bitcoin Core introduce pruning?"\\
Output: \{"has\_temporal\_intent": true, "temporal\_keywords": ["when", "introduce"], \\
~~~~~~~~~"temporal\_focus": "specific\_time"\}

Query: "How long does Bitcoin Core store forked chains?"\\
Output: \{"has\_temporal\_intent": true, "temporal\_keywords": ["how long", "store"], \\
~~~~~~~~~"temporal\_focus": "duration"\}

Query: "What are recent developments in Bitcoin storage?"\\
Output: \{"has\_temporal\_intent": true, "temporal\_keywords": ["recent", "developments"], \\
~~~~~~~~~"temporal\_focus": "recency"\}

Query: "What is Bitcoin Core?"\\
Output: \{"has\_temporal\_intent": false, "temporal\_keywords": [], "temporal\_focus": "none"\}

Now analyze: "\{query\}"
\end{tcolorbox}
\caption{Temporal intent detection prompt. The LLM judge classifies whether a query requires temporal reasoning and identifies specific temporal keywords.}
\label{prompt:intent_detection}
\end{figure*}

GPT-4o classifies the query's temporal intent and extracts temporal keywords. Queries without temporal intent ($\text{has\_temporal\_intent} = \text{false}$) receive NaN for all LLM-based temporal metrics.

\subsubsection{Temporal Relevance Judgment Prompt}
\label{app:temporal_relevance_prompt}

For queries with detected temporal intent, we use GPT-4o to judge whether each retrieved document provides temporal information. The complete prompt is shown in Figure~\ref{prompt:relevance_judgment}:

\begin{figure*}[!t]
\centering
\begin{tcolorbox}[
  colback=gray!5,
  colframe=gray!75,
  title=Prompt for Temporal Relevance Judgment,
  width=\textwidth,
]
\small\ttfamily
Judge if a retrieved document helps answer the TEMPORAL aspects of a query.

\textbf{Query:} "\{query\}"\\
\textbf{Temporal Focus:} \{temporal\_focus\}

\textbf{Document:}\\
\{document\}

\textbf{Question:} Does this document provide information that DIRECTLY helps answer the temporal aspects of the query?

\textbf{Guidelines:}
\begin{itemize}
  \item Verdict = 1 if document contains temporal information (dates, durations, time periods, temporal sequences)
  \item Verdict = 0 if document lacks temporal information even if generally relevant
  \item For "when" queries: document must mention specific times/dates
  \item For "how long" queries: document must mention durations/time periods
  \item For "recent" queries: document must mention recency or recent dates
  \item Be STRICT: generic facts without temporal markers are NOT temporally relevant
\end{itemize}

Respond ONLY with valid JSON:

\{\\
~~~~"verdict": 1 or 0,\\
~~~~"reason": "brief explanation",\\
~~~~"temporal\_contribution": "what temporal information provided, or 'none'"\\
\}
\end{tcolorbox}
\caption{Temporal relevance judgment prompt. Used to compute Temporal Precision and Relevance by determining if a document addresses specific temporal aspects of the query.}
\label{prompt:relevance_judgment}
\end{figure*}

GPT-4o outputs a binary verdict (1 = temporally relevant, 0 = not temporally relevant) with explanation. These verdicts are used to compute Temporal Precision@k and Temporal Relevance@k.

\subsubsection{Temporal Evidence Judgment Prompt}
\label{app:temporal_evidence_prompt}

For cross-period queries requiring comparison across time periods, we use a specialized prompt that evaluates baseline and comparison period coverage. The complete prompt is shown in Figure~\ref{prompt:evidence_judgment}:

\begin{figure*}[!t]
\centering
\begin{tcolorbox}[
  colback=gray!5,
  colframe=gray!75,
  title=Prompt for Temporal Evidence Judgment (Cross-Period Queries),
  width=\textwidth,
]
\small\ttfamily
You are grading retrieved documents for a temporal trend/change query that needs cross-period evidence.

\textbf{Query:} "\{query\}"\\
\textbf{Temporal Focus:} \{temporal\_focus\}

\textbf{Baseline anchor period:} \{baseline\_anchor\}\\
\textbf{Comparison/current anchor period:} \{comparison\_anchor\}

\textbf{Document:}\\
\{document\}

Decide:
\begin{enumerate}
  \item verdict: 1 if the document DIRECTLY helps answer the temporal aspects of the query \\
  ~~~~(contains relevant temporal info), else 0.
  \item covers\_baseline: true if the document contains evidence about the BASELINE anchor period.
  \item covers\_comparison: true if the document contains evidence about the COMPARISON/current \\
  ~~~~anchor period.
\end{enumerate}

\textbf{Strictness rules:}
\begin{itemize}
  \item A random date not related to the anchors does NOT count.
  \item "Currently/as of 2023/recent years" can count for comparison coverage if relevant.
  \item Baseline coverage should connect to the baseline anchor period (e.g., around 2017).
\end{itemize}

Return ONLY valid JSON:

\{\\
~~"verdict": 1 or 0,\\
~~"reason": "brief explanation",\\
~~"temporal\_contribution": "what temporal information provided, or 'none'",\\
~~"covers\_baseline": true/false,\\
~~"covers\_comparison": true/false\\
\}
\end{tcolorbox}
\caption{Cross-period temporal evidence judgment prompt. Used to compute Temporal Coverage by verifying if documents cover both baseline and comparison periods.}
\label{prompt:evidence_judgment}
\end{figure*}

 GPT-4o judges temporal relevance and separately tracks baseline/comparison period coverage. These judgments enable computation of Temporal Coverage@k and NDCG | Full Coverage metrics, which are critical for evaluating cross-period temporal reasoning.

\subsection{Detailed Temporal Metrics}

This section presents detailed per-domain results for the four temporal metrics introduced in Section~\ref{sec:metrics}. All metrics are computed at rank 10 (k=10) and averaged across temporal queries within each domain. Values are reported as percentages, with best results in \textbf{bold} and second-best \underline{underlined}.

\subsubsection{Temporal Precision@10}
Temporal Precision@10 (TP@10) measures the position-weighted precision of temporally relevant documents, rewarding systems that rank temporal evidence higher. \autoref{tab:temporal_tp_10} shows per-domain performance. Reasoning models (DiVeR, ReasonIR) consistently achieve the highest scores across most domains, with particularly strong performance in History (55.8\%, 57.9\%), Politics (57.6\%, 58.2\%), and Law (56.4\%, 65.0\%). BM25 struggles across all domains, with particularly poor performance in Monero (40.0\%) and Quant (29.1\%), domains characterized by technical jargon and evolving terminology that keyword matching cannot capture effectively.

\begin{table*}[t]
\centering

\resizebox{\textwidth}{!}{%
\begin{tabular}{lccccccccccc}
\toprule
\textbf{Domain} & \textbf{BM25} & \textbf{BGE} & \textbf{Contriever} & \textbf{DiVeR} & \textbf{E5} & \textbf{GritLM} & \textbf{Inst-L} & \textbf{Qwen} & \textbf{Rader} & \textbf{ReasonIR} & \textbf{SBERT} \\
\midrule
\rowcolor{gray!12}\multicolumn{12}{c}{\textit{Blockchain}} \\
\midrule
\textbf{Bitcoin} & 26.2 & 52.7 & 42.3 & \textbf{61.2} & 55.9 & 55.6 & 52.8 & 41.9 & 50.5 & \underline{56.4} & 52.1 \\
\textbf{Cardano} & 27.2 & 49.6 & 37.1 & \textbf{56.5} & 49.9 & 50.6 & 44.1 & 44.4 & 46.4 & 50.1 & \underline{52.2} \\
\textbf{Iota} & 10.0 & 21.2 & 21.6 & 19.7 & \underline{29.8} & 19.5 & 23.7 & 9.5 & 12.0 & 27.3 & \textbf{32.9} \\
\textbf{Monero} & 12.7 & 40.0 & 29.9 & \textbf{47.7} & 40.3 & \underline{45.5} & 36.2 & 30.2 & 34.8 & 42.3 & 32.5 \\
\midrule
\rowcolor{gray!12}\multicolumn{12}{c}{\textit{Social Sci.}} \\
\midrule
\textbf{Economics} & 17.5 & 50.3 & 44.7 & \textbf{63.8} & \underline{56.3} & 48.8 & 50.7 & 44.5 & 45.4 & 50.4 & 44.6 \\
\textbf{Law} & 41.9 & 56.4 & 56.5 & \textbf{65.8} & 60.4 & 58.2 & 54.9 & 57.3 & 50.4 & \underline{61.2} & 60.6 \\
\textbf{Politics} & 39.0 & 57.6 & 55.6 & \textbf{64.4} & 61.7 & \underline{62.7} & 55.5 & 56.5 & 55.8 & 57.6 & 59.8 \\
\textbf{History} & 19.8 & 55.8 & 53.3 & \textbf{64.0} & 52.1 & 52.9 & 55.0 & 45.3 & 53.1 & \underline{61.8} & 56.5 \\
\midrule
\rowcolor{gray!12}\multicolumn{12}{c}{\textit{Applied}} \\
\midrule
\textbf{Quant} & 16.9 & 29.1 & 29.3 & 33.8 & \textbf{41.8} & 31.3 & 29.0 & 33.2 & 33.4 & \underline{39.3} & 32.3 \\
\textbf{Travel} & 29.3 & 56.6 & 53.7 & \textbf{64.7} & 55.8 & 56.0 & \underline{62.9} & 45.8 & 51.4 & 57.8 & 54.1 \\
\textbf{Workplace} & 14.3 & 49.6 & 47.2 & 51.6 & \textbf{53.2} & 46.7 & 49.5 & 44.4 & 46.3 & 44.0 & \underline{51.8} \\
\textbf{Genealogy} & 19.6 & 43.3 & 43.4 & \textbf{58.3} & 47.9 & 50.2 & 44.7 & 47.0 & 48.7 & \underline{51.8} & 46.5 \\
\midrule
\rowcolor{gray!12}\multicolumn{12}{c}{\textit{STEM}} \\
\midrule
\textbf{HSM} & 31.6 & 62.7 & 53.1 & \textbf{67.7} & 59.1 & 61.6 & 60.1 & 51.9 & 52.6 & 61.5 & \underline{63.8} \\
\midrule
\textbf{Avg.} & 24.0 & 53.7 & 49.6 & \textbf{62.0} & 53.5 & 53.8 & 53.1 & 46.2 & 50.6 & \underline{57.4} & 54.1 \\
\bottomrule
\end{tabular}}
\caption{TP@10 across all domains (grouped by category). Best in \textbf{bold}, second best \underline{underlined}.}
\label{tab:temporal_tp_10}
\end{table*}

\subsubsection{Temporal Relevance@10}
Temporal Relevance@10 (TR@10) measures the proportion of retrieved documents that contain temporal information relevant to the query. \autoref{tab:temporal_tr_10} reveals substantial variation across domains. DiVeR achieves the highest average at 41.3\%, with particularly strong performance in HSM (39.3\%) and Law (38.5\%). Notably, the Quant domain proves extremely challenging, with even DiVeR achieving only 10.8\% temporal relevance, suggesting that financial queries may require specialized temporal reasoning patterns. The large gap between reasoning models (38--41\%) and sparse retrieval (11.1\%) underscores the importance of understanding query semantics for identifying temporal information needs.

\begin{table*}[t]
\centering

\resizebox{\textwidth}{!}{%
\begin{tabular}{lccccccccccc}
\toprule
\textbf{Domain} & \textbf{BM25} & \textbf{BGE} & \textbf{Contriever} & \textbf{DiVeR} & \textbf{E5} & \textbf{GritLM} & \textbf{Inst-L} & \textbf{Qwen} & \textbf{Rader} & \textbf{ReasonIR} & \textbf{SBERT} \\
\midrule
\rowcolor{gray!12}\multicolumn{12}{c}{\textit{Blockchain}} \\
\midrule
\textbf{Bitcoin} & 14.9 & 38.3 & 30.7 & \textbf{47.3} & 39.5 & 39.8 & 38.0 & 29.8 & 33.7 & \underline{40.1} & 37.6 \\
\textbf{Cardano} & 13.0 & 31.1 & 20.2 & \textbf{39.8} & 35.2 & \underline{35.9} & 29.8 & 29.1 & 32.8 & 33.0 & 31.3 \\
\textbf{Iota} & 1.0 & 6.0 & 5.0 & \underline{7.0} & 7.0 & 6.0 & 5.0 & 5.0 & 2.0 & 7.0 & \textbf{8.0} \\
\textbf{Monero} & 6.4 & 23.7 & 18.0 & \textbf{30.5} & 25.0 & \underline{28.4} & 22.7 & 18.2 & 23.9 & 26.4 & 19.6 \\
\midrule
\rowcolor{gray!12}\multicolumn{12}{c}{\textit{Social Sci.}} \\
\midrule
\textbf{Economics} & 9.2 & 31.1 & 26.4 & \textbf{39.5} & 32.9 & 33.8 & 30.7 & 27.9 & 29.4 & \underline{35.5} & 32.6 \\
\textbf{Law} & 17.9 & 38.5 & 30.9 & \textbf{43.0} & 36.7 & 38.2 & 36.1 & 38.2 & 33.3 & \underline{42.1} & 36.1 \\
\textbf{Politics} & 16.2 & 30.9 & 28.7 & \underline{36.1} & 31.3 & \textbf{36.7} & 31.1 & 29.5 & 30.3 & 33.8 & 31.2 \\
\textbf{History} & 8.9 & 36.9 & 34.5 & \textbf{44.7} & 29.4 & 35.8 & 36.9 & 29.7 & 34.7 & \underline{43.5} & 38.3 \\
\midrule
\rowcolor{gray!12}\multicolumn{12}{c}{\textit{Applied}} \\
\midrule
\textbf{Quant} & 3.5 & 10.8 & 10.8 & 12.7 & \textbf{14.2} & 12.0 & 10.0 & 9.6 & 11.9 & \underline{13.1} & 11.5 \\
\textbf{Travel} & 14.7 & 37.4 & 37.6 & \textbf{45.4} & 40.5 & 39.2 & \underline{40.7} & 28.6 & 35.5 & 37.5 & 37.8 \\
\textbf{Workplace} & 3.7 & \underline{27.8} & 18.5 & \textbf{28.9} & 21.9 & 25.6 & 26.7 & 19.3 & 26.7 & 26.3 & 24.4 \\
\textbf{Genealogy} & 10.5 & 25.8 & 24.6 & \textbf{35.4} & 28.4 & 26.8 & 25.7 & 29.3 & 29.5 & \underline{30.2} & 27.8 \\
\midrule
\rowcolor{gray!12}\multicolumn{12}{c}{\textit{STEM}} \\
\midrule
\textbf{HSM} & 15.1 & 39.3 & 31.7 & \textbf{46.0} & 33.8 & 40.8 & 37.3 & 32.7 & 35.0 & 41.8 & \underline{43.7} \\
\midrule
\textbf{Avg.} & 11.1 & 34.0 & 30.3 & \textbf{41.3} & 31.1 & 35.0 & 33.8 & 28.8 & 32.3 & \underline{38.2} & 35.0 \\
\bottomrule
\end{tabular}}
\caption{TR@10 across all domains (grouped by category). Best in \textbf{bold}, second best \underline{underlined}.}
\label{tab:temporal_tr_10}
\end{table*}

\subsubsection{Temporal Coverage@10}
Temporal Coverage@10 (TC@10) evaluates whether systems retrieve evidence from both the baseline and comparison time periods for cross-period queries. \autoref{tab:temporal_tc_10} shows that ReasonIR and DiVeR achieve the highest average coverage (72.4\%, 71.4\%), yet still fail to provide complete temporal evidence for approximately 30\% of cross-period queries. Domain variation is substantial: Law achieves 84.6\% coverage (ReasonIR), while Quant reaches only 38.2\% (ReasonIR), a 46-point gap. This suggests that legal and policy domains may have more structured temporal discourse patterns that facilitate cross-period retrieval, while financial and quantitative analysis requires more sophisticated temporal reasoning.

\begin{table*}[t]
\centering

\resizebox{\textwidth}{!}{%
\begin{tabular}{lccccccccccc}
\toprule
\textbf{Domain} & \textbf{BM25} & \textbf{BGE} & \textbf{Contriever} & \textbf{DiVeR} & \textbf{E5} & \textbf{GritLM} & \textbf{Inst-L} & \textbf{Qwen} & \textbf{Rader} & \textbf{ReasonIR} & \textbf{SBERT} \\
\midrule
\rowcolor{gray!12}\multicolumn{12}{c}{\textit{Blockchain}} \\
\midrule
\textbf{Bitcoin} & 37.9 & 56.2 & 53.3 & \textbf{67.7} & 54.7 & 61.8 & 53.1 & 40.9 & 57.6 & \underline{67.2} & 60.3 \\
\textbf{Cardano} & 25.0 & 54.5 & 41.7 & \underline{62.5} & 50.0 & 54.5 & 54.2 & 58.3 & 50.0 & 50.0 & \textbf{63.6} \\
\textbf{Iota} & 20.0 & \underline{40.0} & 20.0 & 20.0 & \textbf{50.0} & 20.0 & 20.0 & 20.0 & 20.0 & 20.0 & 20.0 \\
\textbf{Monero} & 28.0 & 42.0 & 32.0 & 44.0 & 34.0 & \textbf{56.0} & 44.0 & 30.0 & 40.0 & \underline{52.0} & 34.0 \\
\midrule
\rowcolor{gray!12}\multicolumn{12}{c}{\textit{Social Sci.}} \\
\midrule
\textbf{Economics} & 36.4 & 68.5 & 64.5 & \textbf{80.9} & 71.8 & 70.5 & \underline{72.7} & 72.7 & 68.5 & 70.0 & 65.2 \\
\textbf{Law} & 57.7 & 84.6 & 73.1 & 79.2 & \textbf{88.5} & 80.8 & \underline{88.5} & 80.8 & 79.2 & 83.3 & 83.3 \\
\textbf{Politics} & 49.5 & 71.3 & 75.0 & \underline{81.2} & 74.0 & \textbf{82.8} & 71.9 & 71.4 & 71.8 & 77.3 & 74.5 \\
\textbf{History} & 26.9 & 72.5 & 71.8 & \underline{75.6} & 64.0 & 71.3 & 72.0 & 61.7 & 70.1 & \textbf{79.6} & 73.4 \\
\midrule
\rowcolor{gray!12}\multicolumn{12}{c}{\textit{Applied}} \\
\midrule
\textbf{Quant} & 21.9 & 38.2 & 32.4 & 44.1 & 46.7 & 46.4 & 34.4 & 41.2 & \textbf{53.1} & \underline{53.1} & 38.2 \\
\textbf{Travel} & 34.8 & 40.9 & 39.6 & 43.8 & 43.5 & 45.7 & 45.7 & 44.0 & 37.5 & \underline{47.6} & \textbf{50.0} \\
\textbf{Workplace} & 0.0 & 36.4 & 31.8 & 36.4 & 36.4 & 27.3 & 25.0 & \underline{40.0} & \textbf{40.9} & 27.8 & 31.8 \\
\textbf{Genealogy} & 17.6 & 45.5 & 50.0 & 55.6 & 51.5 & 50.0 & 54.5 & 48.5 & \textbf{61.8} & 57.4 & \underline{58.8} \\
\midrule
\rowcolor{gray!12}\multicolumn{12}{c}{\textit{STEM}} \\
\midrule
\textbf{HSM} & 41.8 & 82.4 & 75.0 & 83.3 & 74.1 & \textbf{85.5} & 84.3 & 78.2 & 81.2 & \underline{84.5} & 78.3 \\
\midrule
\textbf{Avg.} & 32.8 & 66.1 & 64.1 & \underline{71.4} & 63.2 & 69.1 & 66.9 & 61.0 & 66.1 & \textbf{72.4} & 67.2 \\
\bottomrule
\end{tabular}}
\caption{TC@10 across all domains (grouped by category). Best in \textbf{bold}, second best \underline{underlined}.}
\label{tab:temporal_tc_10}
\end{table*}

\subsubsection{NDCG | Full Coverage@10}
NDCG|Full Coverage@10 evaluates ranking quality specifically for queries where complete temporal evidence (both baseline and comparison periods) is retrieved in the top-10 results. \autoref{tab:temporal_ndcg_fc_10} reveals that E5 achieves the highest average score (33.4\%), closely followed by DiVeR (32.5\%). Interestingly, while reasoning models excel at ensuring temporal coverage (previous subsection), E5 produces superior rankings when all necessary evidence happens to be retrieved. This suggests complementary strengths: reasoning models better identify what temporal evidence is needed across time periods, while dense retrievers may better distinguish relevance gradations among temporally appropriate documents. Domain-specific patterns mirror those in standard NDCG@10, with Law (40.5\%, E5) and Workplace (34.4\%, E5) showing strong performance, while technically complex domains like Monero (7.0\%, DiVeR) remain challenging even with complete temporal coverage.

\begin{table*}[t]
\centering

\resizebox{\textwidth}{!}{%
\begin{tabular}{lccccccccccc}
\toprule
\textbf{Domain} & \textbf{BM25} & \textbf{BGE} & \textbf{Contriever} & \textbf{DiVeR} & \textbf{E5} & \textbf{GritLM} & \textbf{Inst-L} & \textbf{Qwen} & \textbf{Rader} & \textbf{ReasonIR} & \textbf{SBERT} \\
\midrule
\rowcolor{gray!12}\multicolumn{12}{c}{\textit{Blockchain}} \\
\midrule
\textbf{Bitcoin} & 4.3 & 6.4 & 13.0 & 13.7 & \textbf{16.0} & \underline{15.6} & 10.1 & 4.3 & 12.7 & 12.3 & 11.8 \\
\textbf{Cardano} & 0.0 & 22.8 & 4.8 & 27.1 & \textbf{34.0} & 0.0 & 13.8 & \underline{27.1} & 16.3 & 14.2 & 17.1 \\
\textbf{Iota} & 39.0 & 38.4 & 58.6 & \underline{63.7} & 41.4 & 58.6 & 63.7 & 13.0 & 0.0 & \textbf{83.2} & 63.7 \\
\textbf{Monero} & 3.6 & 7.0 & 10.7 & 13.9 & \underline{16.8} & 15.9 & 15.9 & 12.3 & 10.4 & \textbf{17.2} & 14.5 \\
\midrule
\rowcolor{gray!12}\multicolumn{12}{c}{\textit{Social Sci.}} \\
\midrule
\textbf{Economics} & 11.2 & 11.4 & 16.8 & \underline{22.5} & \textbf{28.0} & 18.6 & 14.1 & 16.9 & 17.6 & 18.0 & 11.4 \\
\textbf{Law} & 21.1 & 40.5 & 30.1 & \textbf{48.6} & 31.8 & 35.3 & 32.5 & 35.4 & \underline{40.6} & 35.0 & 36.2 \\
\textbf{Politics} & 43.4 & 30.3 & 36.5 & \underline{43.9} & \textbf{53.7} & 39.8 & 33.7 & 43.3 & 36.3 & 36.4 & 36.8 \\
\textbf{History} & 19.3 & 28.5 & 28.4 & \textbf{33.8} & 29.2 & 26.5 & 27.7 & 32.7 & 27.5 & \underline{33.4} & 28.9 \\
\midrule
\rowcolor{gray!12}\multicolumn{12}{c}{\textit{Applied}} \\
\midrule
\textbf{Quant} & 0.0 & 19.0 & 24.9 & \underline{30.1} & 16.8 & 14.9 & 19.0 & 18.7 & \textbf{34.4} & 18.3 & 22.1 \\
\textbf{Travel} & 0.0 & 21.3 & 17.9 & \underline{24.8} & 20.4 & 21.2 & 21.8 & 18.4 & \textbf{36.7} & 9.5 & 22.7 \\
\textbf{Workplace} & -- & 34.4 & 34.0 & 30.0 & \underline{47.8} & 40.3 & 27.7 & \textbf{56.0} & 32.5 & 19.1 & 38.8 \\
\textbf{Genealogy} & 27.8 & \underline{33.3} & 28.6 & \textbf{34.1} & 33.0 & 20.0 & 30.5 & 27.0 & 21.2 & 32.5 & 26.4 \\
\midrule
\rowcolor{gray!12}\multicolumn{12}{c}{\textit{STEM}} \\
\midrule
\textbf{HSM} & 26.9 & 22.2 & 15.9 & 27.7 & \textbf{33.1} & \underline{28.4} & 24.1 & 24.1 & 19.6 & 20.0 & 24.6 \\
\midrule
\textbf{Avg.} & 22.5 & 25.3 & 26.2 & \underline{32.5} & \textbf{33.4} & 27.2 & 25.8 & 29.9 & 26.1 & 28.8 & 26.9 \\
\bottomrule
\end{tabular}}
\caption{NDCG|FC@10 across all domains (grouped by category). Best in \textbf{bold}, second best \underline{underlined}.}
\label{tab:temporal_ndcg_fc_10}
\end{table*}

\subsubsection{Discussion of Metric Utility}
Standard metrics like NDCG measure topical relevance but are often "temporally blind." For example, in a query regarding Bitcoin's evolution from 2017 to 2024, a BM25 retriever might achieve a high NDCG by returning several documents about Bitcoin from 2015. However, our TC metric would assign a low score (0.0 or 0.5) because the retrieved evidence fails to span the required comparison periods. This meta-evaluation proves that our metrics capture a unique dimension of retrieval quality essential for RAG systems in technical and evolving domains.

\subsection{Metric Validation and Meta-Evaluation}
\label{app:metric_validation}

To validate the scientific rigor of our proposed metrics—Temporal Precision (TP) and Temporal Coverage (TC)—we performed a meta-evaluation using a suite of "golden cases." These cases test the intent detection, temporal synthesis requirements, and mathematical stability of the LLM-as-judge framework.

\subsubsection{Intent and Cross-Period Gating}
Table \ref{tab:intent_val} illustrates the system's ability to distinguish between different temporal needs. Notably, the system correctly identifies atemporal technical definitions (e.g., "What is a Merkle tree?") as having no temporal intent, preventing the inflation of metrics on non-temporal data. It also successfully identifies queries requiring cross-period analysis (e.g., "since 2017").

\begin{table*}[ht]
\centering
\small
\caption{Validation of Temporal Intent and Cross-Period Classification Logic.}
\label{tab:intent_val}
\begin{tabular}{lll}
\toprule
\textbf{Query Example} & \textbf{Detected Intent} & \textbf{Cross-Period} \\ \midrule
How have Bitcoin fees changed since 2017? & \texttt{change\_over\_time} & \textbf{True} \\
What is a Merkle tree? & \texttt{none} & False \\
Who was the president in 1920? & \texttt{specific\_time} & False \\
Recent evolution of the Lightning Network. & \texttt{change\_over\_time} & \textbf{True} \\ \bottomrule
\end{tabular}
\end{table*}

\subsubsection{Temporal Coverage Synthesis}
Table \ref{tab:tc_matrix} demonstrates how the TC metric enforces evidence synthesis. A retriever cannot achieve a perfect score by simply finding historical or current data in isolation; it must provide a set of documents that cover both the baseline and comparison periods.

\begin{table*}[t]
\centering
\small
\caption{Temporal Coverage (TC) Decision Matrix for Trend Analysis Queries.}
\label{tab:tc_matrix}
\begin{tabular}{lccc}
\toprule
\textbf{Document Type} & \textbf{Covers Baseline} & \textbf{Covers Comparison} & \textbf{TC Score} \\ \midrule
Baseline Only & True & False & 0.500 \\
Comparison Only & False & True & 0.500 \\
Both (Full Coverage) & True & True & \textbf{1.000} \\ \bottomrule
\end{tabular}
\end{table*}

\subsubsection{Ranking and Mathematical Stability}
Table \ref{tab:math_weight} confirms the rank-sensitivity of our Temporal Precision metric. By applying a position-weighted calculation, the metric appropriately rewards systems that surface temporally relevant evidence at higher ranks.

\begin{table*}[t]
\centering
\small
\caption{Mathematical Weighting of Temporal Precision (TP) at $k=5$.}
\label{tab:math_weight}
\begin{tabular}{llr}
\toprule
\textbf{Scenario} & \textbf{Verdict Array} & \textbf{Temporal Precision} \\ \midrule
Rank 1 Hit & $[1, 0, 0, 0, 0]$ & \textbf{1.000} \\
Rank 5 Hit & $[0, 0, 0, 0, 1]$ & 0.200 \\
Double Hit (Top) & $[1, 1, 0, 0, 0]$ & 1.000 \\ \bottomrule
\end{tabular}
\end{table*}


\section{Temporal signals Ablation Analysis}
\label{sec:ablation_appendix}

\subsection{Performance Degradation by Domain}

Table~\ref{tab:degradation_analysis} shows the performance change (in NDCG@10 points) when temporal signals are removed from queries. Positive values indicate degradation; negative values (in blue) indicate improvement. Models with smaller positive values demonstrate better robustness to missing temporal information. Notably, several model-domain combinations show improved performance when temporal signals are removed, suggesting that explicit temporal phrases may sometimes introduce noise. BM25 shows remarkably stable performance across domains (average +0.4), while E5 shows the highest average degradation (+3.4), with particularly large drops on Cardano (+10.3) and Genealogy (+6.3).

\begin{table*}[t]
\centering

\resizebox{\textwidth}{!}{%
\begin{tabular}{lcccccccccccc}
\toprule
\textbf{Domain} & \textbf{BM25} & \textbf{BGE} & \textbf{Contriever} & \textbf{DiVeR} & \textbf{E5} & \textbf{GritLM} & \textbf{Inst-L} & \textbf{Qwen} & \textbf{Rader} & \textbf{ReasonIR} & \textbf{SBERT} & \textbf{SFR} \\
\midrule
\rowcolor{gray!12}\multicolumn{13}{c}{\textit{Blockchain}} \\
\midrule
\textbf{Bitcoin} & +0.8 & +2.3 & +1.5 & +0.3 & +0.9 & +2.2 & \textcolor{red}{-0.1} & \textcolor{red}{\textbf{-0.8}} & +0.6 & +2.7 & +2.2 & +0.7 \\
\textbf{Cardano} & +3.8 & \textcolor{red}{-1.2} & +2.8 & +0.7 & \textcolor{red}{+10.3} & +4.7 & +0.2 & +3.5 & \textcolor{red}{\textbf{-2.6}} & \textcolor{red}{-0.7} & +0.0 & +4.8 \\
\textbf{Iota} & \textcolor{red}{\textbf{-7.5}} & +4.2 & +8.7 & +6.8 & +5.8 & +7.1 & +3.8 & +5.9 & \textcolor{red}{-0.8} & +8.9 & +3.3 & +5.7 \\
\textbf{Monero} & \textbf{+0.1} & +1.4 & +0.5 & +0.5 & +2.7 & +0.8 & +2.9 & +0.8 & +5.1 & +2.7 & +1.4 & +2.0 \\
\midrule
\rowcolor{gray!12}\multicolumn{13}{c}{\textit{Social Sci.}} \\
\midrule
\textbf{Economics} & +1.4 & +1.0 & +2.2 & +1.6 & +2.1 & +3.5 & +1.9 & +0.9 & +3.1 & +0.7 & \textbf{+0.3} & +1.5 \\
\textbf{Law} & +3.8 & +8.5 & +6.0 & \textcolor{red}{\textbf{-0.2}} & +4.4 & +0.4 & +3.1 & +3.6 & +4.7 & +0.3 & +3.7 & +3.0 \\
\textbf{Politics} & +1.9 & +1.7 & +1.0 & +0.4 & +2.3 & +0.4 & \textcolor{red}{\textbf{-0.3}} & +0.6 & +1.8 & +2.4 & +1.0 & +1.1 \\
\textbf{History} & \textbf{+0.4} & +2.1 & +1.1 & +2.4 & +1.7 & +1.9 & +0.9 & +1.8 & +1.9 & +1.9 & +1.8 & +2.0 \\
\midrule
\rowcolor{gray!12}\multicolumn{13}{c}{\textit{Applied}} \\
\midrule
\textbf{Quant} & \textcolor{red}{\textbf{-0.1}} & +1.8 & +0.4 & +4.2 & +1.3 & +2.3 & +1.9 & +1.3 & +6.9 & +3.4 & +0.7 & +2.3 \\
\textbf{Travel} & +0.1 & +2.9 & +2.1 & +1.1 & +2.5 & +0.8 & +1.6 & \textcolor{red}{\textbf{-0.0}} & +4.0 & +0.2 & +0.7 & +1.6 \\
\textbf{Workplace} & +0.2 & +2.2 & +1.0 & +1.8 & +1.9 & +2.5 & +0.6 & \textcolor{red}{-0.0} & +3.7 & \textcolor{red}{\textbf{-1.9}} & +3.8 & \textcolor{red}{-0.5} \\
\textbf{Genealogy} & \textcolor{red}{\textbf{-0.3}} & +3.0 & +3.4 & +5.4 & +6.3 & +2.0 & +2.8 & +1.2 & +2.3 & +4.4 & +0.7 & +2.3 \\
\midrule
\rowcolor{gray!12}\multicolumn{13}{c}{\textit{STEM}} \\
\midrule
\textbf{HSM} & \textbf{+0.5} & +2.9 & +2.0 & +2.0 & +2.7 & +2.2 & +1.3 & +2.0 & +0.5 & +2.9 & +1.6 & +1.9 \\
\midrule
\textbf{Avg.} & \textbf{+0.4} & +2.5 & +2.5 & +2.1 & +3.4 & +2.4 & +1.6 & +1.6 & +2.4 & +2.1 & +1.6 & +2.2 \\
\bottomrule
\end{tabular}%
}
\caption{Performance degradation (Baseline - Strip) in NDCG@10 points across domains. Positive values indicate performance drop when temporal signals are removed. Best retention (smallest drop) in \textbf{bold}.}
\label{tab:degradation_analysis}
\end{table*}

\subsection{Per-Domain Results by Query Variant}

This section presents complete per-domain NDCG@10 results for each query variant, enabling analysis of domain-specific sensitivity to temporal information.

\subsubsection{Baseline (Original Queries)}

Results using original TEMPO queries containing natural temporal expressions, topical content, and reasoning requirements. These serve as the reference point for comparing other query variants.

\begin{table*}[t]
\centering

\resizebox{\textwidth}{!}{%
\begin{tabular}{lcccccccccccc}
\toprule
\textbf{Domain} & \textbf{BM25} & \textbf{BGE} & \textbf{Contriever} & \textbf{DiVeR} & \textbf{E5} & \textbf{GritLM} & \textbf{Inst-L} & \textbf{Qwen} & \textbf{Rader} & \textbf{ReasonIR} & \textbf{SBERT} & \textbf{SFR} \\
\midrule
\rowcolor{gray!12}\multicolumn{13}{c}{\textit{Blockchain}} \\
\midrule
\textbf{Bitcoin} & 6.2 & 14.4 & 13.3 & 17.4 & 16.3 & \textbf{19.1} & 15.7 & 11.4 & 14.9 & 16.4 & 14.3 & \underline{17.6} \\
\textbf{Cardano} & 13.4 & 13.1 & 12.1 & \underline{29.3} & \textbf{35.7} & 21.7 & 14.6 & 20.6 & 18.6 & 22.9 & 21.4 & 28.1 \\
\textbf{Iota} & 9.7 & 36.1 & 38.3 & 38.2 & \underline{41.7} & 36.6 & 34.3 & 28.6 & 19.2 & \textbf{41.8} & 33.2 & 37.1 \\
\textbf{Monero} & 2.8 & 14.5 & 9.9 & 20.3 & 20.0 & 14.7 & 16.9 & 11.0 & \underline{21.0} & 19.7 & 15.1 & \textbf{23.7} \\
\midrule
\rowcolor{gray!12}\multicolumn{13}{c}{\textit{Social Sci.}} \\
\midrule
\textbf{Economics} & 5.8 & 12.6 & 16.3 & \textbf{27.8} & \underline{25.0} & 17.2 & 17.5 & 17.1 & 22.7 & 20.0 & 15.3 & 21.9 \\
\textbf{Law} & 12.7 & 31.9 & 28.1 & \underline{40.4} & 34.0 & 38.3 & 37.3 & 32.0 & 33.5 & 38.0 & 33.8 & \textbf{40.8} \\
\textbf{Politics} & 32.7 & 28.2 & 31.6 & \underline{45.5} & \textbf{47.9} & 41.4 & 32.6 & 38.1 & 32.4 & 35.4 & 34.6 & 44.9 \\
\textbf{History} & 9.2 & 27.4 & 26.5 & \textbf{34.5} & 28.7 & 27.3 & 28.5 & 25.6 & 25.8 & \underline{34.4} & 28.7 & 32.4 \\
\midrule
\rowcolor{gray!12}\multicolumn{13}{c}{\textit{Applied}} \\
\midrule
\textbf{Quant} & 2.5 & 11.7 & 11.1 & \underline{27.2} & 13.8 & 21.6 & 14.6 & 12.7 & \textbf{27.8} & 19.5 & 15.7 & 16.8 \\
\textbf{Travel} & 4.6 & 23.8 & 23.7 & 26.8 & \underline{28.3} & 25.0 & 25.0 & 22.0 & 26.1 & 21.5 & 27.3 & \textbf{29.7} \\
\textbf{Workplace} & 6.2 & 27.2 & 23.9 & \textbf{42.6} & 32.9 & 30.8 & 36.2 & 30.3 & \underline{36.6} & 30.0 & 34.6 & 31.6 \\
\textbf{Genealogy} & 13.3 & 22.0 & 24.9 & \textbf{35.6} & \underline{33.5} & 26.9 & 24.6 & 25.3 & 18.7 & 30.3 & 23.5 & 31.7 \\
\midrule
\rowcolor{gray!12}\multicolumn{13}{c}{\textit{STEM}} \\
\midrule
\textbf{HSM} & 21.2 & 23.2 & 18.9 & 31.0 & \textbf{37.7} & 33.4 & 24.4 & 21.3 & 16.9 & 24.7 & 26.1 & \underline{33.5} \\
\midrule
\textbf{Avg.} & 10.8 & 22.0 & 21.4 & \textbf{32.0} & \underline{30.4} & 27.2 & 24.8 & 22.8 & 24.2 & 27.3 & 24.9 & 30.0 \\
\bottomrule
\end{tabular}}
\caption{NDCG@10 for Baseline query variant across all domains. Best in \textbf{bold}, second best \underline{underlined}.}
\label{tab:variant_baseline}
\end{table*}

\subsubsection{Strip (Temporal Signals Removed)}

Results after removing explicit temporal signals (dates, temporal phrases) from queries while preserving topical content. Most models show modest degradation (1.6-3.4 points on average), with some domain-specific improvements indicating temporal signals may introduce noise in certain contexts.

\begin{table*}[t]
\centering
\caption{NDCG@10 for Strip query variant across all domains. Best in \textbf{bold}, second best \underline{underlined}.}
\label{tab:variant_strip}
\resizebox{\textwidth}{!}{%
\begin{tabular}{lcccccccccccc}
\toprule
\textbf{Domain} & \textbf{BM25} & \textbf{BGE} & \textbf{Contriever} & \textbf{DiVeR} & \textbf{E5} & \textbf{GritLM} & \textbf{Inst-L} & \textbf{Qwen} & \textbf{Rader} & \textbf{ReasonIR} & \textbf{SBERT} & \textbf{SFR} \\
\midrule
\rowcolor{gray!12}\multicolumn{13}{c}{\textit{Blockchain}} \\
\midrule
\textbf{Bitcoin} & 5.4 & 12.1 & 11.8 & \textbf{17.1} & 15.5 & 16.9 & 15.7 & 12.2 & 14.3 & 13.6 & 12.0 & \underline{16.9} \\
\textbf{Cardano} & 9.6 & 14.3 & 9.3 & \textbf{28.5} & \underline{25.4} & 17.1 & 14.4 & 17.1 & 21.3 & 23.6 & 21.4 & 23.3 \\
\textbf{Iota} & 17.2 & 31.8 & 29.6 & 31.4 & \textbf{35.9} & 29.5 & 30.5 & 22.7 & 20.0 & \underline{32.9} & 29.9 & 31.4 \\
\textbf{Monero} & 2.7 & 13.1 & 9.4 & \underline{19.8} & 17.3 & 13.9 & 14.0 & 10.2 & 16.0 & 17.0 & 13.6 & \textbf{21.7} \\
\midrule
\rowcolor{gray!12}\multicolumn{13}{c}{\textit{Social Sci.}} \\
\midrule
\textbf{Economics} & 4.4 & 11.6 & 14.1 & \textbf{26.1} & \underline{22.9} & 13.7 & 15.5 & 16.2 & 19.6 & 19.3 & 15.1 & 20.4 \\
\textbf{Law} & 8.9 & 23.4 & 22.1 & \textbf{40.6} & 29.7 & \underline{38.0} & 34.3 & 28.5 & 28.8 & 37.6 & 30.1 & 37.8 \\
\textbf{Politics} & 30.8 & 26.5 & 30.7 & \underline{45.1} & \textbf{45.5} & 40.9 & 32.9 & 37.5 & 30.6 & 33.1 & 33.6 & 43.9 \\
\textbf{History} & 8.8 & 25.3 & 25.4 & \underline{32.1} & 27.0 & 25.4 & 27.6 & 23.8 & 24.0 & \textbf{32.5} & 26.9 & 30.3 \\
\midrule
\rowcolor{gray!12}\multicolumn{13}{c}{\textit{Applied}} \\
\midrule
\textbf{Quant} & 2.6 & 9.9 & 10.7 & \textbf{22.9} & 12.5 & 19.3 & 12.8 & 11.4 & \underline{20.9} & 16.1 & 14.9 & 14.6 \\
\textbf{Travel} & 4.5 & 20.9 & 21.5 & 25.7 & 25.7 & 24.2 & 23.4 & 22.0 & 22.1 & 21.2 & \underline{26.6} & \textbf{28.1} \\
\textbf{Workplace} & 6.0 & 24.9 & 23.0 & \textbf{40.7} & 31.0 & 28.3 & \underline{35.6} & 30.3 & 32.9 & 31.9 & 30.9 & 32.1 \\
\textbf{Genealogy} & 13.6 & 19.1 & 21.5 & \textbf{30.1} & 27.2 & 25.0 & 21.8 & 24.1 & 16.4 & 26.0 & 22.8 & \underline{29.4} \\
\midrule
\rowcolor{gray!12}\multicolumn{13}{c}{\textit{STEM}} \\
\midrule
\textbf{HSM} & 20.7 & 20.2 & 16.9 & 29.0 & \textbf{35.1} & 31.3 & 23.0 & 19.3 & 16.3 & 21.8 & 24.5 & \underline{31.5} \\
\midrule
\textbf{Avg.} & 10.4 & 19.5 & 18.9 & \textbf{29.9} & 27.0 & 24.9 & 23.2 & 21.2 & 21.8 & 25.1 & 23.3 & \underline{27.8} \\
\bottomrule
\end{tabular}}
\end{table*}

\subsubsection{Temporal-Only (Only Temporal Information Retained)}

Results using only temporal information (time anchors, temporal events) with topical content removed. All models experience substantial performance drops (DiVeR: 32.0→17.7), demonstrating that temporal reasoning requires grounding in topical context.

\begin{table*}[t]
\centering
\caption{NDCG@10 for Temporal-Only query variant across all domains. Best in \textbf{bold}, second best \underline{underlined}.}
\label{tab:variant_temporal_only}
\resizebox{\textwidth}{!}{%
\begin{tabular}{lcccccccccccc}
\toprule
\textbf{Domain} & \textbf{BM25} & \textbf{BGE} & \textbf{Contriever} & \textbf{DiVeR} & \textbf{E5} & \textbf{GritLM} & \textbf{Inst-L} & \textbf{Qwen} & \textbf{Rader} & \textbf{ReasonIR} & \textbf{SBERT} & \textbf{SFR} \\
\midrule
\rowcolor{gray!12}\multicolumn{13}{c}{\textit{Blockchain}} \\
\midrule
\textbf{Bitcoin} & 4.3 & 5.9 & 6.1 & \textbf{11.8} & 9.0 & 8.5 & 7.9 & 6.7 & 6.6 & 8.5 & 6.0 & \underline{10.3} \\
\textbf{Cardano} & 14.4 & 7.7 & 7.0 & \underline{15.3} & \textbf{17.8} & 10.4 & 8.3 & 10.6 & 8.9 & 9.5 & 10.1 & 13.9 \\
\textbf{Iota} & 5.3 & 17.6 & 17.8 & \textbf{23.7} & 14.8 & 17.8 & \underline{22.4} & 22.3 & 12.0 & 14.2 & 14.7 & 16.3 \\
\textbf{Monero} & 4.4 & 3.7 & 7.9 & 7.9 & 5.6 & \textbf{12.9} & 8.3 & 8.1 & 4.1 & 10.0 & 6.3 & \underline{10.7} \\
\midrule
\rowcolor{gray!12}\multicolumn{13}{c}{\textit{Social Sci.}} \\
\midrule
\textbf{Economics} & 9.8 & 6.5 & 5.8 & \underline{16.7} & 14.4 & 11.9 & 10.3 & \textbf{17.3} & 11.4 & 12.1 & 5.7 & 15.6 \\
\textbf{Law} & 14.2 & 15.7 & 15.6 & 22.7 & 15.2 & 23.2 & 18.4 & \textbf{27.9} & 13.5 & \underline{27.9} & 13.1 & 20.4 \\
\textbf{Politics} & \underline{24.4} & 9.5 & 9.0 & 23.6 & 22.5 & 22.9 & 16.1 & 22.6 & 9.4 & 13.6 & 9.2 & \textbf{25.3} \\
\textbf{History} & 12.4 & 9.8 & 9.6 & \underline{14.1} & 8.8 & 13.9 & 13.0 & \textbf{16.1} & 6.1 & 12.3 & 9.0 & 12.9 \\
\midrule
\rowcolor{gray!12}\multicolumn{13}{c}{\textit{Applied}} \\
\midrule
\textbf{Quant} & 5.3 & 6.2 & 11.4 & 14.1 & 7.7 & 9.0 & 10.8 & \textbf{15.9} & 10.9 & \underline{14.7} & 5.6 & 9.5 \\
\textbf{Travel} & 9.8 & 13.6 & 17.3 & 17.5 & 12.0 & 15.0 & 14.4 & \underline{17.5} & 9.4 & \textbf{18.1} & 10.2 & 16.5 \\
\textbf{Workplace} & 8.3 & 6.9 & 7.0 & \textbf{29.3} & 19.1 & 15.6 & 15.5 & \underline{22.5} & 12.6 & 17.0 & 3.8 & 21.4 \\
\textbf{Genealogy} & 12.9 & 11.7 & 10.0 & 15.4 & \textbf{17.6} & 15.2 & 14.3 & 16.9 & 9.5 & 9.6 & 8.9 & \underline{17.5} \\
\midrule
\rowcolor{gray!12}\multicolumn{13}{c}{\textit{STEM}} \\
\midrule
\textbf{HSM} & 17.6 & 9.8 & 8.3 & 17.5 & 20.2 & \textbf{21.8} & 15.6 & 19.7 & 7.4 & 11.8 & 10.6 & \underline{21.4} \\
\midrule
\textbf{Avg.} & 11.0 & 9.6 & 10.2 & \textbf{17.7} & 14.2 & 15.2 & 13.5 & \underline{17.2} & 9.4 & 13.8 & 8.7 & 16.3 \\
\bottomrule
\end{tabular}}
\end{table*}

\subsubsection{Normalized (Explicit Temporal Intent Tags)}

Results with explicit temporal intent tags (e.g., "TEMPORAL\_INTENT=when; ANCHORS=[2015, 2023]") appended to original queries. ReasonIR shows dramatic improvement to 35.3 NDCG@10 (+8.0 points), while other models show minimal change, indicating that structured temporal metadata benefits instruction-tuned reasoning models specifically.

\begin{table*}[t]
\centering
\caption{NDCG@10 for Normalized query variant across all domains. Best in \textbf{bold}, second best \underline{underlined}.}
\label{tab:variant_normalized}
\resizebox{\textwidth}{!}{%
\begin{tabular}{lcccccccccccc}
\toprule
\textbf{Domain} & \textbf{BM25} & \textbf{BGE} & \textbf{Contriever} & \textbf{DiVeR} & \textbf{E5} & \textbf{GritLM} & \textbf{Inst-L} & \textbf{Qwen} & \textbf{Rader} & \textbf{ReasonIR} & \textbf{SBERT} & \textbf{SFR} \\
\midrule
\rowcolor{gray!12}\multicolumn{13}{c}{\textit{Blockchain}} \\
\midrule
\textbf{Bitcoin} & 6.6 & 13.0 & 12.3 & 13.7 & 14.4 & \textbf{19.0} & 14.3 & 11.2 & 13.4 & \underline{18.6} & 13.5 & 16.6 \\
\textbf{Cardano} & 13.5 & 15.2 & 15.3 & 25.7 & \underline{36.4} & 19.4 & 12.6 & 19.0 & 20.5 & \textbf{41.6} & 19.9 & 27.4 \\
\textbf{Iota} & 8.3 & 35.9 & 37.3 & 39.0 & \underline{41.7} & 38.5 & 34.6 & 32.8 & 21.1 & \textbf{44.7} & 31.8 & 37.0 \\
\textbf{Monero} & 2.6 & 14.5 & 11.6 & 17.3 & 15.0 & 15.7 & 18.9 & 11.3 & 15.0 & \textbf{25.2} & 15.1 & \underline{22.3} \\
\midrule
\rowcolor{gray!12}\multicolumn{13}{c}{\textit{Social Sci.}} \\
\midrule
\textbf{Economics} & 6.2 & 12.4 & 17.9 & 24.3 & \underline{25.0} & 16.2 & 16.9 & 18.6 & 19.5 & \textbf{32.4} & 15.0 & 22.9 \\
\textbf{Law} & 12.4 & 31.1 & 28.9 & 36.0 & 29.9 & \underline{39.5} & 36.9 & 30.5 & 31.6 & \textbf{45.0} & 29.9 & 38.2 \\
\textbf{Politics} & 32.8 & 29.0 & 33.3 & \underline{46.6} & \textbf{49.1} & 41.5 & 34.0 & 40.1 & 30.9 & 42.6 & 35.2 & 45.9 \\
\textbf{History} & 9.4 & 27.1 & 26.3 & \underline{32.6} & 25.9 & 27.0 & 28.1 & 25.3 & 19.9 & \textbf{39.5} & 28.1 & 30.7 \\
\midrule
\rowcolor{gray!12}\multicolumn{13}{c}{\textit{Applied}} \\
\midrule
\textbf{Quant} & 2.5 & 13.2 & 12.4 & \underline{24.9} & 13.0 & 20.7 & 14.3 & 13.4 & 23.3 & \textbf{26.0} & 14.1 & 15.6 \\
\textbf{Travel} & 4.7 & 23.2 & 23.2 & 24.5 & 25.5 & 24.0 & 23.5 & 23.5 & 20.3 & \textbf{29.0} & 25.8 & \underline{28.8} \\
\textbf{Workplace} & 7.5 & 26.9 & 26.1 & \underline{45.2} & 32.3 & 30.1 & 37.7 & 29.8 & 34.8 & \textbf{47.0} & 30.0 & 31.3 \\
\textbf{Genealogy} & 13.2 & 21.8 & 26.5 & 32.2 & \underline{33.9} & 27.5 & 24.6 & 27.3 & 17.7 & \textbf{35.5} & 22.5 & 32.9 \\
\midrule
\rowcolor{gray!12}\multicolumn{13}{c}{\textit{STEM}} \\
\midrule
\textbf{HSM} & 22.2 & 23.8 & 19.5 & 30.9 & \textbf{36.7} & 33.0 & 24.6 & 19.7 & 13.2 & 31.5 & 24.7 & \underline{34.1} \\
\midrule
\textbf{Avg.} & 10.9 & 22.1 & 22.4 & \underline{30.2} & 29.1 & 27.1 & 24.7 & 23.3 & 21.6 & \textbf{35.3} & 23.5 & 29.5 \\
\bottomrule
\end{tabular}}
\end{table*}

\section{Temporal Reasoning Class Definitions}
\label{app:reasoning_classes}

TEMPO queries are categorized into 10 temporal reasoning classes based on the type of temporal inference required. Each class represents distinct patterns of temporal information needs commonly encountered in technical and academic domains.

\subsection{Event Analysis and Localization (624 queries, 36.1\%)}

Queries requiring identification of when specific events occurred and understanding their temporal context. These queries ask about the timing of technical developments, policy implementations, historical incidents, or procedural changes. Examples: "When did Bitcoin Core introduce transaction pruning?", "When was the first recorded use of proof-of-work in cryptocurrency?"

\subsection{Time Period Contextualization (365 queries, 21.1\%)}

Queries requiring situating phenomena, practices, or concepts within specific historical or contemporary time periods. These queries seek to understand what existed, was practiced, or was true during a particular era. Examples: "What consensus mechanisms were available before 2015?", "How were international borders managed during the Cold War era?"

\subsection{Origins, Evolution, and Comparative Analysis (256 queries, 14.8\%)}

Queries requiring tracking how concepts, technologies, policies, or practices emerged and evolved over time, often comparing earlier and later forms. These queries examine historical development, incremental changes, and evolutionary patterns. Examples: "How has Bitcoin's block size debate evolved since 2015?", "How did voting systems change from the 19th to 21st century?"

\subsection{Trends, Changes, and Cross-Period Comparison (154 queries, 8.9\%)}

Queries requiring comparison of states, statistics, or conditions across distinct time periods, often analyzing trends or identifying changes. These queries explicitly contrast baseline and comparison periods to assess temporal shifts. Examples: "How has cryptocurrency adoption changed since 2017?", "How do current immigration policies differ from those in the 1990s?"

\subsection{Event Verification and Authenticity (115 queries, 6.6\%)}

Queries requiring verification of whether events occurred, validation of temporal claims, or assessment of historical accuracy. These queries seek authoritative evidence to confirm or refute temporal assertions. Examples: "Did Satoshi Nakamoto really propose Bitcoin in 2008?", "Was there actually a treaty signed in 1648 ending the Thirty Years' War?"

\subsection{Materials, Artifacts, and Provenance (89 queries, 5.1\%)}

Queries requiring temporal information about physical or digital artifacts, their creation dates, provenance, or historical authenticity. These queries focus on when objects or documents were produced and their historical chain of custody. Examples: "When was this historical document dated?", "What is the earliest known manuscript of this text?"

\subsection{Other (58 queries, 3.4\%)}

Queries with temporal elements that do not fit the primary categories or combine multiple temporal reasoning patterns in unique ways. These represent edge cases or highly specialized temporal information needs.

\subsection{Sources, Methods, and Documentation (25 queries, 1.4\%)}

Queries requiring temporal information about research methods, documentation practices, or source materials used in different time periods. These queries ask about how information was recorded, preserved, or analyzed historically. Examples: "How were genealogical records maintained in the 18th century?", "What statistical methods were available to economists in the 1960s?"

\subsection{Causation Analysis (23 queries, 1.3\%)}

Queries requiring identification of temporal cause-effect relationships, understanding what events or conditions led to specific outcomes. These queries explicitly probe causal chains with temporal dimensions. Examples: "What caused the Bitcoin price surge in 2017?", "What events led to the 2008 financial crisis?"

\subsection{Historical Attribution and Context (21 queries, 1.2\%)}

Queries requiring attribution of ideas, inventions, or discoveries to specific individuals or groups within temporal context. These queries ask about who did what when and under what historical circumstances. Examples: "Who first proposed the concept of blockchain and when?", "Which mathematician first proved this theorem in the 19th century?"

\clearpage

\clearpage

\section{Detailed Analysis \& Results}
\label{app:DetailedAnalysis}
\subsection{Step-wise Retrieval Results}
\label{sec:stepwise_retrieval}

This section presents detailed results for Task 2: Step-wise Retrieval Planning, which evaluates how effectively retrieval systems can leverage decomposed reasoning steps generated by a large language model. Given the multi-step nature of temporal reasoning queries, we investigate whether explicitly decomposing queries into intermediate reasoning steps can improve retrieval performance compared to using the original query alone.

\subsubsection{Experimental Setup}

We generate reasoning steps for each query using GPT-4o, prompting it to decompose temporal queries into logical retrieval steps. For example, a query asking ``How did Bitcoin's consensus mechanism change after the 2017 fork?'' might be decomposed into steps such as: (1) identify Bitcoin's original consensus mechanism, (2) find information about the 2017 fork event, and (3) locate documentation of post-fork consensus changes.

We evaluate three retrieval strategies that vary in how these reasoning steps are incorporated:

\begin{itemize}
    \item \textbf{Step-Only}: Each reasoning step is used independently as the retrieval query, completely replacing the original question. This tests whether the decomposed steps contain sufficient information for retrieval without the original query context.
    \item \textbf{Query+Step}: The original query is concatenated with each reasoning step sequentially. Retrieval is performed for each step separately, and the final score is computed by averaging NDCG@10 across all steps. This approach preserves query context while focusing on individual reasoning components.
    \item \textbf{Query+All}: The original query is concatenated with all reasoning steps combined into a single augmented query. This creates a comprehensive query that includes both the original information need and all decomposed reasoning components.
\end{itemize}

\subsubsection{Overall Performance Comparison}

Figure~\ref{fig:stepwise_comparison} presents the average NDCG@10 across all 13 domains for each retrieval strategy. The results reveal substantial differences in how retrieval models leverage reasoning steps.

The Step-Only strategy yields the lowest performance across nearly all models, with an average NDCG@10 of 14.6 across all retrievers. This confirms that isolated reasoning steps, while logically sound, lack sufficient context for effective document retrieval. The decomposed steps often contain generic sub-queries that match many irrelevant documents, leading to poor precision. SFR achieves the best Step-Only performance at 20.7, followed closely by DiVeR (20.1) and GritLM (20.0), suggesting that larger embedding models can partially compensate for missing query context.

Both query-augmented strategies achieve substantially higher performance. Query+Step achieves an average of 26.4 NDCG@10, marginally outperforming Query+All at 25.9. This slight advantage suggests that averaging across individual step-augmented retrievals provides more robust results than combining all steps into a single dense query, potentially because the latter can introduce noise from less relevant steps.

ReasonIR demonstrates the most dramatic improvement from step-wise planning, increasing from 17.2 NDCG@10 (Step-Only) to 35.0 (Query+Step) and 35.3 (Query+All)---a gain exceeding 18 absolute points. This indicates that reasoning-enhanced retrievers are specifically designed to exploit structured reasoning information. DiVeR shows similar patterns, improving from 20.1 to 33.3 and 32.0 respectively.

In contrast, BM25 remains largely unaffected by the retrieval strategy, achieving 11.6 (Step-Only), 10.8 (Query+Step), and 10.3 (Query+All). This consistent performance across strategies confirms that lexical matching cannot effectively leverage the semantic information encoded in reasoning steps. The slight performance decrease with query augmentation may result from keyword dilution, where additional terms reduce the relative weight of critical temporal keywords.

\subsubsection{Per-Domain Analysis}

Tables~\ref{tab:task2_step_only}, \ref{tab:task2_query_step}, and \ref{tab:task2_query_all} present comprehensive per-domain results for each strategy. Several domain-specific patterns emerge from this analysis.

\paragraph{Workplace Domain} This domain consistently yields the highest scores across all strategies and models. DiVeR achieves 41.0 NDCG@10 with Step-Only, increasing to 55.4 with Query+Step---the highest single-domain score in our evaluation. The structured nature of workplace-related temporal queries (e.g., policy changes, employment regulations over time) appears particularly amenable to step-wise decomposition. ReasonIR follows closely with 35.8 to 54.6 across strategies.

\paragraph{Blockchain Domains} The four blockchain domains exhibit high variance in step-wise retrieval effectiveness. Iota achieves strong results across all strategies, with ReasonIR reaching 45.0 NDCG@10 on Query+All. This may reflect the relatively focused nature of Iota's temporal evolution as a newer cryptocurrency. In contrast, Monero proves consistently challenging, with the best Query+Step result at only 30.1 (ReasonIR). Bitcoin shows moderate improvement from step augmentation (GritLM: 13.3 Step-Only → 20.2 Query+Step), while Cardano demonstrates substantial gains (ReasonIR: 10.7 → 32.3 → 41.3 across strategies).

\paragraph{Social Science Domains} Politics and Law benefit substantially from query-augmented strategies. Politics reaches 51.2 NDCG@10 with E5 on Query+All, while Law achieves 45.0 with ReasonIR on the same strategy. These domains involve complex temporal reasoning about legislative changes, policy evolution, and historical political events, where explicit reasoning steps help disambiguate the information need. History shows more modest gains, with ReasonIR improving from 15.9 (Step-Only) to 38.7 (Query+Step) to 40.1 (Query+All).

\paragraph{Applied Domains} Quant (quantitative finance) presents an interesting case where Rader achieves its best relative performance, reaching 36.8 NDCG@10 on Query+Step compared to 15.2 on Step-Only. This 21.6-point improvement suggests that financial temporal queries benefit significantly from structured decomposition. Travel shows consistent improvement across strategies, with SFR achieving 31.2 on both Query+Step and Query+All. Genealogy demonstrates moderate gains, with E5 reaching 38.6 on Query+Step.

\paragraph{STEM Domain (HSM)} History of Science and Mathematics shows substantial improvement with query augmentation. E5 achieves 37.9 NDCG@10 on Query+All, a 22.8-point improvement over its Step-Only performance of 15.1. This domain involves tracing the evolution of scientific concepts and mathematical discoveries, where temporal decomposition helps identify relevant historical documents.

\subsubsection{Model Architecture Insights}

The results reveal systematic differences based on model architecture:

\paragraph{Sparse Retrieval (BM25)} Lexical matching shows minimal sensitivity to retrieval strategy, with performance hovering around 10--12 NDCG@10 regardless of how reasoning steps are incorporated. This confirms that BM25 cannot semantically interpret reasoning steps and may suffer from keyword dilution when queries are augmented.

\paragraph{Standard Dense Retrievers} Models like BGE, Contriever, and SBERT show substantial improvements with query augmentation. BGE improves from 8.3 (Step-Only) to 21.5 (Query+Step) and 23.2 (Query+All), demonstrating that even smaller dense models can leverage structured reasoning when combined with the original query context.

\paragraph{Large Dense Retrievers} E5, GritLM, Qwen, and SFR achieve strong performance across strategies. E5 demonstrates consistent improvement from 18.0 to 32.1 to 31.8, while SFR shows a similar pattern from 20.7 to 33.3 to 32.0. These models have sufficient capacity to encode both the original query semantics and the additional reasoning context.

\paragraph{Reasoning-Enhanced Retrievers} ReasonIR and DiVeR, specifically designed for reasoning-intensive retrieval, show the largest absolute gains from step-wise planning. ReasonIR's improvement of 18+ points confirms that these architectures are optimized to exploit structured reasoning information. Interestingly, ReasonIR performs better on Query+All (35.3) than Query+Step (35.0), unlike most other models, suggesting it can effectively synthesize all reasoning steps simultaneously.

\subsubsection{Implications}

These results have several implications for temporal reasoning-intensive retrieval:

First, query context is essential for step-wise retrieval. Isolated reasoning steps, while logically coherent, lack the specificity needed for accurate document retrieval. Systems implementing step-wise retrieval should always maintain the original query alongside decomposed steps.

Second, the choice between Query+Step and Query+All strategies depends on the retrieval model. For most dense retrievers, Query+Step provides marginally better results, possibly because averaging across steps reduces noise from less relevant decompositions. However, reasoning-enhanced models like ReasonIR can effectively leverage all steps simultaneously.

Third, the substantial performance gap between sparse and neural retrieval in step-wise settings (BM25: ~11 vs. ReasonIR: ~35) suggests that semantic understanding is crucial for interpreting reasoning steps. This gap is larger than observed in standard retrieval (Table~\ref{tab:results_all_domains}), indicating that step-wise retrieval amplifies the advantage of neural approaches.

\begin{table*}[t]
\centering
\small
\caption{Task 2 Step-wise Retrieval: NDCG@10 using \textbf{Step-Only} strategy (individual retrieval steps without the original query). Best in \textbf{bold}, second best \underline{underlined}.}
\label{tab:task2_step_only}
\resizebox{\textwidth}{!}{%
\begin{tabular}{lcccccccccccc}
\toprule
\textbf{Domain} & \textbf{BM25} & \textbf{BGE} & \textbf{Contriever} & \textbf{DiVeR} & \textbf{E5} & \textbf{GritLM} & \textbf{Inst-L} & \textbf{Qwen} & \textbf{Rader} & \textbf{ReasonIR} & \textbf{SBERT} & \textbf{SFR} \\
\midrule
\rowcolor{gray!12}\multicolumn{13}{c}{\textit{Blockchain}} \\
\midrule
\textbf{Bitcoin} & 4.1 & 5.4 & 6.1 & 11.5 & 8.9 & \textbf{13.3} & 6.3 & 10.0 & 5.3 & 10.0 & 7.7 & \underline{12.1} \\
\textbf{Cardano} & 9.1 & 4.1 & 1.5 & 11.1 & \textbf{14.9} & 8.4 & 3.3 & 8.6 & 3.6 & 10.7 & 6.6 & \underline{14.2} \\
\textbf{Iota} & 15.1 & 9.7 & 9.4 & 18.5 & \textbf{25.8} & 16.5 & 14.8 & 23.4 & 6.2 & 6.8 & 9.5 & \underline{25.5} \\
\textbf{Monero} & 7.4 & 3.1 & 11.2 & 12.0 & 5.7 & \textbf{21.1} & 9.8 & 10.7 & 3.7 & 13.3 & 9.0 & \underline{13.8} \\
\midrule
\rowcolor{gray!12}\multicolumn{13}{c}{\textit{Social Sci.}} \\
\midrule
\textbf{Economics} & 8.6 & 5.5 & 7.2 & \underline{19.0} & 14.4 & 14.5 & 9.1 & \textbf{19.2} & 8.2 & 13.2 & 7.5 & 18.5 \\
\textbf{Law} & 17.3 & 13.2 & 15.8 & \underline{26.7} & 19.8 & 26.4 & 20.9 & 25.8 & 15.2 & \textbf{29.4} & 12.4 & 26.2 \\
\textbf{Politics} & 18.6 & 7.4 & 9.6 & 22.9 & 24.1 & \textbf{25.5} & 17.8 & 22.2 & 8.9 & 17.1 & 13.5 & \underline{25.0} \\
\textbf{History} & 10.9 & 7.2 & 10.2 & \textbf{19.3} & 10.6 & 17.4 & 14.6 & \underline{17.7} & 6.4 & 15.9 & 10.7 & 15.9 \\
\midrule
\rowcolor{gray!12}\multicolumn{13}{c}{\textit{Applied}} \\
\midrule
\textbf{Quant} & 6.0 & 5.7 & 18.5 & 20.9 & 10.7 & \underline{22.6} & 9.6 & 21.2 & 15.2 & \textbf{26.1} & 9.9 & 14.4 \\
\textbf{Travel} & 6.7 & 9.3 & 15.7 & 19.5 & 17.6 & \textbf{24.4} & 16.7 & 18.2 & 9.3 & 15.3 & 16.1 & \underline{23.4} \\
\textbf{Workplace} & 19.4 & 16.4 & 11.4 & \textbf{41.0} & \underline{38.6} & 28.7 & 22.6 & 38.6 & 15.9 & 35.8 & 18.1 & 38.1 \\
\textbf{Genealogy} & 15.1 & 15.5 & 12.8 & 23.9 & \textbf{27.1} & 23.0 & 20.2 & 23.2 & 12.7 & 19.3 & 14.8 & \underline{26.1} \\
\midrule
\rowcolor{gray!12}\multicolumn{13}{c}{\textit{STEM}} \\
\midrule
\textbf{HSM} & 12.5 & 5.1 & 6.9 & 14.5 & 15.1 & \textbf{18.6} & 11.9 & 13.5 & 3.5 & 11.3 & 10.3 & \underline{16.1} \\
\midrule
\midrule
\textbf{Avg.} & 11.6 & 8.3 & 10.5 & \underline{20.1} & 18.0 & 20.0 & 13.7 & 19.4 & 8.8 & 17.2 & 11.2 & \textbf{20.7} \\
\bottomrule
\end{tabular}}
\end{table*}
\begin{table*}[t]
\centering
\small
\caption{Task 2 Step-wise Retrieval: NDCG@10 using \textbf{Query+Step} strategy (query concatenated with each step sequentially, then averaged across steps). Best in \textbf{bold}, second best \underline{underlined}.}
\label{tab:task2_query_step}
\resizebox{\textwidth}{!}{%
\begin{tabular}{lcccccccccccc}
\toprule
\textbf{Domain} & \textbf{BM25} & \textbf{BGE} & \textbf{Contriever} & \textbf{DiVeR} & \textbf{E5} & \textbf{GritLM} & \textbf{Inst-L} & \textbf{Qwen} & \textbf{Rader} & \textbf{ReasonIR} & \textbf{SBERT} & \textbf{SFR} \\
\midrule
\rowcolor{gray!12}\multicolumn{13}{c}{\textit{Blockchain}} \\
\midrule
\textbf{Bitcoin} & 6.2 & 13.9 & 12.0 & 19.8 & 19.2 & 20.2 & 17.1 & 14.0 & 18.7 & \underline{20.7} & 16.9 & \textbf{20.8} \\
\textbf{Cardano} & 9.7 & 12.6 & 11.0 & 27.1 & \underline{31.5} & 18.0 & 13.9 & 20.5 & 21.5 & \textbf{32.3} & 17.7 & 25.5 \\
\textbf{Iota} & 13.8 & 31.4 & 29.8 & 31.1 & 41.0 & 33.2 & 35.1 & \textbf{42.1} & 16.9 & 38.7 & 28.1 & \textbf{42.1} \\
\textbf{Monero} & 4.2 & 22.4 & 16.0 & 22.9 & 21.1 & 23.7 & 25.2 & 14.2 & 18.5 & \textbf{30.1} & 19.3 & \underline{30.0} \\
\midrule
\rowcolor{gray!12}\multicolumn{13}{c}{\textit{Social Sci.}} \\
\midrule
\textbf{Economics} & 5.8 & 11.1 & 15.6 & \underline{27.7} & 26.1 & 17.9 & 16.5 & 24.8 & 22.9 & \textbf{28.5} & 16.6 & 26.1 \\
\textbf{Law} & 13.8 & 25.4 & 22.3 & 38.1 & 30.8 & 35.6 & 34.9 & 26.1 & 36.3 & \textbf{39.0} & 28.6 & \underline{38.2} \\
\textbf{Politics} & 28.1 & 24.6 & 30.6 & \textbf{48.2} & \underline{47.8} & 40.2 & 35.3 & 41.5 & 32.9 & 40.8 & 35.5 & 45.8 \\
\textbf{History} & 9.3 & 23.1 & 23.5 & \underline{34.5} & 25.7 & 26.5 & 28.8 & 30.8 & 25.2 & \textbf{38.7} & 28.8 & 32.0 \\
\midrule
\rowcolor{gray!12}\multicolumn{13}{c}{\textit{Applied}} \\
\midrule
\textbf{Quant} & 1.9 & 16.0 & 13.3 & 33.0 & 19.1 & 23.2 & 20.8 & 23.1 & \textbf{36.8} & \underline{34.0} & 17.9 & 24.4 \\
\textbf{Travel} & 4.5 & 19.8 & 21.3 & 25.0 & \underline{29.8} & 27.1 & 26.7 & 24.8 & 23.3 & \underline{29.8} & 28.2 & \textbf{31.2} \\
\textbf{Workplace} & 9.4 & 29.4 & 24.3 & \textbf{55.4} & 52.1 & 40.1 & 44.2 & 52.4 & 39.8 & \underline{54.6} & 45.3 & 47.9 \\
\textbf{Genealogy} & 14.1 & 27.2 & 31.1 & \textbf{40.0} & \underline{38.6} & 28.9 & 29.3 & 29.5 & 24.7 & 37.6 & 27.7 & 37.5 \\
\midrule
\rowcolor{gray!12}\multicolumn{13}{c}{\textit{STEM}} \\
\midrule
\textbf{HSM} & 20.2 & 22.0 & 17.6 & 30.1 & \textbf{35.0} & 31.6 & 22.3 & 22.6 & 14.2 & 29.8 & 26.2 & \underline{31.8} \\
\midrule
\midrule
\textbf{Avg.} & 10.8 & 21.5 & 20.7 & \underline{33.3} & 32.1 & 28.2 & 26.9 & 28.2 & 25.5 & \textbf{35.0} & 25.9 & \underline{33.3} \\
\bottomrule
\end{tabular}}
\end{table*}
\begin{table*}[t]
\centering
\small
\caption{Task 2 Step-wise Retrieval: NDCG@10 using \textbf{Query+All} strategy (query concatenated with all steps combined). Best in \textbf{bold}, second best \underline{underlined}.}
\label{tab:task2_query_all}
\resizebox{\textwidth}{!}{%
\begin{tabular}{lcccccccccccc}
\toprule
\textbf{Domain} & \textbf{BM25} & \textbf{BGE} & \textbf{Contriever} & \textbf{DiVeR} & \textbf{E5} & \textbf{GritLM} & \textbf{Inst-L} & \textbf{Qwen} & \textbf{Rader} & \textbf{ReasonIR} & \textbf{SBERT} & \textbf{SFR} \\
\midrule
\rowcolor{gray!12}\multicolumn{13}{c}{\textit{Blockchain}} \\
\midrule
\textbf{Bitcoin} & 5.9 & 15.0 & 13.8 & 17.2 & \underline{18.3} & 18.1 & 15.1 & 11.3 & 17.5 & 17.9 & 14.1 & \textbf{18.6} \\
\textbf{Cardano} & 12.1 & 14.8 & 14.7 & 25.5 & \underline{38.0} & 24.1 & 16.3 & 19.0 & 19.9 & \textbf{41.3} & 22.1 & 30.7 \\
\textbf{Iota} & 8.6 & 33.0 & 39.5 & 37.3 & \underline{41.3} & 37.1 & 31.8 & 35.0 & 19.2 & \textbf{45.0} & 32.1 & 39.0 \\
\textbf{Monero} & 1.7 & 17.7 & 12.4 & 17.6 & 16.2 & 15.6 & 17.6 & 11.1 & 15.5 & \underline{21.4} & 14.4 & \textbf{23.4} \\
\midrule
\rowcolor{gray!12}\multicolumn{13}{c}{\textit{Social Sci.}} \\
\midrule
\textbf{Economics} & 5.5 & 13.8 & 18.9 & \underline{28.9} & 27.3 & 18.0 & 19.4 & 24.6 & 24.6 & \textbf{32.4} & 16.8 & 25.0 \\
\textbf{Law} & 10.5 & 31.8 & 31.0 & \underline{41.8} & 34.3 & 41.3 & 39.8 & 31.5 & 36.2 & \textbf{45.0} & 33.4 & 41.2 \\
\textbf{Politics} & 32.0 & 29.4 & 35.0 & \underline{47.5} & \textbf{51.2} & 42.0 & 34.8 & 42.9 & 33.7 & 43.2 & 35.2 & 47.3 \\
\textbf{History} & 8.2 & 27.8 & 28.0 & \underline{33.8} & 25.7 & 27.4 & 29.7 & 31.1 & 23.9 & \textbf{40.1} & 28.5 & 31.5 \\
\midrule
\rowcolor{gray!12}\multicolumn{13}{c}{\textit{Applied}} \\
\midrule
\textbf{Quant} & 0.6 & 14.6 & 11.5 & \underline{25.0} & 16.0 & 21.2 & 16.3 & 15.8 & \textbf{25.6} & 23.9 & 14.5 & 19.3 \\
\textbf{Travel} & 5.1 & 25.3 & 25.9 & 25.0 & 29.4 & 27.7 & 26.9 & 27.8 & 22.3 & \underline{30.3} & 27.7 & \textbf{31.2} \\
\textbf{Workplace} & 9.3 & 31.3 & 30.6 & \underline{50.3} & 40.1 & 31.6 & 42.5 & 44.7 & 38.4 & \textbf{52.5} & 40.0 & 36.4 \\
\textbf{Genealogy} & 12.7 & 23.2 & 27.6 & 35.3 & \textbf{37.4} & 26.2 & 27.7 & 27.6 & 21.3 & 33.9 & 22.7 & \underline{35.6} \\
\midrule
\rowcolor{gray!12}\multicolumn{13}{c}{\textit{STEM}} \\
\midrule
\textbf{HSM} & 22.1 & 24.4 & 20.9 & 30.9 & \textbf{37.9} & 32.9 & 24.6 & 22.0 & 12.2 & 32.2 & 24.8 & \underline{36.6} \\
\midrule
\midrule
\textbf{Avg.} & 10.3 & 23.2 & 23.8 & \underline{32.0} & 31.8 & 27.9 & 26.3 & 26.5 & 23.9 & \textbf{35.3} & 25.1 & 32.0 \\
\bottomrule
\end{tabular}}
\end{table*}

\clearpage

\subsection{Reasoning Class Analysis}
\label{sec:reasoning_class_appendix}

This section provides detailed results for the reasoning class analysis presented in Section~\ref{sec:reasoning_class_analysis}. We evaluate retrieval performance across 10 temporal reasoning classes derived from our taxonomy.

\subsubsection{Reasoning Class Definitions}

Table~\ref{tab:reasoning_class_definitions} provides definitions for each reasoning class abbreviation used throughout this analysis.

\begin{table*}[h]
\centering
\small
\caption{Temporal reasoning class definitions and query counts.}
\label{tab:reasoning_class_definitions}
\begin{tabular}{llp{5.5cm}r}
\toprule
\textbf{Abbr.} & \textbf{Full Name} & \textbf{Description} & \textbf{Queries} \\
\midrule
EAL & Event Analysis \& Localization & Identifying when specific events occurred or analyzing their temporal properties & 624 \\
TPC & Time Period Contextualization & Understanding practices, norms, or conditions within a specific historical period & 365 \\
OEC & Origins \& Evolution Comparative & Tracing how concepts, technologies, or practices developed over time & 256 \\
TCP & Trends \& Cross-Period Comparison & Comparing information across multiple time periods to identify changes or patterns & 154 \\
EVA & Event Verification \& Authenticity & Confirming whether events occurred and verifying their temporal accuracy & 115 \\
MAP & Materials \& Artifacts Provenance & Determining the origin, age, or temporal history of physical or digital artifacts & 89 \\
OTH & Other & Queries with missing, ambiguous, or rare temporal classifications & 58 \\
SMD & Sources \& Methods Documentation & Locating historical documentation or methodological records from specific periods & 25 \\
CAU & Causation Analysis & Understanding temporal cause-effect relationships between events & 23 \\
HAC & Historical Attribution \& Context & Attributing ideas, inventions, or actions to specific individuals and time periods & 21 \\
\bottomrule
\end{tabular}
\end{table*}

\subsubsection{Performance Heatmap}

\begin{figure*}[h]
    \centering
    \includegraphics[width=\textwidth]{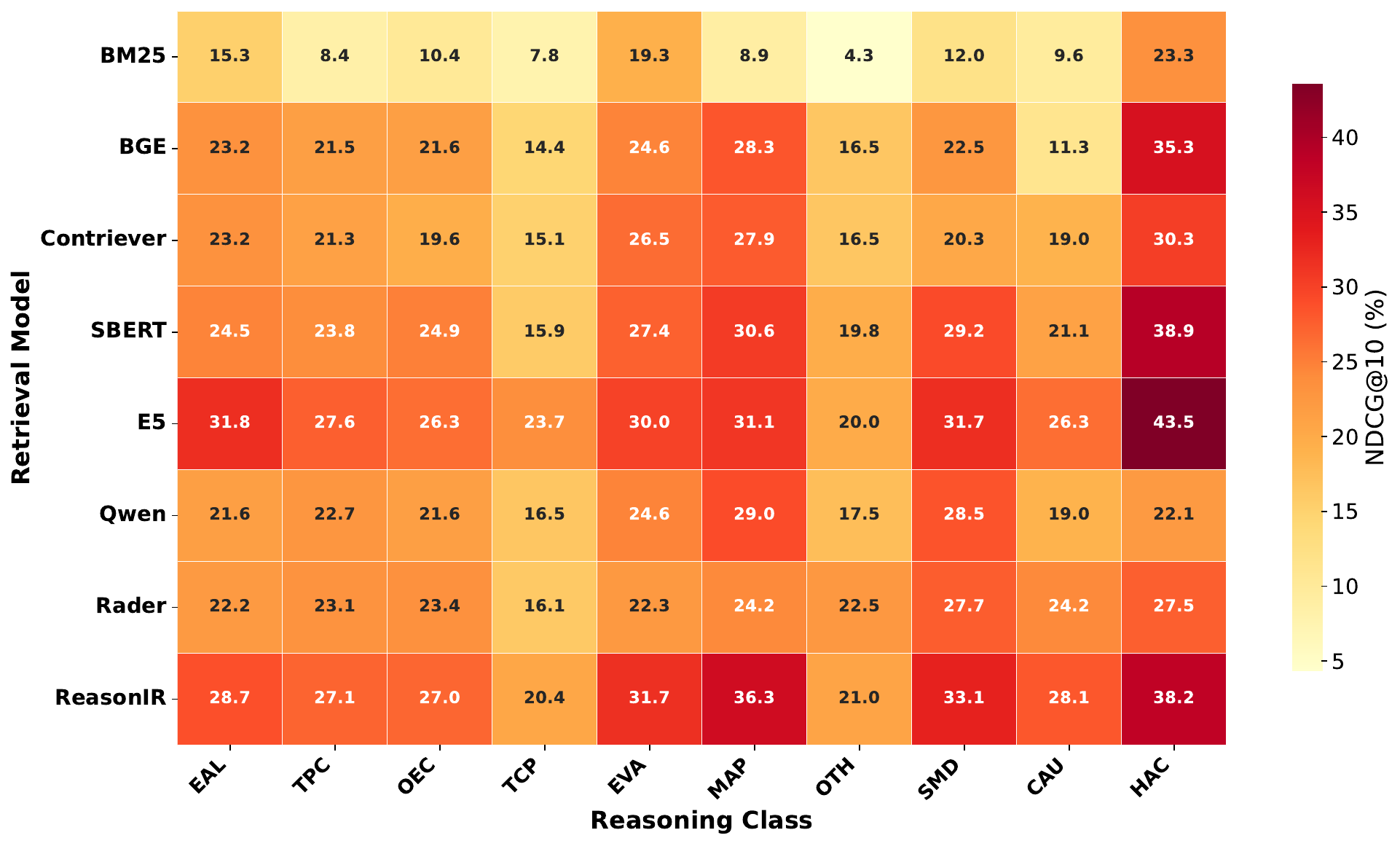}
    \caption{Heatmap of NDCG@10 performance across retrieval models and temporal reasoning classes. Darker colors indicate higher performance. DiVeR and ReasonIR show the most consistent performance across reasoning types, while BM25 exhibits high variance (4.3--23.3 range).}
    \label{fig:reasoning_class_heatmap}
\end{figure*}

Figure~\ref{fig:reasoning_class_heatmap} presents a heatmap visualization of NDCG@10 performance across all model-class combinations. Darker colors indicate higher performance. The visualization reveals clear patterns: (1) the HAC column shows consistently higher scores across most models, (2) the TCP column shows uniformly lower performance, and (3) reasoning-enhanced models (DiVeR, ReasonIR) show more uniform performance across classes compared to the high variance observed in BM25.

\subsubsection{Detailed Results}

Table~\ref{tab:reasoning_class_results} presents complete NDCG@10 scores for all 11 retrieval models across 10 reasoning classes. Key findings include:

\begin{itemize}
    \item \textbf{Best overall}: DiVeR (31.5 avg) achieves the highest average performance, followed by SFR (31.1) and ReasonIR/E5 (29.2).
    \item \textbf{Hardest class}: TCP (Trends \& Cross-Period) with an average of 17.9 across all models.
    \item \textbf{Easiest class}: HAC (Historical Attribution \& Context) with an average of 35.6 across all models.
    \item \textbf{Largest performance gap}: BM25 shows a 19.0-point gap between its best (HAC: 23.3) and worst (OTH: 4.3) classes.
    \item \textbf{Most consistent}: DiVeR shows the smallest coefficient of variation across classes among neural models.
\end{itemize}

\begin{table*}[t]
\centering
\small

\resizebox{\textwidth}{!}{%
\begin{tabular}{lccccccccccc}
\toprule
\textbf{Model} & \textbf{EAL} & \textbf{TPC} & \textbf{OEC} & \textbf{TCP} & \textbf{EVA} & \textbf{MAP} & \textbf{OTH} & \textbf{SMD} & \textbf{CAU} & \textbf{HAC} & \textbf{Avg.} \\
\midrule
\rowcolor{gray!12}\multicolumn{12}{c}{\textit{Sparse}} \\
\midrule
\textbf{BM25} & 15.3 & 8.4 & 10.4 & 7.8 & 19.3 & 8.9 & 4.3 & 12.0 & 9.6 & 23.3 & 11.9 \\
\midrule
\rowcolor{gray!12}\multicolumn{12}{c}{\textit{Dense ($<$1B)}} \\
\midrule
\textbf{BGE} & 23.2 & 21.5 & 21.6 & 14.4 & 24.6 & 28.3 & 16.5 & 22.5 & 11.3 & 35.3 & 21.9 \\
\textbf{Contriever} & 23.2 & 21.3 & 19.6 & 15.1 & 26.5 & 27.9 & 16.5 & 20.3 & 19.0 & 30.3 & 22.0 \\
\textbf{SBERT} & 24.5 & 23.8 & 24.9 & 15.9 & 27.4 & 30.6 & 19.8 & 29.2 & 21.1 & 38.9 & 25.6 \\
\midrule
\rowcolor{gray!12}\multicolumn{12}{c}{\textit{Dense ($>$1B)}} \\
\midrule
\textbf{E5} & \underline{31.8} & 27.6 & 26.3 & \underline{23.7} & 30.0 & 31.1 & 20.0 & 31.7 & 26.3 & \underline{43.5} & 29.2 \\
\textbf{GritLM} & 28.2 & 24.5 & 24.1 & 19.7 & \underline{32.5} & 27.4 & 16.9 & 28.8 & 19.1 & 40.5 & 26.2 \\
\textbf{Qwen} & 21.6 & 22.7 & 21.6 & 16.5 & 24.6 & 29.0 & 17.5 & 28.5 & 19.0 & 22.1 & 22.3 \\
\textbf{SFR} & \textbf{32.3} & \underline{28.4} & 26.9 & 23.4 & 32.4 & 34.8 & 21.6 & \textbf{36.0} & 23.7 & \textbf{51.8} & \underline{31.1} \\
\midrule
\rowcolor{gray!12}\multicolumn{12}{c}{\textit{Reasoning}} \\
\midrule
\textbf{Rader} & 22.2 & 23.1 & 23.4 & 16.1 & 22.3 & 24.2 & \underline{22.5} & 27.7 & 24.2 & 27.5 & 23.3 \\
\textbf{DiVeR} & 31.2 & \textbf{31.8} & \textbf{28.0} & \textbf{23.9} & \textbf{35.8} & \textbf{38.1} & \textbf{24.3} & \underline{33.7} & \underline{27.4} & 40.7 & \textbf{31.5} \\
\textbf{ReasonIR} & 28.7 & 27.1 & \underline{27.0} & 20.4 & 31.7 & \underline{36.3} & 21.0 & 33.1 & \textbf{28.1} & 38.2 & 29.2 \\
\bottomrule
\end{tabular}%
}
\caption{NDCG@10 performance across temporal reasoning classes. Models are evaluated on 10 reasoning classes: EAL=Event Analysis \& Localization (624), TPC=Time Period Contextualization (365), OEC=Origins \& Evolution Comparative (256), TCP=Trends \& Cross-Period Comparison (154), EVA=Event Verification \& Authenticity (115), MAP=Materials \& Artifacts Provenance (89), OTH=Other (58), SMD=Sources \& Methods Documentation (25), CAU=Causation Analysis (23), HAC=Historical Attribution \& Context (21). The ``Other'' category includes queries with missing or rare classifications. Best score per class in \textbf{bold}, second best \underline{underlined}.}
\label{tab:reasoning_class_results}
\end{table*}

\clearpage

\subsection{LLM Query Reformulation Analysis}
\label{sec:reformulation_appendix}

This section provides detailed methodology and results for the query reformulation experiments presented in Section~\ref{sec:reformulation_analysis}.

\subsubsection{Motivation}

Recent work on reasoning-intensive retrieval~\cite{su2024bright} has shown that augmenting queries with LLM-generated reasoning can improve retrieval for complex information needs. We investigate whether this approach benefits temporal reasoning queries, which often require understanding implicit temporal constraints, event relationships, and cross-period comparisons.

\subsubsection{Methodology}

We reformulate each of the 1,730 TEMPO queries using six LLMs of varying architectures and scales: (1) \textbf{GPT-4o}: OpenAI's flagship multimodal model (API-based). (2) \textbf{Llama-3.3-70B-Instruct}: Meta's instruction-tuned model. (3) \textbf{Qwen2.5-72B-Instruct}: Alibaba's large instruction model. (4) \textbf{Qwen2.5-32B-Instruct}: Mid-scale Qwen variant.  (5) \textbf{DeepSeek-R1-Distill-Qwen-32B}: Reasoning-distilled model. 
Each LLM receives the following prompt to generate reasoning about the query's information needs: The generated reasoning is concatenated with the original query to form the reformulated query, which is then used for retrieval with each of the 12 retrieval models evaluated in the main experiments.

\begin{figure}[h]
\centering
\begin{tcolorbox}[
  colback=gray!5,
  colframe=gray!75,
  title=Query Reformulation Prompt,
  width=\columnwidth,
]
\small\ttfamily
Analyze the following question and generate detailed reasoning about what kind of information would help answer it.

\medskip
\textbf{Question:} \{query\}

\medskip
\textbf{Instructions:}
\begin{enumerate}
  \item Identify the core problem or question being asked.
  \item Reason step-by-step about what concepts, knowledge, or solutions would be relevant.
  \item Think about what a helpful document should contain.
\end{enumerate}

\textbf{Provide your analysis:}
\end{tcolorbox}
\caption{Query reformulation prompt. The model generates step-by-step reasoning about the information needs to expand the query before retrieval.}
\label{prompt:reformulation}
\end{figure}

\paragraph{Reasoning-enhanced retrievers benefit most.} ReasonIR shows the largest improvement, gaining +13.7 NDCG@10 with GPT-4o reformulation (27.2$\rightarrow$41.0). This dramatic improvement suggests that ReasonIR's architecture---specifically designed to process reasoning chains---can effectively leverage the step-by-step analysis generated by LLMs. The model appears to extract relevant semantic signals from the expanded context that align with its internal reasoning mechanisms.

\paragraph{Sparse retrieval suffers from context dilution.} BM25 consistently degrades across all reformulation strategies, dropping from 10.8 to 4.3--6.2 NDCG@10. The LLM-generated reasoning introduces many terms that, while semantically relevant, dilute the lexical overlap between query and relevant documents. This confirms that reasoning augmentation is fundamentally incompatible with term-matching approaches.

\paragraph{Some models have internalized reasoning.} DiVeR maintains remarkably stable performance (30.7--32.2) regardless of reformulation strategy, suggesting its reasoning capabilities are already internalized during training and do not benefit from---nor are harmed by---external reasoning augmentation. This architectural difference from ReasonIR highlights distinct approaches to reasoning-enhanced retrieval.

\paragraph{LLM choice matters, but not always scale.} GPT-4o and DeepSeek-32B produce the most effective reformulations, while larger models do not consistently outperform smaller ones. Qwen-72B underperforms Qwen-32B for several retrievers, suggesting that reformulation quality depends more on reasoning style than raw model capacity.

\paragraph{Domain-specific patterns.} Per-domain analysis (Tables~\ref{tab:reformulation_gpt_4o}--\ref{tab:reformulation_qwen_32b}) reveals that reformulation benefits vary by domain. Technical domains (Blockchain, Quant) show more consistent improvements with reasoning augmentation, while social science domains exhibit higher variance. This may reflect differences in how temporal reasoning manifests across subject areas.

\subsubsection{Implications}

These findings have practical implications for temporal retrieval system design:

\begin{enumerate}
    \item \textbf{Architecture selection}: When LLM-based query expansion is available, reasoning-enhanced retrievers like ReasonIR should be preferred over sparse methods.
    \item \textbf{Hybrid approaches}: Systems could route queries to different retrievers based on whether reasoning augmentation is applied---using BM25 for original queries and ReasonIR for reformulated ones.
    \item \textbf{Cost-benefit tradeoff}: The computational cost of LLM reformulation may be justified for reasoning-intensive queries but unnecessary for simple temporal lookups.
\end{enumerate}

\subsubsection{Detailed Results}

Tables~\ref{tab:reformulation_gpt_4o}--\ref{tab:reformulation_qwen_32b} present per-domain NDCG@10 scores for each reformulation strategy across all 12 retrieval models and 13 domains.

\begin{table*}[t]
\centering
\small

\resizebox{\textwidth}{!}{%
\begin{tabular}{lcccccccccccc}
\toprule
\textbf{Domain} & \textbf{BM25} & \textbf{BGE} & \textbf{Contriever} & \textbf{DiVeR} & \textbf{E5} & \textbf{GritLM} & \textbf{Inst-L} & \textbf{Qwen} & \textbf{Rader} & \textbf{ReasonIR} & \textbf{SBERT} & \textbf{SFR} \\
\midrule
\rowcolor{gray!12}\multicolumn{13}{c}{\textit{Blockchain}} \\
\midrule
\textbf{Bitcoin} & 1.4 & 17.0 & 10.8 & 15.8 & 16.0 & 20.2 & 10.6 & 13.3 & 12.0 & \textbf{22.5} & 14.7 & \underline{20.9} \\
\textbf{Cardano} & 3.2 & 26.2 & 16.6 & 31.9 & \underline{34.5} & 25.4 & 17.6 & 26.6 & 23.5 & \textbf{40.9} & 17.4 & 33.8 \\
\textbf{Iota} & 0.0 & 30.2 & 32.7 & 31.2 & 25.4 & \underline{33.7} & 20.3 & 23.8 & 20.5 & \textbf{38.5} & 23.2 & 29.1 \\
\textbf{Monero} & 0.0 & \underline{17.6} & 7.4 & 14.0 & 9.3 & 16.9 & 10.3 & 11.9 & 11.2 & \textbf{21.9} & 12.8 & 16.8 \\
\midrule
\rowcolor{gray!12}\multicolumn{13}{c}{\textit{Social Sci.}} \\
\midrule
\textbf{Economics} & 3.7 & 19.3 & 22.8 & \underline{28.9} & 25.5 & 19.8 & 24.4 & 26.9 & 21.1 & \textbf{45.9} & 16.5 & 26.3 \\
\textbf{Law} & 3.8 & 34.2 & 30.3 & 37.2 & 35.0 & \underline{43.1} & 32.5 & 34.2 & 28.9 & \textbf{43.7} & 33.0 & 40.0 \\
\textbf{Politics} & 29.2 & 32.9 & 35.2 & \underline{50.3} & 48.1 & 44.2 & 47.3 & 41.1 & 31.8 & \textbf{57.1} & 35.9 & 46.9 \\
\textbf{History} & 1.6 & 28.3 & 28.3 & \underline{30.3} & 21.8 & 27.4 & 22.8 & 28.0 & 18.4 & \textbf{43.3} & 27.0 & 28.6 \\
\midrule
\rowcolor{gray!12}\multicolumn{13}{c}{\textit{Applied}} \\
\midrule
\textbf{Quant} & 0.0 & 12.0 & 15.6 & \underline{24.9} & 16.2 & 24.2 & 7.2 & 18.9 & 20.2 & \textbf{38.3} & 18.1 & 22.7 \\
\textbf{Travel} & 1.1 & 23.9 & 20.1 & 23.1 & 20.6 & 26.9 & 17.7 & 27.0 & 19.0 & \textbf{32.1} & 22.5 & \underline{28.3} \\
\textbf{Workplace} & 10.1 & 42.6 & 43.4 & 47.6 & 51.7 & 45.6 & \underline{52.9} & 52.3 & 44.7 & \textbf{70.3} & 41.0 & 48.2 \\
\textbf{Genealogy} & 4.4 & 25.4 & 31.2 & 32.9 & \underline{37.2} & 24.8 & 31.3 & 28.6 & 16.0 & \textbf{39.3} & 20.6 & 36.1 \\
\midrule
\rowcolor{gray!12}\multicolumn{13}{c}{\textit{STEM}} \\
\midrule
\textbf{HSM} & 22.1 & 25.6 & 26.2 & 31.2 & 35.3 & 34.6 & \underline{37.8} & 27.1 & 12.3 & \textbf{39.0} & 27.2 & 35.2 \\
\midrule
\midrule
\textbf{Avg.} & 8.0 & 25.8 & 24.7 & 30.7 & 29.0 & 29.7 & 25.6 & 27.7 & 21.5 & \textbf{41.0} & 23.8 & \underline{31.8} \\
\bottomrule
\end{tabular}}
\caption{NDCG@10 performance using queries reformulated by \textbf{GPT-4o}. The LLM generates reasoning about information needs before retrieval. Best in \textbf{bold}, second best \underline{underlined}.}
\label{tab:reformulation_gpt_4o}
\end{table*}

\begin{table*}[t]
\centering
\small

\resizebox{\textwidth}{!}{%
\begin{tabular}{lcccccccccccc}
\toprule
\textbf{Domain} & \textbf{BM25} & \textbf{BGE} & \textbf{Contriever} & \textbf{DiVeR} & \textbf{E5} & \textbf{GritLM} & \textbf{Inst-L} & \textbf{Qwen} & \textbf{Rader} & \textbf{ReasonIR} & \textbf{SBERT} & \textbf{SFR} \\
\midrule
\rowcolor{gray!12}\multicolumn{13}{c}{\textit{Blockchain}} \\
\midrule
\textbf{Bitcoin} & 1.0 & 15.4 & 11.1 & 18.4 & 18.9 & 18.5 & 10.8 & 16.1 & 18.4 & \textbf{22.7} & 15.1 & \underline{19.4} \\
\textbf{Cardano} & 2.3 & 19.7 & 22.9 & 28.3 & \underline{29.6} & 22.6 & 13.8 & 18.5 & 20.9 & \textbf{34.7} & 19.2 & 27.9 \\
\textbf{Iota} & 0.0 & \underline{37.5} & 35.4 & 27.3 & 28.4 & 29.0 & 23.1 & 27.9 & 29.7 & \textbf{44.8} & 27.6 & 28.1 \\
\textbf{Monero} & 0.0 & 17.7 & 9.7 & \underline{18.3} & 10.8 & 17.3 & 8.6 & 15.3 & 13.3 & \textbf{23.6} & 14.9 & 18.1 \\
\midrule
\rowcolor{gray!12}\multicolumn{13}{c}{\textit{Social Sci.}} \\
\midrule
\textbf{Economics} & 2.1 & 16.5 & 22.4 & \underline{32.1} & 24.3 & 20.8 & 18.8 & 26.6 & 24.9 & \textbf{43.5} & 17.3 & 24.9 \\
\textbf{Law} & 4.4 & 39.4 & 29.8 & 41.5 & 36.4 & \underline{41.5} & 32.2 & 35.1 & 35.7 & \textbf{43.5} & 33.6 & 39.1 \\
\textbf{Politics} & 27.5 & 33.6 & 39.5 & \underline{49.2} & 48.9 & 42.9 & 45.2 & 41.2 & 34.6 & \textbf{55.5} & 35.2 & 46.9 \\
\textbf{History} & 1.7 & 30.5 & 31.5 & \underline{33.9} & 24.4 & 27.6 & 20.7 & 30.9 & 23.6 & \textbf{43.5} & 27.9 & 29.5 \\
\midrule
\rowcolor{gray!12}\multicolumn{13}{c}{\textit{Applied}} \\
\midrule
\textbf{Quant} & 0.0 & 14.0 & 18.3 & \underline{27.4} & 19.9 & 27.1 & 7.7 & 20.0 & 26.9 & \textbf{34.4} & 21.9 & 22.7 \\
\textbf{Travel} & 0.5 & 25.2 & 21.9 & 23.9 & 19.3 & 26.1 & 14.8 & \textbf{28.2} & 19.3 & \underline{27.1} & 26.2 & 24.1 \\
\textbf{Workplace} & 4.8 & 38.6 & 43.1 & \underline{50.2} & 45.1 & 36.3 & 46.0 & 47.7 & 40.1 & \textbf{65.0} & 41.2 & 43.3 \\
\textbf{Genealogy} & 2.5 & 23.5 & 28.3 & 33.4 & \textbf{39.0} & 25.9 & 23.4 & 29.2 & 17.3 & 36.1 & 18.5 & \underline{36.3} \\
\midrule
\rowcolor{gray!12}\multicolumn{13}{c}{\textit{STEM}} \\
\midrule
\textbf{HSM} & 22.0 & 30.7 & 30.4 & 34.2 & 37.7 & 34.2 & \underline{39.1} & 29.8 & 18.0 & \textbf{41.0} & 27.9 & 36.6 \\
\midrule
\midrule
\textbf{Avg.} & 6.9 & 26.3 & 26.5 & \underline{32.2} & 29.4 & 28.5 & 23.4 & 28.2 & 24.8 & \textbf{39.6} & 25.1 & 30.5 \\
\bottomrule
\end{tabular}}
\caption{NDCG@10 performance using queries reformulated by \textbf{DeepSeek-32B}. The LLM generates reasoning about information needs before retrieval. Best in \textbf{bold}, second best \underline{underlined}.}
\label{tab:reformulation_deepseek_32b}
\end{table*}

\begin{table*}[t]
\centering
\small

\resizebox{\textwidth}{!}{%
\begin{tabular}{lcccccccccccc}
\toprule
\textbf{Domain} & \textbf{BM25} & \textbf{BGE} & \textbf{Contriever} & \textbf{DiVeR} & \textbf{E5} & \textbf{GritLM} & \textbf{Inst-L} & \textbf{Qwen} & \textbf{Rader} & \textbf{ReasonIR} & \textbf{SBERT} & \textbf{SFR} \\
\midrule
\rowcolor{gray!12}\multicolumn{13}{c}{\textit{Blockchain}} \\
\midrule
\textbf{Bitcoin} & 1.2 & 12.9 & 6.5 & 15.2 & 15.8 & 19.8 & 12.5 & 13.9 & 11.9 & \textbf{20.7} & 10.2 & \underline{20.0} \\
\textbf{Cardano} & 1.9 & 13.0 & 10.6 & 26.1 & 26.6 & 23.9 & 17.2 & 21.4 & 11.8 & \underline{28.2} & 19.5 & \textbf{29.0} \\
\textbf{Iota} & 2.3 & 25.2 & 31.1 & 33.3 & 22.3 & 32.3 & 23.4 & \underline{37.5} & 23.7 & \textbf{37.9} & 15.6 & 32.4 \\
\textbf{Monero} & 0.0 & 11.4 & 4.6 & 15.3 & 6.2 & \underline{17.4} & 13.6 & 12.4 & 9.6 & \textbf{20.2} & 10.5 & 13.9 \\
\midrule
\rowcolor{gray!12}\multicolumn{13}{c}{\textit{Social Sci.}} \\
\midrule
\textbf{Economics} & 3.2 & 14.5 & 18.4 & \underline{28.8} & 20.9 & 20.9 & 22.6 & 25.3 & 21.6 & \textbf{43.6} & 14.6 & 23.5 \\
\textbf{Law} & 1.8 & 27.7 & 25.8 & 41.1 & 31.6 & \underline{42.7} & 35.3 & 37.5 & 32.4 & \textbf{46.1} & 37.8 & 37.6 \\
\textbf{Politics} & 20.7 & 26.8 & 28.2 & \underline{47.6} & 43.7 & 42.1 & 41.3 & 39.5 & 30.3 & \textbf{54.9} & 32.5 & 43.5 \\
\textbf{History} & 0.9 & 24.1 & 22.7 & \underline{30.8} & 18.3 & 27.9 & 26.0 & 27.3 & 18.8 & \textbf{40.6} & 24.6 & 25.2 \\
\midrule
\rowcolor{gray!12}\multicolumn{13}{c}{\textit{Applied}} \\
\midrule
\textbf{Quant} & 0.0 & 12.9 & 19.1 & \underline{27.1} & 16.4 & 25.6 & 9.2 & 18.6 & 22.5 & \textbf{36.8} & 14.7 & 21.4 \\
\textbf{Travel} & 1.7 & 21.2 & 17.9 & 24.0 & 16.9 & \textbf{28.2} & 20.0 & 24.8 & 16.6 & \underline{26.7} & 21.3 & 22.5 \\
\textbf{Workplace} & 9.3 & 40.5 & 40.6 & 49.4 & 50.2 & 47.4 & \underline{52.8} & 50.3 & 41.9 & \textbf{66.1} & 40.5 & 46.1 \\
\textbf{Genealogy} & 2.2 & 22.4 & 25.5 & 31.2 & \underline{33.5} & 26.2 & 29.0 & 29.3 & 17.0 & \textbf{37.4} & 21.2 & 32.3 \\
\midrule
\rowcolor{gray!12}\multicolumn{13}{c}{\textit{STEM}} \\
\midrule
\textbf{HSM} & 15.4 & 21.9 & 18.4 & 31.7 & 30.5 & \textbf{34.3} & 32.1 & 25.8 & 8.3 & \underline{32.8} & 25.6 & 30.3 \\
\midrule
\midrule
\textbf{Avg.} & 5.5 & 21.1 & 20.7 & \underline{30.9} & 25.6 & 29.9 & 25.8 & 28.0 & 20.5 & \textbf{37.8} & 22.2 & 29.0 \\
\bottomrule
\end{tabular}}
\caption{NDCG@10 performance using queries reformulated by \textbf{Llama-70B}. The LLM generates reasoning about information needs before retrieval. Best in \textbf{bold}, second best \underline{underlined}.}
\label{tab:reformulation_llama_70b}
\end{table*}

\begin{table*}[t]
\centering
\small

\resizebox{\textwidth}{!}{%
\begin{tabular}{lcccccccccccc}
\toprule
\textbf{Domain} & \textbf{BM25} & \textbf{BGE} & \textbf{Contriever} & \textbf{DiVeR} & \textbf{E5} & \textbf{GritLM} & \textbf{Inst-L} & \textbf{Qwen} & \textbf{Rader} & \textbf{ReasonIR} & \textbf{SBERT} & \textbf{SFR} \\
\midrule
\rowcolor{gray!12}\multicolumn{13}{c}{\textit{Blockchain}} \\
\midrule
\textbf{Bitcoin} & 1.1 & 16.9 & 8.9 & 17.0 & 16.6 & 18.4 & 12.3 & 14.1 & 14.1 & \textbf{22.9} & 10.9 & \underline{21.2} \\
\textbf{Cardano} & 0.5 & 19.1 & 14.8 & 27.5 & 26.3 & 23.2 & 18.7 & 18.1 & 14.8 & \textbf{32.8} & 16.4 & \underline{28.2} \\
\textbf{Iota} & 0.0 & 27.0 & \underline{34.5} & 29.0 & 21.7 & 32.0 & 21.4 & 33.8 & 22.4 & \textbf{45.8} & 14.3 & 27.1 \\
\textbf{Monero} & 0.0 & 14.9 & 8.1 & \underline{17.3} & 5.1 & 16.8 & 8.9 & 9.3 & 10.3 & \textbf{19.5} & 9.5 & 11.8 \\
\midrule
\rowcolor{gray!12}\multicolumn{13}{c}{\textit{Social Sci.}} \\
\midrule
\textbf{Economics} & 2.2 & 19.3 & 21.4 & \underline{29.1} & 18.9 & 19.3 & 21.8 & 25.1 & 18.8 & \textbf{43.8} & 14.0 & 23.2 \\
\textbf{Law} & 2.5 & 36.8 & 27.1 & 38.9 & 32.5 & 37.5 & 33.8 & 34.6 & 32.6 & \textbf{46.8} & 33.5 & \underline{39.6} \\
\textbf{Politics} & 20.4 & 27.8 & 30.4 & \underline{50.5} & 42.2 & 43.0 & 41.1 & 36.3 & 25.9 & \textbf{53.5} & 32.3 & 42.8 \\
\textbf{History} & 1.2 & 24.8 & 24.3 & \underline{30.2} & 17.5 & 26.2 & 25.6 & 26.6 & 15.1 & \textbf{41.5} & 23.4 & 24.4 \\
\midrule
\rowcolor{gray!12}\multicolumn{13}{c}{\textit{Applied}} \\
\midrule
\textbf{Quant} & 0.0 & 12.6 & 19.5 & \underline{24.3} & 15.3 & 23.2 & 8.5 & 17.2 & 21.1 & \textbf{32.7} & 17.7 & 19.7 \\
\textbf{Travel} & 1.1 & 23.6 & 17.6 & 25.5 & 17.1 & \underline{26.1} & 20.4 & 25.4 & 14.0 & \textbf{31.6} & 19.8 & 23.5 \\
\textbf{Workplace} & 9.8 & 41.2 & 35.1 & 48.9 & 42.8 & 44.1 & 49.1 & \underline{50.1} & 37.8 & \textbf{66.3} & 40.5 & 42.5 \\
\textbf{Genealogy} & 2.0 & 26.4 & 29.4 & 31.2 & 32.5 & 26.3 & 30.9 & 25.9 & 14.7 & \textbf{37.3} & 19.2 & \underline{32.6} \\
\midrule
\rowcolor{gray!12}\multicolumn{13}{c}{\textit{STEM}} \\
\midrule
\textbf{HSM} & 17.8 & 25.0 & 21.7 & 31.7 & 30.6 & 34.6 & \underline{35.5} & 26.7 & 8.4 & \textbf{35.6} & 25.7 & 32.4 \\
\midrule
\midrule
\textbf{Avg.} & 5.9 & 24.3 & 22.5 & \underline{30.9} & 24.6 & 28.5 & 25.2 & 26.4 & 19.2 & \textbf{39.2} & 21.3 & 28.4 \\
\bottomrule
\end{tabular}}
\caption{NDCG@10 performance using queries reformulated by \textbf{Qwen-72B}. The LLM generates reasoning about information needs before retrieval. Best in \textbf{bold}, second best \underline{underlined}.}
\label{tab:reformulation_qwen_72b}
\end{table*}

\begin{table*}[t]
\centering
\small

\resizebox{\textwidth}{!}{%
\begin{tabular}{lcccccccccccc}
\toprule
\textbf{Domain} & \textbf{BM25} & \textbf{BGE} & \textbf{Contriever} & \textbf{DiVeR} & \textbf{E5} & \textbf{GritLM} & \textbf{Inst-L} & \textbf{Qwen} & \textbf{Rader} & \textbf{ReasonIR} & \textbf{SBERT} & \textbf{SFR} \\
\midrule
\rowcolor{gray!12}\multicolumn{13}{c}{\textit{Blockchain}} \\
\midrule
\textbf{Bitcoin} & 0.0 & 14.5 & 9.6 & 15.7 & 15.8 & \textbf{19.9} & 12.8 & 14.1 & 13.5 & \underline{19.8} & 9.3 & 18.8 \\
\textbf{Cardano} & 0.6 & 15.3 & 11.6 & 27.9 & \underline{31.5} & 22.0 & 17.2 & 18.9 & 14.7 & \textbf{35.1} & 21.3 & 31.3 \\
\textbf{Iota} & 0.0 & 23.0 & 29.7 & \underline{31.6} & 24.5 & 31.3 & 21.5 & 23.9 & 15.2 & \textbf{37.6} & 26.7 & 27.3 \\
\textbf{Monero} & 0.0 & 14.4 & 6.5 & \underline{19.0} & 9.3 & 16.8 & 15.0 & 11.7 & 13.3 & \textbf{19.1} & 10.4 & 14.5 \\
\midrule
\rowcolor{gray!12}\multicolumn{13}{c}{\textit{Social Sci.}} \\
\midrule
\textbf{Economics} & 3.6 & 15.6 & 16.7 & \underline{31.3} & 18.7 & 20.7 & 23.8 & 22.9 & 18.8 & \textbf{39.1} & 15.3 & 21.5 \\
\textbf{Law} & 3.2 & 32.6 & 23.2 & 38.8 & 36.6 & \underline{42.4} & 38.1 & 36.7 & 29.0 & 36.6 & 32.4 & \textbf{43.4} \\
\textbf{Politics} & 20.6 & 27.0 & 27.8 & \underline{46.7} & 44.6 & 41.2 & 38.4 & 37.2 & 29.5 & \textbf{49.4} & 32.0 & 44.5 \\
\textbf{History} & 2.0 & 23.1 & 21.9 & \underline{30.5} & 19.9 & 25.6 & 26.9 & 27.1 & 18.7 & \textbf{37.6} & 23.2 & 26.1 \\
\midrule
\rowcolor{gray!12}\multicolumn{13}{c}{\textit{Applied}} \\
\midrule
\textbf{Quant} & 0.0 & 15.3 & 14.1 & \underline{28.9} & 19.3 & 26.0 & 11.7 & 19.3 & 25.5 & \textbf{34.4} & 18.6 & 20.6 \\
\textbf{Travel} & 2.8 & 20.0 & 16.8 & 23.7 & 21.7 & \textbf{26.0} & 19.6 & 23.4 & 17.7 & \underline{25.1} & 20.7 & 24.3 \\
\textbf{Workplace} & 6.5 & 36.5 & 37.7 & \underline{47.7} & 41.1 & 37.7 & \textbf{48.0} & 41.1 & 37.4 & 47.6 & 39.4 & 40.5 \\
\textbf{Genealogy} & 3.4 & 22.6 & 23.6 & 32.8 & \underline{33.6} & 25.7 & 28.3 & 26.3 & 17.5 & \textbf{34.8} & 21.2 & 32.4 \\
\midrule
\rowcolor{gray!12}\multicolumn{13}{c}{\textit{STEM}} \\
\midrule
\textbf{HSM} & 13.0 & 23.5 & 19.1 & 31.0 & 31.0 & \underline{32.3} & 31.3 & 24.6 & 11.3 & \textbf{33.1} & 22.1 & 31.0 \\
\midrule
\midrule
\textbf{Avg.} & 6.2 & 21.8 & 19.9 & \underline{31.2} & 26.7 & 28.3 & 25.6 & 25.2 & 20.1 & \textbf{34.6} & 22.5 & 28.9 \\
\bottomrule
\end{tabular}}
\caption{NDCG@10 performance using queries reformulated by \textbf{Qwen-32B}. The LLM generates reasoning about information needs before retrieval. Best in \textbf{bold}, second best \underline{underlined}.}
\label{tab:reformulation_qwen_32b}
\end{table*}

\subsection{Additional Retrieval Metrics}
\label{app:additional_metrices}
In this section, we provide a multidimensional view of model performance on the TEMPO benchmark using standard IR metrics. Table~\ref{tab:model_comparison} presents the global averages across all 13 domains, while Tables~\ref{tab:detailed_sfr} through \ref{tab:detailed_bge} offer fine-grained per-domain breakdowns for key representative models.

The supplemental results reveal several critical patterns regarding temporal retrieval capability:

\begin{itemize}
    \item \textbf{High-Speed Convergence:} DiVeR achieves the highest overall Mean Reciprocal Rank (MRR) of 39.8, indicating that its reasoning-enhanced architecture is highly effective at surfacing a relevant document at the first position. This is further supported by its leading Recall@10 (40.1\%) and Precision@10 (11.1\%).
    
    \item \textbf{The Parameter Tier:} A clear performance gap exists between standard dense retrievers and large models ($>1$B). E5 and SFR outperform smaller models by significant margins in MAP and Recall. This suggests that representing the nuanced temporal dependencies in our dataset requires the expanded capacity of larger encoders.
    
    \item \textbf{Domain Robustness:} Domain-specific analysis (Tables~\ref{tab:detailed_sfr}--\ref{tab:detailed_bge}) shows that models generally struggle more with \textit{Quantitative Finance} and \textit{Blockchain} domains compared to \textit{History} or \textit{Politics}. For instance, SFR's MRR drops to 20.4 in the Quant domain, suggesting that numeric temporal reasoning remains a significant bottleneck.
    
    \item \textbf{Lexical Vulnerability:} BM25 exhibits a massive performance drop-off, with a MAP@10 of only 7.7. This demonstrates that temporal relevance is fundamentally a semantic task that cannot be solved by keyword overlap alone.
\end{itemize}

\begin{table}[h!]
\centering
\small
\caption{Comparison of retrieval models on TEMPO (averaged across 13 domains). Best in \textbf{bold}, second best \underline{underlined}.}
\label{tab:model_comparison}
\resizebox{0.50\textwidth}{!}{%
\begin{tabular}{lcccccccccc}
\toprule
& \multicolumn{3}{c}{\textbf{MAP}} & \multicolumn{3}{c}{\textbf{Recall}} & \multicolumn{3}{c}{\textbf{Precision}} & \\
\cmidrule(lr){2-4} \cmidrule(lr){5-7} \cmidrule(lr){8-10}
\textbf{Model} & \textbf{@1} & \textbf{@10} & \textbf{@25} & \textbf{@1} & \textbf{@10} & \textbf{@25} & \textbf{@1} & \textbf{@10} & \textbf{@25} & \textbf{MRR} \\
\midrule
\rowcolor{gray!12}\multicolumn{11}{c}{\textit{Sparse}} \\
\midrule
BM25 & 4.0 & 7.7 & 8.3 & 4.0 & 13.7 & 19.0 & 9.4 & 3.5 & 2.0 & 14.5 \\
\midrule
\rowcolor{gray!12}\multicolumn{11}{c}{\textit{Dense (<1B)}} \\
\midrule
BGE & 8.0 & 15.4 & 16.7 & 8.0 & 27.5 & 39.3 & 20.7 & 7.4 & 4.3 & 30.3 \\
Contriever & 7.6 & 15.2 & 16.5 & 7.6 & 26.9 & 39.4 & 19.2 & 7.2 & 4.3 & 28.8 \\
Inst-L & 8.6 & 17.3 & 18.7 & 8.6 & 31.7 & 44.3 & 21.2 & 8.5 & 4.9 & 32.6 \\
SBERT & 8.9 & 17.4 & 18.8 & 8.9 & 31.6 & 43.4 & 22.9 & 8.4 & 4.8 & 33.5 \\
\midrule
\rowcolor{gray!12}\multicolumn{11}{c}{\textit{Dense (>1B)}} \\
\midrule
E5 & \textbf{11.5} & \underline{22.4} & \underline{23.8} & \textbf{11.5} & 37.0 & 48.7 & \textbf{27.4} & 10.1 & 5.4 & \underline{38.9} \\
GritLM & 10.2 & 19.7 & 21.4 & 10.2 & 34.8 & 49.6 & 23.4 & 9.1 & 5.4 & 34.5 \\
Qwen & 8.0 & 16.1 & 17.4 & 8.0 & 28.1 & 40.1 & 20.4 & 8.1 & 4.7 & 30.5 \\
SFR & 11.0 & 21.7 & 23.3 & 11.0 & \underline{37.9} & \underline{50.9} & 26.2 & \underline{10.2} & \underline{5.6} & 37.8 \\
\midrule
\rowcolor{gray!12}\multicolumn{11}{c}{\textit{Reasoning}} \\
\midrule
DiVeR & \underline{11.2} & \textbf{23.5} & \textbf{25.1} & \underline{11.2} & \textbf{40.1} & \textbf{52.6} & \underline{27.4} & \textbf{11.1} & \textbf{5.9} & \textbf{39.8} \\
Rader & 8.4 & 17.0 & 18.3 & 8.4 & 31.8 & 43.6 & 20.7 & 8.3 & 4.7 & 31.0 \\
ReasonIR & 9.2 & 19.6 & 21.2 & 9.2 & 35.1 & 48.7 & 22.7 & 9.5 & 5.4 & 34.2 \\
\bottomrule
\end{tabular}}
\end{table}

\begin{table}[h!]
\centering
\small
\caption{Detailed retrieval metrics for \textbf{SFR} on TEMPO across all domains.}
\label{tab:detailed_sfr}
\resizebox{0.5\textwidth}{!}{%
\begin{tabular}{lcccccccccc}
\toprule
& \multicolumn{3}{c}{\textbf{MAP}} & \multicolumn{3}{c}{\textbf{Recall}} & \multicolumn{3}{c}{\textbf{Precision}} & \\
\cmidrule(lr){2-4} \cmidrule(lr){5-7} \cmidrule(lr){8-10}
\textbf{Domain} & \textbf{@1} & \textbf{@10} & \textbf{@25} & \textbf{@1} & \textbf{@10} & \textbf{@25} & \textbf{@1} & \textbf{@10} & \textbf{@25} & \textbf{MRR} \\
\midrule
\rowcolor{gray!12}\multicolumn{11}{c}{\textit{Blockchain}} \\
\midrule
\textbf{Bitcoin} & 7.7 & 12.5 & 12.9 & 7.7 & 20.6 & 24.8 & 21.0 & 5.4 & 2.9 & 27.2 \\
\textbf{Cardano} & 8.6 & 19.2 & 20.7 & 8.6 & 42.4 & 57.8 & 19.6 & 9.0 & 4.9 & 31.7 \\
\textbf{Iota} & 11.7 & 26.8 & 28.3 & 11.7 & 39.5 & 50.3 & 40.0 & 13.0 & 6.4 & 52.6 \\
\textbf{Monero} & 9.7 & 16.8 & 17.7 & 9.7 & 28.8 & 40.2 & 23.1 & 7.7 & 4.1 & 32.8 \\
\midrule
\rowcolor{gray!12}\multicolumn{11}{c}{\textit{Social Sci.}} \\
\midrule
\textbf{Economics} & 8.3 & 14.4 & 16.2 & 8.3 & 26.2 & 41.8 & 24.1 & 9.2 & 6.0 & 33.8 \\
\textbf{Law} & 16.4 & 29.5 & 32.2 & 16.4 & 55.5 & 72.0 & 28.6 & 14.3 & 8.1 & 45.7 \\
\textbf{Politics} & 17.2 & 35.6 & 37.4 & 17.2 & 54.0 & 66.3 & 42.0 & 13.7 & 7.0 & 53.2 \\
\textbf{History} & 11.1 & 22.8 & 24.8 & 11.1 & 39.0 & 53.1 & 32.0 & 12.6 & 7.1 & 43.9 \\
\midrule
\rowcolor{gray!12}\multicolumn{11}{c}{\textit{Applied}} \\
\midrule
\textbf{Quant} & 5.4 & 11.7 & 12.0 & 5.4 & 22.0 & 26.7 & 8.8 & 5.6 & 2.6 & 20.4 \\
\textbf{Travel} & 8.9 & 21.1 & 22.8 & 8.9 & 41.1 & 54.4 & 18.0 & 10.1 & 5.6 & 33.0 \\
\textbf{Workplace} & 12.2 & 22.4 & 24.8 & 12.2 & 43.9 & 63.9 & 25.0 & 11.9 & 6.9 & 35.2 \\
\textbf{Genealogy} & 13.2 & 24.0 & 26.3 & 13.2 & 36.2 & 52.5 & 32.2 & 10.6 & 6.3 & 41.9 \\
\midrule
\rowcolor{gray!12}\multicolumn{11}{c}{\textit{STEM}} \\
\midrule
\textbf{HSM} & 12.9 & 24.9 & 26.4 & 12.9 & 43.7 & 58.2 & 26.7 & 9.9 & 5.4 & 39.5 \\
\midrule
\midrule
\textbf{Average} & \textbf{11.0} & \textbf{21.7} & \textbf{23.3} & \textbf{11.0} & \textbf{37.9} & \textbf{50.9} & \textbf{26.2} & \textbf{10.2} & \textbf{5.6} & \textbf{37.8} \\
\bottomrule
\end{tabular}}
\end{table}

\begin{table}[h!]
\centering
\small
\caption{Detailed retrieval metrics for \textbf{SBERT} on TEMPO across all domains.}
\label{tab:detailed_sbert}
\resizebox{0.5\textwidth}{!}{%
\begin{tabular}{lcccccccccc}
\toprule
& \multicolumn{3}{c}{\textbf{MAP}} & \multicolumn{3}{c}{\textbf{Recall}} & \multicolumn{3}{c}{\textbf{Precision}} & \\
\cmidrule(lr){2-4} \cmidrule(lr){5-7} \cmidrule(lr){8-10}
\textbf{Domain} & \textbf{@1} & \textbf{@10} & \textbf{@25} & \textbf{@1} & \textbf{@10} & \textbf{@25} & \textbf{@1} & \textbf{@10} & \textbf{@25} & \textbf{MRR} \\
\midrule
\rowcolor{gray!12}\multicolumn{11}{c}{\textit{Blockchain}} \\
\midrule
\textbf{Bitcoin} & 5.2 & 9.3 & 10.0 & 5.2 & 16.9 & 24.1 & 16.0 & 5.1 & 3.0 & 23.1 \\
\textbf{Cardano} & 6.0 & 14.0 & 15.1 & 6.0 & 32.7 & 45.5 & 15.7 & 7.3 & 4.1 & 25.8 \\
\textbf{Iota} & 15.0 & 23.6 & 25.7 & 15.0 & 31.2 & 44.5 & 50.0 & 10.0 & 5.6 & 57.2 \\
\textbf{Monero} & 3.7 & 9.0 & 9.7 & 3.7 & 22.2 & 30.6 & 10.8 & 5.5 & 3.1 & 21.0 \\
\midrule
\rowcolor{gray!12}\multicolumn{11}{c}{\textit{Social Sci.}} \\
\midrule
\textbf{Economics} & 2.4 & 9.1 & 10.6 & 2.4 & 21.3 & 32.3 & 12.0 & 7.3 & 4.8 & 22.4 \\
\textbf{Law} & 18.8 & 24.7 & 26.4 & 18.8 & 42.5 & 55.4 & 34.3 & 10.3 & 6.1 & 44.4 \\
\textbf{Politics} & 12.3 & 26.3 & 27.4 & 12.3 & 42.7 & 52.6 & 30.0 & 11.1 & 5.5 & 41.6 \\
\textbf{History} & 10.4 & 19.9 & 21.9 & 10.4 & 33.9 & 48.6 & 29.8 & 10.8 & 6.5 & 40.8 \\
\midrule
\rowcolor{gray!12}\multicolumn{11}{c}{\textit{Applied}} \\
\midrule
\textbf{Quant} & 6.1 & 12.0 & 12.8 & 6.1 & 19.0 & 30.1 & 11.8 & 5.0 & 3.1 & 20.7 \\
\textbf{Travel} & 7.5 & 18.5 & 20.4 & 7.5 & 38.9 & 53.4 & 18.0 & 9.1 & 5.4 & 32.4 \\
\textbf{Workplace} & 10.7 & 25.0 & 27.5 & 10.7 & 44.9 & 67.6 & 27.8 & 11.1 & 6.6 & 43.0 \\
\textbf{Genealogy} & 8.8 & 16.8 & 18.0 & 8.8 & 27.7 & 35.7 & 23.5 & 8.5 & 4.4 & 31.7 \\
\midrule
\rowcolor{gray!12}\multicolumn{11}{c}{\textit{STEM}} \\
\midrule
\textbf{HSM} & 8.5 & 18.1 & 19.0 & 8.5 & 36.4 & 44.4 & 18.7 & 8.1 & 4.0 & 31.0 \\
\midrule
\midrule
\textbf{Average} & \textbf{8.9} & \textbf{17.4} & \textbf{18.8} & \textbf{8.9} & \textbf{31.6} & \textbf{43.4} & \textbf{22.9} & \textbf{8.4} & \textbf{4.8} & \textbf{33.5} \\
\bottomrule
\end{tabular}}
\end{table}

\begin{table}[h!]
\centering
\small
\caption{Detailed retrieval metrics for \textbf{ReasonIR} on TEMPO across all domains.}
\label{tab:detailed_reasonir}
\resizebox{0.5\textwidth}{!}{%
\begin{tabular}{lcccccccccc}
\toprule
& \multicolumn{3}{c}{\textbf{MAP}} & \multicolumn{3}{c}{\textbf{Recall}} & \multicolumn{3}{c}{\textbf{Precision}} & \\
\cmidrule(lr){2-4} \cmidrule(lr){5-7} \cmidrule(lr){8-10}
\textbf{Domain} & \textbf{@1} & \textbf{@10} & \textbf{@25} & \textbf{@1} & \textbf{@10} & \textbf{@25} & \textbf{@1} & \textbf{@10} & \textbf{@25} & \textbf{MRR} \\
\midrule
\rowcolor{gray!12}\multicolumn{11}{c}{\textit{Blockchain}} \\
\midrule
\textbf{Bitcoin} & 5.9 & 11.5 & 12.1 & 5.9 & 21.6 & 29.6 & 13.0 & 5.6 & 3.3 & 21.6 \\
\textbf{Cardano} & 6.3 & 15.4 & 17.1 & 6.3 & 35.5 & 52.1 & 13.7 & 7.5 & 4.4 & 25.3 \\
\textbf{Iota} & 13.3 & 31.3 & 31.9 & 13.3 & 47.3 & 51.0 & 40.0 & 15.0 & 6.8 & 54.3 \\
\textbf{Monero} & 7.7 & 13.7 & 15.0 & 7.7 & 23.2 & 36.5 & 20.0 & 6.6 & 3.9 & 28.9 \\
\midrule
\rowcolor{gray!12}\multicolumn{11}{c}{\textit{Social Sci.}} \\
\midrule
\textbf{Economics} & 5.8 & 13.0 & 15.2 & 5.8 & 26.1 & 43.6 & 18.1 & 8.3 & 6.0 & 28.5 \\
\textbf{Law} & 14.1 & 27.9 & 30.5 & 14.1 & 47.3 & 67.9 & 31.4 & 13.1 & 7.4 & 47.7 \\
\textbf{Politics} & 12.4 & 26.9 & 28.1 & 12.4 & 46.1 & 55.9 & 29.3 & 11.5 & 5.8 & 40.1 \\
\textbf{History} & 12.0 & 24.3 & 26.6 & 12.0 & 41.9 & 57.4 & 33.3 & 13.3 & 7.7 & 45.6 \\
\midrule
\rowcolor{gray!12}\multicolumn{11}{c}{\textit{Applied}} \\
\midrule
\textbf{Quant} & 5.2 & 14.1 & 15.3 & 5.2 & 26.2 & 37.2 & 11.8 & 6.2 & 3.8 & 23.7 \\
\textbf{Travel} & 7.0 & 15.7 & 17.1 & 7.0 & 29.9 & 44.2 & 11.0 & 7.1 & 4.4 & 23.3 \\
\textbf{Workplace} & 8.4 & 20.8 & 22.8 & 8.4 & 42.9 & 62.1 & 19.4 & 11.7 & 6.4 & 33.6 \\
\textbf{Genealogy} & 12.1 & 22.8 & 24.8 & 12.1 & 34.6 & 49.3 & 33.0 & 10.7 & 6.1 & 40.9 \\
\midrule
\rowcolor{gray!12}\multicolumn{11}{c}{\textit{STEM}} \\
\midrule
\textbf{HSM} & 9.1 & 17.6 & 19.0 & 9.1 & 33.0 & 47.0 & 20.7 & 7.6 & 4.5 & 30.9 \\
\midrule
\midrule
\textbf{Average} & \textbf{9.2} & \textbf{19.6} & \textbf{21.2} & \textbf{9.2} & \textbf{35.1} & \textbf{48.7} & \textbf{22.7} & \textbf{9.5} & \textbf{5.4} & \textbf{34.2} \\
\bottomrule
\end{tabular}}
\end{table}

\begin{table}[h!]
\centering
\small
\caption{Detailed retrieval metrics for \textbf{Rader} on TEMPO across all domains.}
\label{tab:detailed_rader}
\resizebox{0.5\textwidth}{!}{%
\begin{tabular}{lcccccccccc}
\toprule
& \multicolumn{3}{c}{\textbf{MAP}} & \multicolumn{3}{c}{\textbf{Recall}} & \multicolumn{3}{c}{\textbf{Precision}} & \\
\cmidrule(lr){2-4} \cmidrule(lr){5-7} \cmidrule(lr){8-10}
\textbf{Domain} & \textbf{@1} & \textbf{@10} & \textbf{@25} & \textbf{@1} & \textbf{@10} & \textbf{@25} & \textbf{@1} & \textbf{@10} & \textbf{@25} & \textbf{MRR} \\
\midrule
\rowcolor{gray!12}\multicolumn{11}{c}{\textit{Blockchain}} \\
\midrule
\textbf{Bitcoin} & 6.8 & 10.7 & 11.7 & 6.8 & 17.2 & 27.3 & 16.0 & 4.8 & 3.1 & 22.8 \\
\textbf{Cardano} & 3.5 & 11.4 & 12.3 & 3.5 & 31.2 & 41.0 & 9.8 & 6.1 & 3.5 & 21.4 \\
\textbf{Iota} & 7.5 & 11.9 & 13.0 & 7.5 & 25.8 & 36.7 & 20.0 & 8.0 & 4.4 & 27.9 \\
\textbf{Monero} & 7.4 & 15.3 & 16.0 & 7.4 & 26.3 & 34.5 & 16.9 & 6.6 & 3.5 & 27.1 \\
\midrule
\rowcolor{gray!12}\multicolumn{11}{c}{\textit{Social Sci.}} \\
\midrule
\textbf{Economics} & 6.1 & 15.0 & 16.6 & 6.1 & 30.0 & 42.9 & 20.5 & 9.5 & 5.6 & 31.0 \\
\textbf{Law} & 10.6 & 22.8 & 24.9 & 10.6 & 47.4 & 62.3 & 25.7 & 11.1 & 6.4 & 41.1 \\
\textbf{Politics} & 10.9 & 24.0 & 25.4 & 10.9 & 40.2 & 51.4 & 28.7 & 10.7 & 5.5 & 39.4 \\
\textbf{History} & 9.0 & 17.4 & 19.4 & 9.0 & 31.1 & 45.7 & 26.3 & 9.9 & 6.2 & 37.7 \\
\midrule
\rowcolor{gray!12}\multicolumn{11}{c}{\textit{Applied}} \\
\midrule
\textbf{Quant} & 9.7 & 20.1 & 20.9 & 9.7 & 36.3 & 45.9 & 23.5 & 8.2 & 4.4 & 35.5 \\
\textbf{Travel} & 11.8 & 19.8 & 20.9 & 11.8 & 32.2 & 42.9 & 23.0 & 7.9 & 4.4 & 32.4 \\
\textbf{Workplace} & 15.1 & 27.3 & 29.5 & 15.1 & 48.6 & 66.7 & 30.6 & 11.9 & 6.9 & 41.6 \\
\textbf{Genealogy} & 6.1 & 13.4 & 14.7 & 6.1 & 23.4 & 36.2 & 16.5 & 7.0 & 4.2 & 24.6 \\
\midrule
\rowcolor{gray!12}\multicolumn{11}{c}{\textit{STEM}} \\
\midrule
\textbf{HSM} & 5.1 & 11.4 & 12.3 & 5.1 & 24.0 & 33.6 & 11.3 & 5.9 & 3.3 & 20.2 \\
\midrule
\midrule
\textbf{Average} & \textbf{8.4} & \textbf{17.0} & \textbf{18.3} & \textbf{8.4} & \textbf{31.8} & \textbf{43.6} & \textbf{20.7} & \textbf{8.3} & \textbf{4.7} & \textbf{31.0} \\
\bottomrule
\end{tabular}}
\end{table}

\begin{table}[h!]
\centering
\small
\caption{Detailed retrieval metrics for \textbf{Qwen} on TEMPO across all domains.}
\label{tab:detailed_qwen}
\resizebox{0.5\textwidth}{!}{%
\begin{tabular}{lcccccccccc}
\toprule
& \multicolumn{3}{c}{\textbf{MAP}} & \multicolumn{3}{c}{\textbf{Recall}} & \multicolumn{3}{c}{\textbf{Precision}} & \\
\cmidrule(lr){2-4} \cmidrule(lr){5-7} \cmidrule(lr){8-10}
\textbf{Domain} & \textbf{@1} & \textbf{@10} & \textbf{@25} & \textbf{@1} & \textbf{@10} & \textbf{@25} & \textbf{@1} & \textbf{@10} & \textbf{@25} & \textbf{MRR} \\
\midrule
\rowcolor{gray!12}\multicolumn{11}{c}{\textit{Blockchain}} \\
\midrule
\textbf{Bitcoin} & 4.6 & 7.5 & 7.9 & 4.6 & 12.7 & 17.5 & 14.0 & 3.9 & 2.2 & 19.3 \\
\textbf{Cardano} & 5.7 & 13.9 & 15.0 & 5.7 & 29.8 & 42.5 & 13.7 & 6.9 & 3.8 & 24.3 \\
\textbf{Iota} & 10.8 & 18.2 & 20.6 & 10.8 & 32.3 & 52.2 & 30.0 & 11.0 & 7.2 & 44.9 \\
\textbf{Monero} & 2.9 & 6.7 & 7.4 & 2.9 & 14.6 & 23.2 & 7.7 & 4.3 & 2.6 & 15.8 \\
\midrule
\rowcolor{gray!12}\multicolumn{11}{c}{\textit{Social Sci.}} \\
\midrule
\textbf{Economics} & 4.6 & 11.3 & 12.4 & 4.6 & 21.8 & 30.7 & 12.0 & 7.6 & 4.6 & 23.3 \\
\textbf{Law} & 14.2 & 25.2 & 27.3 & 14.2 & 37.1 & 56.9 & 28.6 & 10.6 & 6.6 & 39.7 \\
\textbf{Politics} & 14.4 & 28.8 & 30.2 & 14.4 & 46.5 & 57.3 & 36.0 & 12.1 & 6.1 & 47.0 \\
\textbf{History} & 8.9 & 17.5 & 19.2 & 8.9 & 30.5 & 43.5 & 26.0 & 10.0 & 5.9 & 36.5 \\
\midrule
\rowcolor{gray!12}\multicolumn{11}{c}{\textit{Applied}} \\
\midrule
\textbf{Quant} & 2.8 & 8.3 & 9.1 & 2.8 & 15.8 & 22.5 & 8.8 & 4.1 & 2.5 & 19.4 \\
\textbf{Travel} & 9.4 & 16.0 & 17.4 & 9.4 & 28.8 & 41.5 & 18.0 & 6.9 & 4.2 & 26.8 \\
\textbf{Workplace} & 10.9 & 21.7 & 24.0 & 10.9 & 36.9 & 54.1 & 30.6 & 11.4 & 6.3 & 39.5 \\
\textbf{Genealogy} & 7.5 & 18.3 & 19.2 & 7.5 & 30.6 & 38.3 & 22.6 & 9.7 & 4.8 & 33.2 \\
\midrule
\rowcolor{gray!12}\multicolumn{11}{c}{\textit{STEM}} \\
\midrule
\textbf{HSM} & 7.7 & 15.1 & 16.3 & 7.7 & 28.4 & 40.9 & 17.3 & 6.9 & 4.0 & 26.4 \\
\midrule
\midrule
\textbf{Average} & \textbf{8.0} & \textbf{16.1} & \textbf{17.4} & \textbf{8.0} & \textbf{28.1} & \textbf{40.1} & \textbf{20.4} & \textbf{8.1} & \textbf{4.7} & \textbf{30.5} \\
\bottomrule
\end{tabular}}
\end{table}

\begin{table}[h!]
\centering
\small
\caption{Detailed retrieval metrics for \textbf{Inst-L} on TEMPO across all domains.}
\label{tab:detailed_inst_l}
\resizebox{0.5\textwidth}{!}{%
\begin{tabular}{lcccccccccc}
\toprule
& \multicolumn{3}{c}{\textbf{MAP}} & \multicolumn{3}{c}{\textbf{Recall}} & \multicolumn{3}{c}{\textbf{Precision}} & \\
\cmidrule(lr){2-4} \cmidrule(lr){5-7} \cmidrule(lr){8-10}
\textbf{Domain} & \textbf{@1} & \textbf{@10} & \textbf{@25} & \textbf{@1} & \textbf{@10} & \textbf{@25} & \textbf{@1} & \textbf{@10} & \textbf{@25} & \textbf{MRR} \\
\midrule
\rowcolor{gray!12}\multicolumn{11}{c}{\textit{Blockchain}} \\
\midrule
\textbf{Bitcoin} & 5.6 & 10.3 & 10.9 & 5.6 & 20.5 & 26.6 & 15.0 & 5.3 & 3.2 & 23.3 \\
\textbf{Cardano} & 1.3 & 8.4 & 9.6 & 1.3 & 25.0 & 39.8 & 3.9 & 5.9 & 3.7 & 16.6 \\
\textbf{Iota} & 16.7 & 23.9 & 25.5 & 16.7 & 32.8 & 44.0 & 50.0 & 11.0 & 6.0 & 58.3 \\
\textbf{Monero} & 3.7 & 10.2 & 11.5 & 3.7 & 23.9 & 38.4 & 10.8 & 6.5 & 4.0 & 23.3 \\
\midrule
\rowcolor{gray!12}\multicolumn{11}{c}{\textit{Social Sci.}} \\
\midrule
\textbf{Economics} & 4.3 & 11.1 & 12.5 & 4.3 & 22.8 & 34.3 & 13.3 & 7.8 & 4.9 & 24.5 \\
\textbf{Law} & 16.6 & 26.6 & 28.7 & 16.6 & 47.2 & 61.4 & 31.4 & 12.0 & 6.9 & 47.6 \\
\textbf{Politics} & 11.0 & 24.4 & 25.6 & 11.0 & 41.2 & 53.1 & 25.3 & 10.5 & 5.5 & 38.6 \\
\textbf{History} & 9.8 & 20.0 & 22.0 & 9.8 & 34.4 & 48.2 & 27.7 & 11.1 & 6.6 & 38.7 \\
\midrule
\rowcolor{gray!12}\multicolumn{11}{c}{\textit{Applied}} \\
\midrule
\textbf{Quant} & 5.6 & 10.2 & 11.3 & 5.6 & 17.6 & 30.8 & 14.7 & 4.1 & 2.8 & 22.9 \\
\textbf{Travel} & 6.8 & 17.2 & 18.5 & 6.8 & 37.1 & 47.7 & 13.0 & 9.2 & 4.9 & 25.6 \\
\textbf{Workplace} & 10.8 & 27.4 & 29.4 & 10.8 & 46.1 & 64.8 & 27.8 & 11.7 & 6.3 & 42.8 \\
\textbf{Genealogy} & 10.6 & 17.7 & 19.5 & 10.6 & 30.4 & 41.6 & 24.3 & 8.6 & 5.0 & 32.0 \\
\midrule
\rowcolor{gray!12}\multicolumn{11}{c}{\textit{STEM}} \\
\midrule
\textbf{HSM} & 8.4 & 17.6 & 18.8 & 8.4 & 32.9 & 45.2 & 18.0 & 7.4 & 4.2 & 29.2 \\
\midrule
\midrule
\textbf{Average} & \textbf{8.6} & \textbf{17.3} & \textbf{18.7} & \textbf{8.6} & \textbf{31.7} & \textbf{44.3} & \textbf{21.2} & \textbf{8.5} & \textbf{4.9} & \textbf{32.6} \\
\bottomrule
\end{tabular}}
\end{table}

\begin{table}[h!]
\centering
\small
\caption{Detailed retrieval metrics for \textbf{GritLM} on TEMPO across all domains.}
\label{tab:detailed_gritlm}
\resizebox{0.5\textwidth}{!}{%
\begin{tabular}{lcccccccccc}
\toprule
& \multicolumn{3}{c}{\textbf{MAP}} & \multicolumn{3}{c}{\textbf{Recall}} & \multicolumn{3}{c}{\textbf{Precision}} & \\
\cmidrule(lr){2-4} \cmidrule(lr){5-7} \cmidrule(lr){8-10}
\textbf{Domain} & \textbf{@1} & \textbf{@10} & \textbf{@25} & \textbf{@1} & \textbf{@10} & \textbf{@25} & \textbf{@1} & \textbf{@10} & \textbf{@25} & \textbf{MRR} \\
\midrule
\rowcolor{gray!12}\multicolumn{11}{c}{\textit{Blockchain}} \\
\midrule
\textbf{Bitcoin} & 10.3 & 14.1 & 15.2 & 10.3 & 20.8 & 30.5 & 24.0 & 5.5 & 3.4 & 29.2 \\
\textbf{Cardano} & 5.1 & 14.0 & 15.5 & 5.1 & 35.9 & 53.0 & 9.8 & 6.7 & 4.4 & 23.6 \\
\textbf{Iota} & 11.7 & 27.8 & 29.1 & 11.7 & 40.8 & 52.0 & 40.0 & 12.0 & 6.8 & 50.6 \\
\textbf{Monero} & 5.6 & 10.3 & 12.2 & 5.6 & 16.8 & 40.1 & 13.8 & 4.9 & 4.2 & 22.7 \\
\midrule
\rowcolor{gray!12}\multicolumn{11}{c}{\textit{Social Sci.}} \\
\midrule
\textbf{Economics} & 4.6 & 10.3 & 11.5 & 4.6 & 23.7 & 34.1 & 13.3 & 8.2 & 4.9 & 24.0 \\
\textbf{Law} & 19.2 & 28.6 & 31.7 & 19.2 & 48.6 & 70.3 & 34.3 & 12.3 & 7.8 & 47.0 \\
\textbf{Politics} & 15.0 & 32.4 & 34.0 & 15.0 & 50.6 & 65.0 & 35.3 & 13.0 & 6.8 & 48.2 \\
\textbf{History} & 9.6 & 18.7 & 20.7 & 9.6 & 33.9 & 49.2 & 25.3 & 10.6 & 6.5 & 36.9 \\
\midrule
\rowcolor{gray!12}\multicolumn{11}{c}{\textit{Applied}} \\
\midrule
\textbf{Quant} & 9.8 & 16.4 & 17.2 & 9.8 & 26.4 & 37.1 & 14.7 & 6.5 & 3.4 & 25.7 \\
\textbf{Travel} & 6.3 & 16.7 & 18.0 & 6.3 & 37.0 & 49.9 & 16.0 & 9.0 & 4.9 & 28.9 \\
\textbf{Workplace} & 12.0 & 22.6 & 24.9 & 12.0 & 42.5 & 64.4 & 25.0 & 10.6 & 6.4 & 35.7 \\
\textbf{Genealogy} & 11.5 & 19.9 & 21.5 & 11.5 & 32.0 & 44.7 & 26.1 & 8.9 & 5.3 & 35.5 \\
\midrule
\rowcolor{gray!12}\multicolumn{11}{c}{\textit{STEM}} \\
\midrule
\textbf{HSM} & 11.8 & 24.9 & 26.3 & 11.8 & 42.8 & 55.1 & 26.7 & 9.9 & 5.3 & 40.6 \\
\midrule
\midrule
\textbf{Average} & \textbf{10.2} & \textbf{19.7} & \textbf{21.4} & \textbf{10.2} & \textbf{34.8} & \textbf{49.6} & \textbf{23.4} & \textbf{9.1} & \textbf{5.4} & \textbf{34.5} \\
\bottomrule
\end{tabular}}
\end{table}

\begin{table}[h!]
\centering
\small
\caption{Detailed retrieval metrics for \textbf{E5} on TEMPO across all domains.}
\label{tab:detailed_e5}
\resizebox{0.5\textwidth}{!}{%
\begin{tabular}{lcccccccccc}
\toprule
& \multicolumn{3}{c}{\textbf{MAP}} & \multicolumn{3}{c}{\textbf{Recall}} & \multicolumn{3}{c}{\textbf{Precision}} & \\
\cmidrule(lr){2-4} \cmidrule(lr){5-7} \cmidrule(lr){8-10}
\textbf{Domain} & \textbf{@1} & \textbf{@10} & \textbf{@25} & \textbf{@1} & \textbf{@10} & \textbf{@25} & \textbf{@1} & \textbf{@10} & \textbf{@25} & \textbf{MRR} \\
\midrule
\rowcolor{gray!12}\multicolumn{11}{c}{\textit{Blockchain}} \\
\midrule
\textbf{Bitcoin} & 6.9 & 11.5 & 12.3 & 6.9 & 18.4 & 27.0 & 18.0 & 5.2 & 3.1 & 25.6 \\
\textbf{Cardano} & 12.5 & 25.5 & 26.2 & 12.5 & 51.3 & 59.1 & 23.5 & 10.8 & 4.9 & 38.5 \\
\textbf{Iota} & 16.7 & 31.5 & 32.9 & 16.7 & 41.2 & 47.0 & 50.0 & 13.0 & 6.0 & 61.8 \\
\textbf{Monero} & 6.5 & 13.4 & 14.5 & 6.5 & 26.7 & 37.7 & 15.4 & 7.1 & 4.1 & 25.9 \\
\midrule
\rowcolor{gray!12}\multicolumn{11}{c}{\textit{Social Sci.}} \\
\midrule
\textbf{Economics} & 7.7 & 16.5 & 18.2 & 7.7 & 30.2 & 43.5 & 22.9 & 10.7 & 6.4 & 36.1 \\
\textbf{Law} & 13.4 & 26.0 & 29.2 & 13.4 & 41.0 & 67.7 & 25.7 & 11.7 & 7.8 & 41.3 \\
\textbf{Politics} & 18.8 & 38.1 & 39.8 & 18.8 & 56.7 & 68.6 & 45.3 & 14.6 & 7.2 & 56.8 \\
\textbf{History} & 10.2 & 20.2 & 21.6 & 10.2 & 32.8 & 42.5 & 31.2 & 11.1 & 6.0 & 41.0 \\
\midrule
\rowcolor{gray!12}\multicolumn{11}{c}{\textit{Applied}} \\
\midrule
\textbf{Quant} & 4.7 & 9.3 & 9.8 & 4.7 & 17.6 & 24.4 & 8.8 & 4.4 & 2.4 & 18.5 \\
\textbf{Travel} & 9.2 & 20.2 & 21.5 & 9.2 & 38.2 & 47.9 & 19.0 & 9.2 & 4.8 & 32.4 \\
\textbf{Workplace} & 12.7 & 24.1 & 26.3 & 12.7 & 41.8 & 59.5 & 27.8 & 11.9 & 6.7 & 38.9 \\
\textbf{Genealogy} & 14.7 & 25.5 & 27.3 & 14.7 & 38.1 & 51.1 & 34.8 & 11.2 & 6.1 & 44.0 \\
\midrule
\rowcolor{gray!12}\multicolumn{11}{c}{\textit{STEM}} \\
\midrule
\textbf{HSM} & 15.8 & 29.2 & 30.3 & 15.8 & 46.7 & 57.5 & 33.3 & 10.9 & 5.4 & 44.9 \\
\midrule
\midrule
\textbf{Average} & \textbf{11.5} & \textbf{22.4} & \textbf{23.8} & \textbf{11.5} & \textbf{37.0} & \textbf{48.7} & \textbf{27.4} & \textbf{10.1} & \textbf{5.4} & \textbf{38.9} \\
\bottomrule
\end{tabular}}
\end{table}

\begin{table}[h!]
\centering
\small
\caption{Detailed retrieval metrics for \textbf{DiVeR} on TEMPO across all domains.}
\label{tab:detailed_diver}
\resizebox{0.5\textwidth}{!}{%
\begin{tabular}{lcccccccccc}
\toprule
& \multicolumn{3}{c}{\textbf{MAP}} & \multicolumn{3}{c}{\textbf{Recall}} & \multicolumn{3}{c}{\textbf{Precision}} & \\
\cmidrule(lr){2-4} \cmidrule(lr){5-7} \cmidrule(lr){8-10}
\textbf{Domain} & \textbf{@1} & \textbf{@10} & \textbf{@25} & \textbf{@1} & \textbf{@10} & \textbf{@25} & \textbf{@1} & \textbf{@10} & \textbf{@25} & \textbf{MRR} \\
\midrule
\rowcolor{gray!12}\multicolumn{11}{c}{\textit{Blockchain}} \\
\midrule
\textbf{Bitcoin} & 5.9 & 11.7 & 12.5 & 5.9 & 22.2 & 31.3 & 17.0 & 5.9 & 3.6 & 25.5 \\
\textbf{Cardano} & 12.3 & 20.8 & 22.5 & 12.3 & 38.3 & 52.7 & 25.5 & 8.4 & 4.7 & 37.9 \\
\textbf{Iota} & 8.3 & 26.5 & 27.0 & 8.3 & 46.5 & 49.0 & 30.0 & 15.0 & 6.4 & 49.7 \\
\textbf{Monero} & 8.2 & 14.5 & 15.5 & 8.2 & 26.2 & 35.1 & 18.5 & 6.9 & 3.9 & 25.5 \\
\midrule
\rowcolor{gray!12}\multicolumn{11}{c}{\textit{Social Sci.}} \\
\midrule
\textbf{Economics} & 6.0 & 19.0 & 21.4 & 6.0 & 35.7 & 51.3 & 21.7 & 12.2 & 7.3 & 35.9 \\
\textbf{Law} & 11.3 & 28.5 & 31.1 & 11.3 & 57.1 & 74.9 & 25.7 & 15.7 & 8.6 & 43.1 \\
\textbf{Politics} & 18.8 & 36.6 & 38.5 & 18.8 & 51.4 & 66.8 & 44.0 & 13.3 & 7.0 & 56.2 \\
\textbf{History} & 11.9 & 24.7 & 26.8 & 11.9 & 41.1 & 55.1 & 34.6 & 13.3 & 7.6 & 46.1 \\
\midrule
\rowcolor{gray!12}\multicolumn{11}{c}{\textit{Applied}} \\
\midrule
\textbf{Quant} & 11.6 & 19.8 & 20.8 & 11.6 & 34.2 & 45.2 & 20.6 & 8.5 & 4.6 & 33.4 \\
\textbf{Travel} & 10.4 & 19.5 & 21.3 & 10.4 & 36.2 & 50.9 & 20.0 & 9.0 & 5.1 & 30.5 \\
\textbf{Workplace} & 16.9 & 34.0 & 35.7 & 16.9 & 49.8 & 65.4 & 41.7 & 13.9 & 7.1 & 51.1 \\
\textbf{Genealogy} & 13.2 & 26.8 & 28.1 & 13.2 & 41.9 & 50.9 & 33.9 & 12.0 & 6.0 & 45.7 \\
\midrule
\rowcolor{gray!12}\multicolumn{11}{c}{\textit{STEM}} \\
\midrule
\textbf{HSM} & 10.3 & 22.9 & 24.5 & 10.3 & 40.0 & 55.7 & 22.7 & 9.8 & 5.4 & 36.8 \\
\midrule
\midrule
\textbf{Average} & \textbf{11.2} & \textbf{23.5} & \textbf{25.1} & \textbf{11.2} & \textbf{40.1} & \textbf{52.6} & \textbf{27.4} & \textbf{11.1} & \textbf{5.9} & \textbf{39.8} \\
\bottomrule
\end{tabular}}
\end{table}

\begin{table}[h!]
\centering
\small
\caption{Detailed retrieval metrics for \textbf{Contriever} on TEMPO across all domains.}
\label{tab:detailed_contriever}
\resizebox{0.5\textwidth}{!}{%
\begin{tabular}{lcccccccccc}
\toprule
& \multicolumn{3}{c}{\textbf{MAP}} & \multicolumn{3}{c}{\textbf{Recall}} & \multicolumn{3}{c}{\textbf{Precision}} & \\
\cmidrule(lr){2-4} \cmidrule(lr){5-7} \cmidrule(lr){8-10}
\textbf{Domain} & \textbf{@1} & \textbf{@10} & \textbf{@25} & \textbf{@1} & \textbf{@10} & \textbf{@25} & \textbf{@1} & \textbf{@10} & \textbf{@25} & \textbf{MRR} \\
\midrule
\rowcolor{gray!12}\multicolumn{11}{c}{\textit{Blockchain}} \\
\midrule
\textbf{Bitcoin} & 3.1 & 8.7 & 9.4 & 3.1 & 18.0 & 25.3 & 9.0 & 4.7 & 2.8 & 18.6 \\
\textbf{Cardano} & 4.6 & 8.3 & 9.9 & 4.6 & 18.3 & 35.9 & 7.8 & 3.3 & 2.7 & 15.2 \\
\textbf{Iota} & 16.7 & 28.7 & 29.5 & 16.7 & 39.5 & 46.5 & 50.0 & 12.0 & 6.0 & 55.9 \\
\textbf{Monero} & 2.1 & 5.6 & 6.7 & 2.1 & 13.8 & 26.2 & 6.2 & 3.7 & 2.8 & 16.1 \\
\midrule
\rowcolor{gray!12}\multicolumn{11}{c}{\textit{Social Sci.}} \\
\midrule
\textbf{Economics} & 5.7 & 10.7 & 12.0 & 5.7 & 18.6 & 30.6 & 19.3 & 6.1 & 4.1 & 26.8 \\
\textbf{Law} & 9.0 & 19.9 & 21.6 & 9.0 & 40.8 & 57.9 & 14.3 & 10.6 & 6.5 & 29.6 \\
\textbf{Politics} & 11.6 & 24.3 & 25.3 & 11.6 & 38.6 & 48.6 & 29.3 & 9.7 & 5.0 & 39.1 \\
\textbf{History} & 9.4 & 18.2 & 20.1 & 9.4 & 31.0 & 45.4 & 28.6 & 10.2 & 6.2 & 38.6 \\
\midrule
\rowcolor{gray!12}\multicolumn{11}{c}{\textit{Applied}} \\
\midrule
\textbf{Quant} & 5.0 & 8.3 & 9.0 & 5.0 & 13.3 & 21.2 & 8.8 & 3.2 & 2.1 & 15.6 \\
\textbf{Travel} & 7.1 & 16.3 & 18.0 & 7.1 & 32.8 & 46.9 & 17.0 & 8.0 & 4.7 & 28.5 \\
\textbf{Workplace} & 5.8 & 17.1 & 19.0 & 5.8 & 29.5 & 49.5 & 16.7 & 8.6 & 5.1 & 31.9 \\
\textbf{Genealogy} & 12.3 & 18.5 & 19.9 & 12.3 & 29.1 & 40.5 & 28.7 & 8.1 & 4.7 & 34.5 \\
\midrule
\rowcolor{gray!12}\multicolumn{11}{c}{\textit{STEM}} \\
\midrule
\textbf{HSM} & 6.1 & 13.1 & 14.1 & 6.1 & 26.3 & 37.1 & 14.0 & 5.6 & 3.3 & 23.5 \\
\midrule
\midrule
\textbf{Average} & \textbf{7.6} & \textbf{15.2} & \textbf{16.5} & \textbf{7.6} & \textbf{26.9} & \textbf{39.4} & \textbf{19.2} & \textbf{7.2} & \textbf{4.3} & \textbf{28.8} \\
\bottomrule
\end{tabular}}
\end{table}

\begin{table}[h!]
\centering
\small
\caption{Detailed retrieval metrics for \textbf{BM25} on TEMPO across all domains.}
\label{tab:detailed_bm25}
\resizebox{0.5\textwidth}{!}{%
\begin{tabular}{lcccccccccc}
\toprule
& \multicolumn{3}{c}{\textbf{MAP}} & \multicolumn{3}{c}{\textbf{Recall}} & \multicolumn{3}{c}{\textbf{Precision}} & \\
\cmidrule(lr){2-4} \cmidrule(lr){5-7} \cmidrule(lr){8-10}
\textbf{Domain} & \textbf{@1} & \textbf{@10} & \textbf{@25} & \textbf{@1} & \textbf{@10} & \textbf{@25} & \textbf{@1} & \textbf{@10} & \textbf{@25} & \textbf{MRR} \\
\midrule
\rowcolor{gray!12}\multicolumn{11}{c}{\textit{Blockchain}} \\
\midrule
\textbf{Bitcoin} & 2.5 & 4.3 & 4.4 & 2.5 & 8.5 & 10.1 & 5.0 & 1.7 & 0.8 & 8.2 \\
\textbf{Cardano} & 4.9 & 9.6 & 10.0 & 4.9 & 20.6 & 25.5 & 5.9 & 3.5 & 1.9 & 13.8 \\
\textbf{Iota} & 2.5 & 5.4 & 5.4 & 2.5 & 11.7 & 11.7 & 10.0 & 4.0 & 1.6 & 17.3 \\
\textbf{Monero} & 0.8 & 1.5 & 1.7 & 0.8 & 4.7 & 7.8 & 1.5 & 0.9 & 0.7 & 3.9 \\
\midrule
\rowcolor{gray!12}\multicolumn{11}{c}{\textit{Social Sci.}} \\
\midrule
\textbf{Economics} & 2.2 & 4.4 & 4.8 & 2.2 & 6.2 & 10.3 & 6.0 & 2.7 & 1.7 & 8.8 \\
\textbf{Law} & 4.0 & 8.8 & 9.9 & 4.0 & 16.5 & 25.0 & 8.6 & 4.9 & 2.9 & 16.2 \\
\textbf{Politics} & 12.8 & 25.6 & 26.5 & 12.8 & 38.3 & 47.2 & 32.7 & 9.5 & 4.7 & 41.5 \\
\textbf{History} & 4.0 & 6.7 & 7.2 & 4.0 & 10.4 & 14.1 & 10.6 & 3.3 & 1.9 & 13.6 \\
\midrule
\rowcolor{gray!12}\multicolumn{11}{c}{\textit{Applied}} \\
\midrule
\textbf{Quant} & 1.5 & 1.8 & 2.1 & 1.5 & 2.9 & 5.9 & 2.9 & 0.6 & 0.6 & 4.2 \\
\textbf{Travel} & 1.3 & 3.2 & 3.8 & 1.3 & 7.1 & 15.0 & 2.0 & 1.6 & 1.2 & 5.6 \\
\textbf{Workplace} & 3.1 & 4.2 & 4.8 & 3.1 & 9.3 & 16.2 & 5.6 & 1.9 & 1.4 & 9.8 \\
\textbf{Genealogy} & 4.8 & 9.7 & 10.6 & 4.8 & 15.2 & 21.8 & 12.2 & 4.9 & 2.8 & 18.1 \\
\midrule
\rowcolor{gray!12}\multicolumn{11}{c}{\textit{STEM}} \\
\midrule
\textbf{HSM} & 7.7 & 15.4 & 16.3 & 7.7 & 26.9 & 36.8 & 18.7 & 6.5 & 3.4 & 27.2 \\
\midrule
\midrule
\textbf{Average} & \textbf{4.0} & \textbf{7.7} & \textbf{8.3} & \textbf{4.0} & \textbf{13.7} & \textbf{19.0} & \textbf{9.4} & \textbf{3.5} & \textbf{2.0} & \textbf{14.5} \\
\bottomrule
\end{tabular}}
\end{table}
\begin{table}[h!]
\centering
\small
\caption{Detailed retrieval metrics for \textbf{BGE} on TEMPO across all domains.}
\label{tab:detailed_bge}
\resizebox{0.5\textwidth}{!}{%
\begin{tabular}{lcccccccccc}
\toprule
& \multicolumn{3}{c}{\textbf{MAP}} & \multicolumn{3}{c}{\textbf{Recall}} & \multicolumn{3}{c}{\textbf{Precision}} & \\
\cmidrule(lr){2-4} \cmidrule(lr){5-7} \cmidrule(lr){8-10}
\textbf{Domain} & \textbf{@1} & \textbf{@10} & \textbf{@25} & \textbf{@1} & \textbf{@10} & \textbf{@25} & \textbf{@1} & \textbf{@10} & \textbf{@25} & \textbf{MRR} \\
\midrule
\rowcolor{gray!12}\multicolumn{11}{c}{\textit{Blockchain}} \\
\midrule
\textbf{Bitcoin} & 6.0 & 9.6 & 10.6 & 6.0 & 17.5 & 27.3 & 16.0 & 4.6 & 3.1 & 23.0 \\
\textbf{Cardano} & 1.3 & 7.4 & 9.4 & 1.3 & 22.1 & 42.3 & 3.9 & 4.9 & 3.7 & 16.2 \\
\textbf{Iota} & 15.0 & 26.1 & 27.4 & 15.0 & 34.5 & 42.8 & 50.0 & 11.0 & 5.6 & 59.7 \\
\textbf{Monero} & 5.1 & 9.3 & 10.2 & 5.1 & 16.8 & 26.6 & 15.4 & 4.9 & 3.0 & 23.6 \\
\midrule
\rowcolor{gray!12}\multicolumn{11}{c}{\textit{Social Sci.}} \\
\midrule
\textbf{Economics} & 3.0 & 7.6 & 9.1 & 3.0 & 15.5 & 31.2 & 13.3 & 6.1 & 4.2 & 20.5 \\
\textbf{Law} & 16.1 & 23.3 & 25.3 & 16.1 & 40.2 & 55.5 & 31.4 & 10.0 & 6.3 & 41.4 \\
\textbf{Politics} & 10.2 & 21.7 & 22.7 & 10.2 & 33.8 & 44.1 & 24.7 & 8.5 & 4.5 & 35.8 \\
\textbf{History} & 10.2 & 19.1 & 20.8 & 10.2 & 32.8 & 45.5 & 28.2 & 10.4 & 6.1 & 38.7 \\
\midrule
\rowcolor{gray!12}\multicolumn{11}{c}{\textit{Applied}} \\
\midrule
\textbf{Quant} & 1.5 & 7.1 & 7.8 & 1.5 & 18.9 & 26.9 & 2.9 & 4.4 & 2.8 & 13.8 \\
\textbf{Travel} & 7.7 & 17.0 & 18.2 & 7.7 & 32.0 & 42.6 & 18.0 & 7.8 & 4.3 & 28.8 \\
\textbf{Workplace} & 9.9 & 18.8 & 20.2 & 9.9 & 36.8 & 48.8 & 25.0 & 8.6 & 4.8 & 35.7 \\
\textbf{Genealogy} & 9.3 & 16.2 & 17.2 & 9.3 & 26.0 & 34.8 & 21.7 & 7.6 & 4.1 & 29.2 \\
\midrule
\rowcolor{gray!12}\multicolumn{11}{c}{\textit{STEM}} \\
\midrule
\textbf{HSM} & 9.1 & 16.9 & 18.1 & 9.1 & 30.8 & 42.6 & 18.7 & 7.1 & 4.0 & 27.9 \\
\midrule
\midrule
\textbf{Average} & \textbf{8.0} & \textbf{15.4} & \textbf{16.7} & \textbf{8.0} & \textbf{27.5} & \textbf{39.3} & \textbf{20.7} & \textbf{7.4} & \textbf{4.3} & \textbf{30.3} \\
\bottomrule
\end{tabular}}
\end{table}

\section{Quality Assessment \& RAG}
\label{app:Quality}
\subsection{Dataset Quality Validation}
\label{app:quality_validation}

To ensure the quality of TEMPO annotations, we employ Qwen-72B as an independent LLM judge to evaluate the alignment between queries, retrieval plans, and gold documents. This validation is performed separately from our main evaluation pipeline (which uses GPT-4o) to provide an unbiased assessment.

\subsubsection{Evaluation Protocol}

For each query in the dataset, we construct an evaluation instance containing: (1) the original temporal query, (2) the annotated retrieval plan with sequential steps, and (3) the gold documents mapped to the query. The LLM judge evaluates each instance on a 0--100 scale based on five criteria:

\begin{itemize}
    \item \textbf{Temporal Relevance}: Whether gold documents contain the temporal information needed to answer the query.
    \item \textbf{Plan-Document Alignment}: Whether gold documents align with the retrieval steps specified in the plan.
    \item \textbf{Temporal Coverage}: Whether documents cover the required time periods (baseline and comparison periods if applicable).
    \item \textbf{Completeness}: Whether documents provide sufficient evidence to comprehensively answer the query.
    \item \textbf{Authority}: Whether documents are from reliable sources appropriate for the domain.
\end{itemize}

\subsubsection{Results}

Table~\ref{tab:quality_validation_full} presents the quality validation results across all 13 domains. TEMPO achieves an overall average quality score of  86.7, with domain scores ranging from 84.3 (Bitcoin) to 89.1 (HSM). The consistently high scores across diverse domains validate that our annotation process produces high-quality query-document alignments suitable for evaluating temporal reasoning in retrieval systems.

\begin{table}[h]
\centering
\small

\begin{tabular}{llc}
\toprule
\textbf{Category} & \textbf{Domain}  & \textbf{Score} \\
\midrule
\multirow{3}{*}{Blockchain} 
& Bitcoin& 84.3 \\
& Cardano &  87.3 \\
& Iota &  85.7 \\
& Monero & 85.5 \\
\midrule
\multirow{3}{*}{Social Sciences} 
& Economics  & 86.8 \\
& Law &  87.1 \\
& Politics  & 86.1 \\
& History  & 88.7 \\
\midrule
\multirow{3}{*}{Applied} 
& Quant & 87.3 \\
& Travel  & 86.3 \\
& Workplace  & 85.7 \\
& Genealogy & 86.6 \\
\midrule
STEM & HSM  & 89.1 \\
\midrule
\midrule
\multicolumn{2}{l}{\textbf{Overall}} & \textbf{87.0} \\
\bottomrule
\end{tabular}
\caption{Dataset quality validation results using Qwen-72B as LLM judge across all domains.}
\label{tab:quality_validation_full}
\end{table}

\subsubsection{Evaluation Prompt}

The following prompt is used for quality validation with Qwen-72B:

\begin{figure*}[t]
\centering
\begin{tcolorbox}[
  colback=gray!5,
  colframe=gray!75,
  title=Prompt for Dataset Quality Validation,
  width=\textwidth,
]
\small\ttfamily
You are an expert evaluator assessing the quality of dataset annotations for a temporal retrieval benchmark.\\[0.5em]
\textbf{=== QUERY ===}\\
\{query\}\\[0.5em]
\textbf{=== RETRIEVAL PLAN ===}\\
\{retrieval\_plan\}\\[0.5em]
\textbf{=== GOLD DOCUMENTS ===}\\
\{gold\_documents\}\\[0.5em]
\textbf{=== TASK ===}\\
Evaluate how well the retrieval plan and gold documents serve to answer the temporal query. Consider:\\
1. \textbf{Temporal Relevance}: Do the gold documents contain the temporal information needed?\\
2. \textbf{Plan-Document Alignment}: Do the gold documents align with the retrieval steps?\\
3. \textbf{Temporal Coverage}: Do the documents cover the required time periods?\\
4. \textbf{Completeness}: Would these documents provide sufficient evidence to answer the query?\\
5. \textbf{Authority}: Are the documents from reliable sources appropriate for the domain?\\[0.5em]
\textbf{=== SCORING CRITERIA (0-100) ===}\\
\begin{itemize}
\item 90-100: Excellent - Documents perfectly match plan, comprehensive temporal coverage
\item 80-89: Strong - Documents highly relevant, minor gaps in temporal coverage
\item 70-79: Good - Documents mostly aligned, some temporal aspects not fully covered
\item 60-69: Adequate - Documents partially relevant, noticeable gaps
\item 50-59: Weak - Documents loosely related, significant misalignment
\item 0-49: Poor - Documents minimally helpful or irrelevant
\end{itemize}
~\\
\textbf{=== OUTPUT FORMAT ===}\\
REASONING: [Provide detailed assessment]\\
SCORE: [A single integer between 0 and 100]\\[0.5em]
Do not include anything after the score.
\end{tcolorbox}
\caption{Dataset quality validation prompt. An independent judge (Qwen-72B) evaluates the alignment between the query, the retrieval plan, and the gold documents.}
\label{prompt:quality_validation}
\end{figure*}

\clearpage

\subsection{Temporal Distribution by Domain} \label{app:domain_temporal} 
\begin{figure*}[t]
        \centering
        \includegraphics[width=.8\linewidth]{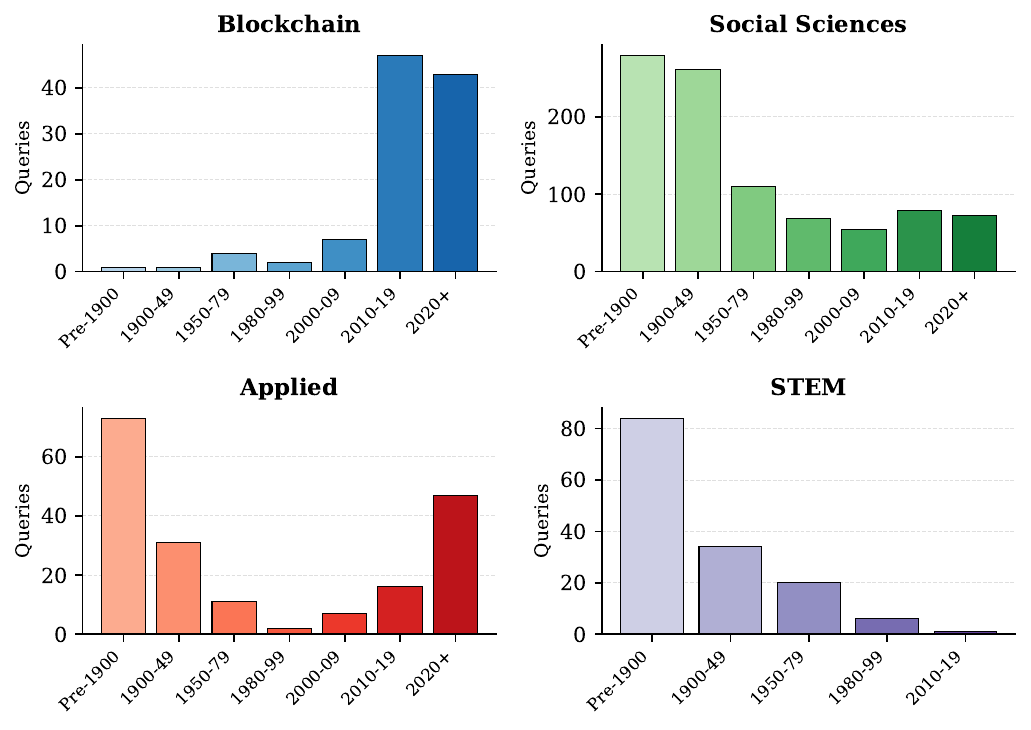}

    \caption{Temporal anchor distribution decomposed by major domain categories. Note the varying Y-axis scales and distinct "temporal signatures" for each field.}
    \label{fig:domain_dist}
\end{figure*}

We observe distinct temporal patterns when decomposing the query distribution across our primary domain categories: Blockchain, Social Sciences, Applied, and STEM. As illustrated in Figure \ref{fig:domain_dist}, each domain exhibits a unique temporal signature that reflects its inherent characteristics: \begin{itemize} \item \textbf{Blockchain}: Exclusively focused on modern history, with the vast majority of queries occurring after 2010. This aligns with the technological birth and rapid evolution of decentralized protocols. \item \textbf{Social Sciences}: Provides the strongest historical grounding, peaking at nearly 300 queries in the Pre-1900 category. This enables rigorous testing of reasoning over deep historical records and long-term societal shifts. \item \textbf{Applied}: Displays a bimodal distribution; while it includes significant historical data (Pre-1900, ~75 queries) related to fields like Genealogy, it also shows a strong recent resurgence in the 2020+ period (~50 queries) driven by modern workplace and travel dynamics. \item \textbf{STEM}: Almost exclusively focuses on the foundations of scientific and mathematical thought, with over 80 queries anchored Pre-1900 and a sharp decline as questions reach the 21st century. \end{itemize}

\clearpage

\subsection{RAG Evaluation Details}
\label{app:rag_eval_prompt}

\subsubsection{Full Per-Domain Results}
\label{app:rag_full_results}

Table~\ref{tab:rag_results_full} presents the complete RAG evaluation results across all 13 domains.

\begin{table}[t]
\centering

\resizebox{0.5\textwidth}{!}{%
\begin{tabular}{lccccc}
\toprule
\textbf{Domain} & \textbf{None} & \textbf{Oracle} & \textbf{BM25} & \textbf{DiVeR} & \textbf{BGE} \\
\midrule
\rowcolor{gray!12}\multicolumn{6}{c}{\textit{Blockchain}} \\
\midrule
Bitcoin & 79.4 & \textbf{82.7} & 79.3 & 79.3 & 78.1 \\
Cardano & 77.5 & \textbf{80.1} & 69.8 & 73.9 & 69.2 \\
Iota & 76.0 & \textbf{78.0} & 67.0 & 70.0 & 72.0 \\
Monero & 76.3 & \textbf{80.2} & 70.3 & 70.5 & 72.0 \\
\midrule
\rowcolor{gray!12}\multicolumn{6}{c}{\textit{Social Sciences}} \\
\midrule
Economics & 80.8 & \textbf{82.6} & 79.0 & 80.7 & 80.0 \\
Law & 80.9 & \textbf{83.1} & 76.3 & 78.0 & 74.0 \\
Politics & 71.8 & \textbf{76.4} & 68.8 & 69.7 & 72.1 \\
History & 73.3 & \textbf{76.9} & 70.5 & 72.1 & 72.2 \\
\midrule
\rowcolor{gray!12}\multicolumn{6}{c}{\textit{Applied}} \\
\midrule
Quant & 79.1 & \textbf{83.9} & 77.9 & 79.1 & 76.5 \\
Travel & 79.5 & \textbf{82.8} & 76.2 & 78.5 & 77.9 \\
Workplace & 81.4 & \textbf{84.0} & 78.9 & 80.8 & 80.0 \\
Genealogy & 80.1 & \textbf{82.0} & 78.9 & 79.8 & 79.0 \\
\midrule
\rowcolor{gray!12}\multicolumn{6}{c}{\textit{STEM}} \\
\midrule
HSM & 69.5 & \textbf{74.2} & 66.6 & 70.9 & 69.2 \\
\midrule
\midrule
\textbf{Average} & 77.3 & \textbf{80.5} & 73.8 & 75.6 & 74.8 \\
\bottomrule
\end{tabular}}
\caption{RAG performance (answer correctness score, 0--100) across all domains and retrievers. \textbf{None}: no retrieval (parametric knowledge only); \textbf{Oracle}: gold documents provided. Best scores per domain in \textbf{bold}.}
\label{tab:rag_results_full}
\end{table}

\subsubsection{Answer Generation Prompt}

For answer generation, we prompt Llama-3-70B-Instruct with retrieved documents as context:

\begin{figure*}[!t]
\centering
\begin{tcolorbox}[
  colback=gray!5,
  colframe=gray!75,
  title=Prompt for Answer Generation,
  width=\textwidth,
]
\small\ttfamily
You are a helpful assistant. Answer the following question based on the provided context documents.\\[0.5em]
\textbf{Context Documents:}\\
\{retrieved\_documents\}\\[0.5em]
\textbf{Question:}\\
\{query\}\\[0.5em]
\textbf{Answer:}
\end{tcolorbox}
\caption{RAG answer generation prompt. The generator is instructed to answer the user query based strictly on the provided retrieved context documents.}
\label{prompt:rag_generation}
\end{figure*}

\subsubsection{Answer Evaluation Prompt}

Following BRIGHT~\citep{su2024bright}, we use GPT-4o as an evaluator to score answer correctness on a 0--100 scale based on coverage of the reference answer.

\begin{figure*}[!t]
\centering
\begin{tcolorbox}[
  colback=gray!5,
  colframe=gray!75,
  title=Prompt for Answer Evaluation (GPT-4o),
  width=\textwidth,
]
\small\ttfamily
You are a teacher to judge student's answer.\\[0.5em]
---------- PROBLEM START ----------\\
\{query\}\\
---------- PROBLEM END ----------\\[0.5em]
---------- STUDENT ANSWER START ----------\\
\{predicted\_answer\}\\
---------- STUDENT ANSWER END ----------\\[0.5em]
---------- REFERENCE ANSWER START ----------\\
\{gold\_answer\}\\
---------- REFERENCE ANSWER END ----------\\[0.5em]
\textbf{Criteria:}\\
\begin{itemize}
\item 0  The student's answer is completely irrelevant or blank.
\item 10 The student's answer addresses about 10\% of the reference content.
\item 20 The student's answer addresses about 20\% of the reference content.
\item 30 The student's answer addresses about 30\% of the reference content.
\item 40 The student's answer addresses about 40\% of the reference content.
\item 50 The student's answer addresses about 50\% of the reference content.
\item 60 The student's answer addresses about 60\% of the reference content.
\item 70 The student's answer addresses about 70\% of the reference content.
\item 80 The student's answer addresses about 80\% of the reference content.
\item 90 The student's answer addresses about 90\% of the reference content.
\item 100 The student's answer addresses about 100\% of the reference content.
\end{itemize}
~\\
Use the following format to give a score:\\[0.5em]
\textbf{REASON:}\\
Describe why you give a specific score\\[0.5em]
\textbf{SCORE:}\\
The score you give, e.g., 60\\[0.5em]
Do not say anything after the score.
\end{tcolorbox}
\caption{RAG answer evaluation prompt. The judge scores the generated answer (0--100) based on semantic coverage of the reference gold answer.}
\label{prompt:rag_evaluation}
\end{figure*}

\clearpage

\section{Extended Related Work}
\label{app:benchmark_comparison_1}

This appendix provides a comprehensive review of the literature surrounding temporal information retrieval, temporal question answering (QA), and reasoning-intensive retrieval. We contextualize TEMPO within the recent surge of interest in temporal reasoning for Large Language Models (LLMs) and Retrieval-Augmented Generation (RAG).

\subsection{Temporal Information Retrieval}
Temporal Information Retrieval (TIR) has long been recognized as a critical subfield of IR. Early surveys~\citep{campos2014survey} and tasks like NTCIR Temporalia~\citep{joho2014ntcir} established the importance of classifying queries by temporal intent (e.g., past, future, atemporal). However, these benchmarks primarily relied on news and blog corpora where temporal reasoning was limited to identifying explicit timestamps or "recency" signals~\citep{joho2016building}. 

A recent survey by \citet{piryani2025s} highlights that while traditional TIR focused on document re-ranking based on publication dates, modern challenges require "content-aware" temporal reasoning that goes beyond metadata. TEMPO addresses this by requiring systems to retrieve documents based on semantic temporal alignment (e.g., "post-GDPR era" vs. "pre-2018") rather than simple timestamp filtering.

\subsection{Temporal Question Answering}
The field of Temporal QA has seen rapid development, shifting from simple factoid questions to complex reasoning tasks.
\begin{itemize}
    \item \textbf{Fact-Centric Benchmarks:} Early datasets like TempQuestions~\citep{jia2018tempquestions} and TimeQA~\citep{chen2021dataset} focused on knowledge base (KB) lookup. More recent large-scale efforts include ComplexTempQA~\citep{gruber-etal-2025-complextempqa}, which offers 100 million synthetic questions over Wikipedia, and HistoryBankQA~\citep{mandal2025historybankqa}, which targets historical events across multiple languages.
    \item \textbf{LLM-Oriented Benchmarks:} The TIME benchmark~\citep{wei2025time} evaluates LLMs on explicit temporal reasoning tasks across text, news, and dialogue. Similarly, TimelineQA~\citep{tan2023timelineqa} tests an agent's ability to construct timelines from single-document contexts.
\end{itemize}
\textbf{Limitation:} As noted in recent systematic reviews~\citep{brown2025systematic}, these benchmarks primarily evaluate the final generated answer. They do not penalize systems that hallucinate the correct date or retrieve temporally irrelevant documents (e.g., retrieving 2020 data to answer a 2010 question) as long as the final output is correct. TEMPO uniquely evaluates the intermediate retrieval step, ensuring that the evidence itself is temporally valid.

\subsection{Reasoning-Intensive Retrieval}
A parallel trend is the emergence of "Reasoning-Intensive Retrieval," where the difficulty lies in the logic rather than the keyword overlap.
\begin{itemize}
    \item \textbf{Atemporal Reasoning:} BRIGHT~\citep{su2024bright} and RAR-b~\citep{xiao2024rar} demonstrated that state-of-the-art retrievers fail on queries requiring logical deduction, coding, or economic reasoning.
    \item \textbf{Reasoning Models:} New architectures like ReasonIR~\citep{shao2025reasonir}, RaDeR~\citep{das2025rader}, and query-reasoning reinforcers~\citep{qin2025reinforced} have been proposed to bridge this gap.
\end{itemize}
However, these benchmarks lack a temporal dimension. A query in BRIGHT might ask about a static code function, whereas a query in TEMPO asks how that function's security vulnerabilities evolved over five years. TEMPO is the first to combine the logical complexity of BRIGHT with the temporal constraints of TimeQA.

\subsection{Temporal RAG and Knowledge Graphs}
The intersection of Temporal Reasoning and RAG is a nascent field. ChronoQA~\citep{chen2025question} specifically targets "temporal-sensitive RAG," but it is restricted to Chinese news articles and focuses on recency bias updates. In the Knowledge Graph (KG) domain, architectures like TimeR4~\citep{qian2024timer4} and TempoQR~\citep{mavromatis2022tempoqr} attempt to embed temporal scopes into graph nodes. Recent frameworks like CRP-RAG~\citep{xu2024crp} attempt to plan "knowledge actions" to answer complex logical queries.

TEMPO complements these works by providing a domain-diverse, English-language testbed that requires cross-period analysis—synthesizing evidence from a baseline period (e.g., "pre-2008 crisis") and a comparison period (e.g., "post-2010 recovery")—a capability not explicitly measured by existing temporal RAG benchmarks.

\section{Dataset Examples}
\label{app:dataset_examples}

\begin{table*}[t]
\caption{\textbf{History example.} A randomly sampled query with one positive and one negative document.}
\centering
\begin{tabular}{p{0.97\textwidth}}
\toprule
\textit{\textbf{Query}} \\
\midrule
Was the Tsar's property separated from state property in Russian Empire in early 19th century (regarding land)?\\ 
Was the Tsar's property on land separated from state property in Russian Empire in early 19th century? I mean, were there Tsar's serfs who were not state serfs? \\
\midrule
\textit{\textbf{Example positive document}} \\
\midrule
The appanage peasants lived on the personal properties of the Romanov family; Alexander II granted them personal freedom in 1863. They received land allotments in 1863 and were placed on forty-nine-year redemption payments in 1865. The state peasants lived on state lands under state administrators; they received freedom in 1866.\\ 
The core "freedom" the peasants received was the elimination of the personal, arbitrary, and capricious power of their noble and state masters. Members of the noble landowning estate and the tsar's agents could no longer buy and sell peasants, mortgage them for cash, order their daily labors, determine whom and when they married, move them from one estate to another, break up families, beat them, claim sexual rights over them, exile them to Siberia, impose both police and judicial authority over them, demand that they gather forest products such as berries for their masters' larders, or decide who would enter military service for virtually their entire adult lives.\\ 
The emancipation legislation involved a land reform that transferred as much as half of the nobility's land to the peasants. The reformers tried to design this transfer so that it would not cause dangerous instability in the countryside. They also tried to soften the economic blows to the nobility and to guarantee that peasants would continue to produce crops and pay their taxes. These aims\\ 
...\\ 
 \\
\midrule
\textit{\textbf{Example negative document}} \\
\midrule
By an odd twist of fate, defeat in the war proved of value to the new Tsar. Although he had been trained for government from an early age, foreign observers had remarked on how diffident and unsure he appeared. The war changed all that. Coming to the throne in 1855 in the middle of the conflict, Alexander II was unable to save Russia from military failure, but the humiliation convinced him that, if his nation was to have stability and peace at home and be honoured abroad, military and domestic reforms were vitally necessary. The first step on that path would be the removal of serfdom, whose manifest inefficiency benefited neither lord, peasant, nor nation. Alexander declared that, despite Russia’s defeat, the end of the war marked a golden moment in the nation’s history. Now was the hour when every Russian, under the protection of the law, could begin to enjoy ‘the fruits of his own labours’. Alexander was right in thinking the time was propitious. It had long been appreciated that some land reform was necessary. To the social and economic arguments were now added powerful military ones. The army was the great symbol of Russia’s worth. As long as its army remained strong Russia could afford to ignore its backwardness as a nation. But the Crimean defeat had undermined this notion of Russia’s invincibility. Few now had reasoned objections to reform. Serfdom was manifestly not wor\\ 
...\\ 
 \\
\bottomrule
\end{tabular}
\label{tab:history_example}
\end{table*}

\begin{table*}[t]
\caption{\textbf{Hsm example.} A randomly sampled query with one positive and one negative document.}
\centering
\begin{tabular}{p{0.97\textwidth}}
\toprule
\textit{\textbf{Query}} \\
\midrule
What is the most ancient milestone of mathematical reasoning or mathematical knowledge?\\ 
I know about the Plimpton 322 tablet and Pythagorean triples. But can we be sure that this is the first instance of mathematical reasoning? I am talking about notions, propositions, or  questions about mathematics.\\ 
I ask you if you can say what the milestone of mathematical reasoning or mathematical knowledge is, as accepted by the scientific community.\\ 
I have two ideas as to what ancient notions of mathematics could be like. Let's say that humans were a band of Paleolithic hunters or gatherers. If the group gets \$N\$ items of food, then, leaving out social synergies, the appropriate distribution is the Euclidean division between the participants. If there are paintings of two animals in a cave, does this mean that Paleolithic men/women had the notion of the integer \$2\$? Are there scientific thoughts\\ 
...\\ 
 \\
\midrule
\textit{\textbf{Example positive document}} \\
\midrule
It consists of 29 distinct notches that were deliberately cut into a baboon’s fibula.\\ 
The bone is between 44,200 and 43,000 years old, according to 24 radiocarbon datings. This is far older than the Ishango bone with which it is sometimes confused. Other notched bones are 80,000 years old but it is unclear if the notches are merely decorative or if they bear a functional meaning.\\ 
According to The Universal Book of Mathematics, the Lebombo bone’s 29 notches “may have been used as a lunar phase counter, in which case African women may have been the first mathematicians, because keeping track of menstrual cycles requires a lunar calendar.” However, the bone is clearly broken at one end, so the 29 notches may or may not be a minimum number. In the cases of other notched bones since found globally, there has been no consistent notch tally, many being in the 1–10 range. The Lebombo bone resembles a calendar used by the early men of the area, coming from the San clans of Namibia; this way of making tallies is still used by the San people today.\\ 
Lebombo Ishango bones\\ 
Top image: Lebombo bone. Bottom: Ishango bone with prime numbers engraving (J.D. Loreto and D.H. Hurlbert Smithsonian)\\ 
According to The Universal Book of Mathematics, the Lebombo bone’s 29 notches “may have been used as a lunar phase counter, in which case African women may have been the first mathematicians, because ke\\ 
...\\ 
 \\
\midrule
\textit{\textbf{Example negative document}} \\
\midrule
The discovery made by Otto Neugebauer and his assistant in the 1940s was an important one. The numbers in Plimpton 322 are what are now called Pythagorean triples. It gives the short side and the diagonal (hypotenuse) of 15 right triangles. The long sides of the right triangles are not shown. As we will see below, the 15 right triangles have steadily decreasing slopes. The Sumerians in the Old Babylonian period knew about the Pythagorean theorem over 1,000 years before the time of Pythagoras! \\

Since the discovery made by Otto Neugebauer, Plimpton 322 was a subject of extensive research by mathematicians. Obviously mathematicians are intrigued by the connection of a 4000-year tablet with modern mathematics. Because of the intricate mathematical interpretations they made of the tablet, many mathematicians thought highly of the tablet. For example, the author of the tablet must be a mathematical prodigy or a professional mathematician, doing high level research in the Old Babylonian Period.\\

\bottomrule
\end{tabular}
\label{tab:hsm_example}
\end{table*}

\begin{table*}[t]
\caption{\textbf{Politics example.} A randomly sampled query with one positive and one negative document.}
\centering
\begin{tabular}{p{0.97\textwidth}}
\toprule
\textit{\textbf{Query}} \\
\midrule
What was the most succesful military dictatorship in the last 200 years in terms of economic growth?\\ 
Military dictatorships are controlled by military officers. I am wondering which military dictatorship was the most economically successful, you can use any figure like GDP growth, GDP per capita, GDP (PPP), etc. to base your argument. I don't think the current Chinese government was a military dictatorship. Maybe Mao's government, but not the government that followed, so I am guessing China wouldn't count. \\
\midrule
\textit{\textbf{Example positive document}} \\
\midrule
Park began a series of economic reforms that eventually led to rapid and unprecedented economic growth and industrialization, a phenomenon that is now known as the Miracle on the Han River. This made South Korea one of the fastest growing economies of the 1960s and 1970s, albeit with costs to labor rights. This era also saw the formation of chaebols: family companies supported by the state similar to the Japanese zaibatsu. Examples of significant chaebols include Hyundai, LG, and Samsung.\\ 
Although popular during the 1960s, Park's popularity started to plateau by the 1970s, with closer than expected victories during the 1971 presidential election and the subsequent legislative elections. In 1972, Park declared martial law after carrying out a self-coup. He then introduced the highly authoritarian Yushin Constitution, ushering in the Fourth Republic. Now ruling as a dictator, he constantly repressed political opposition and dissent and completely controlled the military. He also had much control over the media and expressions of art. In 1979, Park was assassinated by his close friend Kim Jae-gyu, director of the KCIA, following the Busan–Masan Uprising.[2] Whether the assassination was spontaneous or premeditated remains unclear to this day. Economic growth continued in spite of the 1979 coup d'état and considerable political turmoil in the wake of his assassination. He was soon\\ 
...\\ 
 \\
\midrule
\textit{\textbf{Example negative document}} \\
\midrule
Determining adequate levels of military spending and sustaining the burden of conflicts have been among key fiscal problems in history. Ancient societies were usually less complicated in terms of the administrative, fiscal, technological, and material demands of warfare. The most pressing problem was frequently the adequate maintenance of supply routes for the armed forces. On the other hand, these societies were by and large subsistence societies, so they could not extract massive resources for such ventures, at least until the arrival of the Roman and Byzantine Empires. The emerging nation states of the early modern period were much better equipped to fight wars. On the one hand, the frequent wars, new gunpowder technologies, and the commercialization of warfare forced them to consolidate resources for the needs of warfare. On the other hand, the rulers had to – slowly but surely – give up some of their sovereignty to be able to secure required credit both domestically and abroad. The Dutch and the British were masters at this, with the latter amassing an empire that spanned the globe at the eve of the First World War.
 \\
\bottomrule
\end{tabular}
\label{tab:politics_example}
\end{table*}

\begin{table*}[t]
\caption{\textbf{Hsm example.} A randomly sampled query with one positive and one negative document.}
\centering
\begin{tabular}{p{0.97\textwidth}}
\toprule
\textit{\textbf{Query}} \\
\midrule
How accurate was the measurement of the period of Earth's orbit in the 19th Century?\\ 
There was a section on my textbook on history of theories of sun's energy source.\\ 
It talks about how the Meteorite Theory was dismissed, as it would decrease the period of Earth's orbit by 2 seconds per year due to increased mass of the Sun.\\ 
This theory was dismissed due to disagreeing with observation. And the textbook says the change in period is "easily measurable"\\ 
My question is how is 2 seconds difference easily measurable in Nineteenth Century? \\
\midrule
\textit{\textbf{Example positive document}} \\
\midrule
In the nineteenth century, scientists thought that the source of the Sun’s heat might be the mechanical motion of meteorites falling into it. Their calculations showed, however, that in order to produce the total amount of energy emitted by the Sun, the mass in meteorites that would have to fall into the Sun every 100 years would equal the mass of Earth. The resulting increase in the Sun’s mass would, according to Kepler’s third law, change the period of Earth’s orbit by 2 seconds per year. Such a change would be easily measurable and was not, in fact, occurring. Scientists could then disprove this as the source of the Sun’s energy.\\ 
Gravitational Contraction as a Source of Energy\\ 
Proposing an alternative explanation, British physicist Lord Kelvin and German scientist Hermann von Helmholtz (Figure 16.2), in about the middle of the nineteenth century, proposed that the Sun might produce energy by the conversion of gravitational energy into heat. They suggested that the outer layers of the Sun might be “falling” inward because of the force of gravity. In other words, they proposed that the Sun could be shrinking in size, staying hot and bright as a result.\\ 
Kelvin (1824–1907) and Helmholtz (1821–1894).\\ 
Left: photograph of William Thomson (Lord Kelvin). Right: photograph of Hermann von Helmholtz.\\ 
Figure 16.2. (a) British physicist William Thomson (Lord Kelvin) and (b) German scien\\ 
...\\ 
 \\
\midrule
\textit{\textbf{Example negative document}} \\
\midrule
The Principia is the founding document of physics and astronomy as we know them, and it played a key role in the scientific revolution four centuries ago. For two centuries afterward, mathematicians worked out the details of Newtonian mechanics, which led to determinism. The rise of modern physics in the early twentieth century undermined determinism, leading to indeterminism. In the seventeenth century, the so-called Enlightenment hijacked science, robbing it of its foundation in a worldview that had biblical elements, replacing it with a foundation of humanism. This has led to a growing hostility toward any concern of theism in scientific endeavors. The trends undermine the worldview that created science in the first place. Consequently, the future of science may be in question. Keywords: philosophy of science, determinism, deism, positivism, conflict thesis, special relativity, general relativity, quantum mechanics, Copenhagen interpretation \\
\bottomrule
\end{tabular}
\label{tab:hsm_example2}
\end{table*}

\clearpage

\section{Annotation Guidelines}
\label{app:annotation_guidelines}

This section provides detailed guidelines for annotators constructing the \textsc{TEMPO} dataset. The annotation process involves selecting StackExchange posts with temporal reasoning requirements, identifying temporally relevant documents, mining hard negatives, and annotating temporal metadata at multiple levels.

\subsection{Query Selection and Filtering}
\label{app:query_selection}

Annotators browse Stack Exchange posts from newest to oldest within their assigned domain and select posts meeting ALL of the following criteria:

\textbf{Required Criteria:}
\begin{enumerate}
    \item \textbf{High-quality answer}: The post must have at least one answer that is either:
    \begin{itemize}
        \item Accepted by the question author (marked with green checkmark), OR
        \item Has received more than 10 upvotes
    \end{itemize}
    
    \item \textbf{Temporal reasoning requirement}: The post must require temporal reasoning to answer. This includes questions that:
    \begin{itemize}
        \item Track changes or evolution over time (e.g., ``How has X changed since Y?'')
        \item Compare historical baselines with current states (e.g., ``What was different before/after Z?'')
        \item Require understanding temporal dependencies or causation
        \item Ask about trends, patterns, or developments across time periods
        \item Need cross-period evidence synthesis to answer
    \end{itemize}
    
    \item \textbf{Technical complexity}: The question requires reasoning beyond simple date lookup or keyword matching. Avoid questions that can be answered by retrieving a single date or simple fact.
    
    \item \textbf{Temporal signals}: The post should contain explicit or implicit temporal signals such as:
    \begin{itemize}
        \item Specific dates, years, or time periods (e.g., ``in 1914'', ``during the 1990s'')
        \item Relative temporal references (e.g., ``before the war'', ``after GDPR'', ``since 2017'')
        \item Temporal keywords (e.g., ``evolution'', ``history'', ``change'', ``trend'', ``origin'')
    \end{itemize}
\end{enumerate}

\textbf{Exclusion Criteria:}
\begin{itemize}
    \item Posts with only simple fact-seeking queries (e.g., ``When did X happen?'')
    \item Questions where temporal aspects are merely supplementary, not central to the answer
    \item Opinion-based or subjective questions without temporal grounding
    \item Posts where answers rely purely on speculation without temporal evidence
    \item Duplicate or near-duplicate questions
    \item Questions that can be answered by retrieving a single Wikipedia date entry
\end{itemize}

\subsection{Constructing Queries}
\label{app:constructing_queries}

For each selected post, construct the temporal query as follows:

\textbf{Step 1: Extract text content}
\begin{itemize}
    \item Combine the post title and body text
    \item Preserve HTML formatting where it aids readability (lists, emphasis)
    \item Retain technical terminology, historical references, and temporal expressions
    \item Remove broken links and irrelevant HTML artifacts
\end{itemize}

\textbf{Step 2: Verify temporal complexity}
\begin{itemize}
    \item Confirm the query requires multi-step temporal reasoning
    \item Identify the temporal scope (specific years, decades, centuries, or relative periods)
    \item Determine if cross-period analysis is needed (comparing multiple time periods)
    \item Note key temporal anchors mentioned in the query
\end{itemize}

\textbf{Step 3: Classify temporal characteristics}
\begin{itemize}
    \item Identify the primary temporal intent (when/duration/order/before\_after/ongoing\allowbreak\_status/period\_definition/timeline)
    \item Assign the temporal reasoning class (see \S\ref{app:reasoning_classes})
    \item Extract explicit temporal signals and events from the query text
\end{itemize}

\subsection{Positive Document Construction}
\label{app:positive_docs}

Positive documents must provide \textit{temporal evidence} that helps reason through the query's temporal aspects. Follow these steps:

\textbf{Step 1: Discover candidate documents}

Use TWO methods to find candidate documents:

\textbf{Method A - Answer links:}
\begin{itemize}
    \item Visit all external URLs linked in accepted or highly-voted answers
    \item Check if the linked page contains temporal information relevant to the query
    \item Prioritize sources that discuss the time periods mentioned in the query
\end{itemize}

\textbf{Method B - AI-assisted discovery:}
\begin{itemize}
    \item Use Gemini (Google AI) with the following prompt template:
    \item[] \textit{``Give me articles from the internet to answer this temporal query: [paste full question text]. Focus on sources that discuss the time periods and temporal evolution mentioned.''}
    \item Visit suggested web pages and evaluate their temporal relevance
\end{itemize}

\textbf{Step 2: Evaluate temporal relevance}

For each candidate web page, extract passages that meet the temporal relevance criteria:

\textbf{A document/passage is POSITIVE if it:}
\begin{itemize}
    \item \textbf{Provides baseline temporal evidence}: Contains information about the historical state or starting point referenced in the query
    \item \textbf{Provides comparison temporal evidence}: Contains information about the later state or endpoint for cross-period queries
    \item \textbf{Explains temporal evolution}: Describes how phenomena changed, developed, or evolved over the relevant time periods
    \item \textbf{Contains temporal context}: Provides historical background necessary to understand temporal relationships
    \item \textbf{Includes temporal synthesis}: Helps connect evidence across multiple time periods to form a complete answer
\end{itemize}

\textbf{A document/passage is NOT positive if it:}
\begin{itemize}
    \item Discusses the topic but lacks temporal information or covers wrong time periods
    \item Only mentions dates without explaining temporal relationships or changes
    \item Provides general background without addressing the specific temporal scope of the query
    \item Covers only one time period when the query requires cross-period comparison
    \item Contains anachronistic information that doesn't match the query's temporal focus
\end{itemize}

\textbf{Step 3: Extract passages with temporal annotations}
\begin{itemize}
    \item For each positive web page, identify and extract temporally relevant passages
    \item Each passage should be self-contained (typically 1--5 paragraphs)
    \item Annotate each passage with:
    \begin{itemize}
        \item Temporal signals present in the passage
        \item Time scope (start and end dates in ISO format when determinable)
        \item Temporal events mentioned
        \item Dominant tense (past/present/future/mixed)
    \end{itemize}
    \item Record confidence score (0--1) for temporal annotations
\end{itemize}

\textbf{Step 4: Ensure temporal coverage}
\begin{itemize}
    \item For cross-period queries, ensure positive documents cover BOTH baseline and comparison periods
    \item Verify that the combined positive documents provide sufficient temporal evidence to answer the query
    \item If temporal coverage is incomplete, search for additional documents
\end{itemize}

\textbf{Step 5: Record metadata}
\begin{itemize}
    \item Source URL of the document
    \item Type of source (Wikipedia, academic article, news archive, official documentation, blog, etc.)
    \item Date accessed
    \item Temporal scope covered by the document
    \item Brief justification for temporal relevance
\end{itemize}

\subsection{Hard Negative Mining}
\label{app:hard_negative_mining}

Hard negatives for TEMPO are documents that are \textit{topically related but temporally incomplete or irrelevant}. These prevent models from relying on simple semantic matching without temporal understanding.

\textbf{Types of temporal hard negatives:}
\begin{enumerate}
    \item \textbf{Wrong time period}: Documents discussing the same topic but in a different time period than the query requires
    \item \textbf{Missing temporal coverage}: Documents covering only one time period when cross-period analysis is needed
    \item \textbf{Temporally vague}: Documents discussing the topic without specific temporal grounding
    \item \textbf{Anachronistic}: Documents with temporal information that doesn't align with the query's temporal scope
\end{enumerate}

\textbf{Step 1: Generate hard negative search query}

\begin{itemize}
    \item Use the GPT-4o prompt (Appendix~\ref{app:negative_mining_prompt}) to generate:
    \begin{enumerate}
        \item A search query designed to find semantically similar but temporally incomplete content
        \item List of entities, events, and temporal anchors from the post
    \end{enumerate}
    \item The LLM outputs JSON with \texttt{llm\_summary} and \texttt{entities\_events}
\end{itemize}

\textbf{Step 2: Collect hard negative URLs}

\begin{itemize}
    \item Use the generated \texttt{llm\_summary} as your Google search query
    \item Additionally search using combinations of \texttt{entities\_events} WITHOUT temporal qualifiers
    \item Collect URLs that are:
    \begin{itemize}
        \item Topically related to the query domain
        \item Semantically similar to the query content
        \item BUT missing the specific temporal information or time periods needed
    \end{itemize}
\end{itemize}

\textbf{Step 3: Extract hard negative passages}

For each hard negative URL:
\begin{itemize}
    \item Extract passages that are topically related but temporally inadequate
    \item Hard negatives should be challenging---they might discuss the same general topic but:
    \begin{itemize}
        \item Cover a different time period than required
        \item Lack temporal specificity needed to answer the query
        \item Miss one of the required time periods for cross-period queries
        \item Discuss temporal aspects tangentially without depth
    \end{itemize}
    \item Avoid completely unrelated content
\end{itemize}

\textbf{Step 4: Verify hard negative quality}
\begin{itemize}
    \item Confirm hard negatives are semantically similar (would rank highly with keyword matching)
    \item Verify they fail temporal requirements (wrong period, incomplete coverage, or temporally vague)
    \item Ensure a mix of hard negative types for diversity
\end{itemize}

\subsection{Step-wise Retrieval Planning}
\label{app:stepwise_planning}

For Task 2 evaluation, decompose each query into sequential retrieval steps:

\textbf{Step 1: Analyze temporal structure}
\begin{itemize}
    \item Identify distinct temporal aspects or time periods in the query
    \item Determine the logical order of retrieval (e.g., baseline first, then comparison)
    \item Note dependencies between retrieval steps
\end{itemize}

\textbf{Step 2: Create retrieval steps}
\begin{itemize}
    \item Each step should target a specific temporal aspect or time period
    \item Steps should be concrete and actionable (e.g., ``Retrieve historical statistics from 2013--2017'')
    \item Typically 2--4 steps per query
    \item Example step structure:
    \begin{itemize}
        \item Step 1: Retrieve baseline evidence from [time period 1]
        \item Step 2: Retrieve comparison evidence from [time period 2]
        \item Step 3: Retrieve documents explaining the change/evolution
    \end{itemize}
\end{itemize}

\textbf{Step 3: Map gold documents to steps}
\begin{itemize}
    \item Assign each positive document to the retrieval step(s) it satisfies
    \item A document may satisfy multiple steps if it covers multiple time periods
    \item Ensure each step has at least one mapped gold document
\end{itemize}

\subsection{Temporal Reasoning Classes}
\label{app:reasoning_classes}

Assign each query to one primary temporal reasoning class:

\begin{enumerate}
    \item \textbf{Event Analysis \& Localization (EAL)}: Pinpointing when events occurred and understanding their temporal context
    
    \item \textbf{Time Period Contextualization (TPC)}: Situating phenomena within specific historical periods
    
    \item \textbf{Origins \& Evolution Comparative (OEC)}: Tracking how concepts evolved over time
    
    \item \textbf{Trends \& Cross-Period Comparison (TCP)}: Comparing states across multiple time periods
    
    \item \textbf{Event Verification \& Authenticity (EVA)}: Verifying temporal claims or dating artifacts
    
    \item \textbf{Materials \& Artifacts Provenance (MAP)}: Dating and tracing origins of physical items
    
    \item \textbf{Sources \& Methods Documentation (SMD)}: Understanding historical methodology and sources
    
    \item \textbf{Causation Analysis (CAU)}: Analyzing temporal cause-effect relationships
    
    \item \textbf{Historical Attribution \& Context (HAC)}: Attributing ideas or events to correct time periods
\end{enumerate}

\subsection{Quality Control and Review}
\label{app:qc_review}

\textbf{Self-check before submission:}
\begin{enumerate}
    \item Does the selected post genuinely require temporal reasoning (not just date lookup)?
    \item Did you verify that answers have $>10$ votes or are accepted?
    \item Do positive documents provide temporal evidence for the required time periods?
    \item For cross-period queries, do positive documents cover BOTH baseline and comparison periods?
    \item Are hard negatives temporally inadequate but semantically similar?
    \item Is the retrieval plan logical and are gold documents correctly mapped to steps?
    \item Are all temporal annotations (signals, events, time scope) accurate?
\end{enumerate}

\textbf{Annotation review process:}
\begin{itemize}
    \item Initial annotations are reviewed by two PhD students with domain expertise
    \item Reviewers check: (1) temporal relevance of positive documents, (2) temporal inadequacy of hard negatives, (3) accuracy of temporal annotations, (4) validity of retrieval plans
    \item Inter-annotator agreement measured using Cohen's Kappa
    \item Only annotations with \textbf{unanimous approval} from all reviewers are retained
    \item Disagreements are resolved through discussion or the example is discarded
\end{itemize}

\textbf{Common Mistakes to Avoid:}

\begin{enumerate}
    \item \textbf{Selecting simple date-lookup queries}: The query must require temporal reasoning, not just retrieving when something happened
    
    \item \textbf{Incomplete temporal coverage}: For cross-period queries, ensure positive documents cover all required time periods
    
    \item \textbf{Too-easy temporal negatives}: Hard negatives should be semantically similar but fail on temporal grounds---don't include completely off-topic documents
    
    \item \textbf{Ignoring temporal scope}: Ensure positive documents cover the specific time periods in the query, not just any temporal information
    
    \item \textbf{Incorrect temporal annotations}: Double-check time scope annotations, especially ISO date formats
    
    \item \textbf{Illogical retrieval plans}: Steps should follow a natural temporal progression and be independently actionable
    
    \item \textbf{Missing step-document mappings}: Every retrieval step must have at least one gold document mapped to it
\end{enumerate}

\subsection{Domain-Specific Annotation Notes}
\label{app:domain_notes}

\textbf{For Blockchain domains (Bitcoin, Cardano, Iota, Monero):}
\begin{itemize}
    \item Temporal focus typically on protocol evolution, market changes, and technology updates
    \item Time periods often span 2009--present with rapid changes
    \item Hard negatives can discuss similar blockchain concepts but from different protocol versions or time periods
    \item Sources include whitepapers, technical documentation, and cryptocurrency news archives
\end{itemize}

\textbf{For Social Sciences (Economics, Law, Politics, History):}
\begin{itemize}
    \item Temporal scope may span centuries; ensure positive documents match the query's era
    \item Cross-period queries common (e.g., comparing policies before/after major events)
    \item Prefer academic sources, government documents, and authoritative analyses
    \item Hard negatives often discuss same topic but in wrong historical period
\end{itemize}

\textbf{For Applied domains (Quant, Travel, Workplace, Genealogy):}
\begin{itemize}
    \item Temporal aspects often relate to policy changes, regulation updates, or practice evolution
    \item For Genealogy, temporal accuracy is critical---verify historical dates carefully
    \item Hard negatives may discuss similar practices but from different eras
    \item Balance between academic sources and practical documentation
\end{itemize}

\textbf{For STEM (History of Science and Mathematics):}
\begin{itemize}
    \item Focus on evolution of scientific ideas and mathematical concepts
    \item Attribution to correct time periods is crucial
    \item Positive documents should explain how ideas developed temporally
    \item Hard negatives might discuss same concepts but misattribute time periods
    \item Prefer peer-reviewed history of science literature
\end{itemize}

\clearpage



\end{document}